\documentclass[12pt]{article}
\usepackage{amssymb}
\usepackage{graphicx,epstopdf}
\usepackage{amsthm}
\usepackage{tikz}
\usepackage{subfig}
\RequirePackage[hyphens]{url}
\RequirePackage[colorlinks,citecolor=blue,urlcolor=blue]{hyperref}
\usepackage[authoryear]{natbib}
\usepackage[lined,algonl,boxed]{algorithm2e}
\usepackage{paralist}
\usepackage{url}
\usepackage{amsmath}
\usepackage{amsfonts}
\usepackage{color}
\usepackage{afterpage}
\usepackage{lineno}
\usepackage{mathtools}
\usepackage[utf8]{inputenc}
\usepackage[english]{babel}
\usepackage{enumitem}
 \usepackage{multirow}
 
\newtheorem{theorem}{Theorem}[section]
\newtheorem{corollary}{Corollary}[theorem]
\newtheorem{lemma}{Lemma}[section]
\newcommand{\nti}{n\rightarrow\infty}
\def\qmq#1{{\quad\mbox{#1}\quad}}
\newcommand{\raw}{\rightarrow}
\newcommand{\qb}{{\hfill$\Box$}}
\newcommand{\ind}{{1\!\!1}}
\newcommand{\sa}{^{0}}
\newcommand{\ci}{\iota}
\newcommand{\aal}{\gamma}
\newcommand{\om}{\omega}
\newcommand{\bbr}{{\mathbb  R}}
\newcommand{\bfc}{{\bf c}}
\newcommand{\bfd}{{\bf d}}
\newcommand{\bfZ}{{\bf Z}}
\newcommand{\C}{{\cal C}}
\newcommand{\F}{{\cal F}}
\newcommand{\G}{{\cal G}}
\newcommand{\hc}{\hat{c}}

\newcommand{\hn}{\hat{n}}
\newcommand{\hP}{\hat{P}}

\newcommand{\al}{\alpha}
\newcommand{\ep}{\epsilon}
\newcommand{\ttau}{\tilde{\tau}}

\newtheorem{remark}{\n{Remark}}[section]

\font\n=cmcsc10

\usepackage{url} % not crucial - just used below for the URL 

\newcommand{\blind}{1}

\begin{document}

\def\spacingset#1{\renewcommand{\baselinestretch}%
{#1}\small\normalsize} \spacingset{1}

%%%%%%%%%%%%%%%%%%%%%%%%%%%%%%%%%%%%%%%%%%%%%%%%%%%%%%%%%%%%%%%%%%%%%%%%%%%%%%

\if1\blind
{
  \title{\bf A Bootstrap-based Method for Testing Similarity of Matched Networks}
  \author{Somnath Bhadra\hspace{.2cm}\\
  Department of Statistics, University of Florida\\
  Kaustav Chakraborty\\
  Department of Statistics, University of Illinois at Urbana-Champaign\\
  Srijan Sengupta\\
    Department of Statistics, North Carolina State University\\
    Soumendra N. Lahiri\\
    Department of Mathematics and Statistics, Washington University in St. Louis
    }
  \maketitle
} \fi

\if0\blind
{
  \bigskip
  \bigskip
  \bigskip
  \begin{center}
    {\LARGE\bf A Bootstrap-based Method for Testing Network Similarity}
\end{center}
  \medskip
} \fi

\bigskip
\begin{abstract}
This paper studies the matched network inference problem, where the goal is to determine if two networks, defined on a common set of nodes, exhibit a specific form of stochastic similarity. Two notions of similarity are considered: (i) equality, i.e., testing whether the networks arise from the same random graph model, and (ii) scaling, i.e., testing whether their probability matrices are proportional for some unknown scaling constant.
We develop a testing framework based on a parametric bootstrap approach and a Frobenius norm-based test statistic.
The proposed approach is highly versatile as it covers both the equality and scaling problems, and ensures adaptability under various model settings, including stochastic blockmodels, Chung-Lu models, and random dot product graph models.
We establish theoretical consistency of the proposed tests and demonstrate their empirical performance through extensive simulations under a wide range of model classes.
Our results establish the flexibility and computational efficiency of the proposed method compared to existing approaches. 
We also report a real-world application involving the Aarhus network dataset, which reveals meaningful sociological patterns across different communication layers. 

\end{abstract}
\noindent%

\vfill

\newpage
\spacingset{1.9} % DON'T change the spacing!

\section{Introduction}
\label{sec:intro}

Consider $n$ entities labeled $1, \ldots, n$, and two undirected networks (without multiple edges or self-loops) represented by $n \times n$ symmetric adjacency matrices $A_1$ and $A_2$. Here, $A_1(i,j) = 1$ if entities $i$ and $j$ interact in the first network, and $A_1(i,j) = 0$ otherwise; similarly for $A_2$. The statistical model assumes $A_1 \sim P_1$ and $A_2 \sim P_2$, where $A_k(i,j) \sim \text{Bernoulli}(P_k(i,j))$ for $1 \leq i < j \leq n$, $k=1,2$. This setup aligns with multilayer or multiplex networks studied in the literature \citep{paul2016consistent,macdonald2022latent,wang2023multilayer}.

Let $\tau(P)$ denote some network feature, i.e., a function of the probability matrix, which represents the model parameter of interest. The inference task of interest is to test whether the networks are similar in terms of $\tau(\cdot)$, i.e., whether $\tau(P_1) = \tau(P_2)$.
We call this problem the \textit{matched} network inference problem to emphasize that the nodes in the two networks are matched, meaning the $i^{th}$ rows of $A_1$ and $A_2$ correspond to the same entity. 

This framework encompasses various problems depending on $\tau$.
In this paper, we concern ourselves with two inference problems: equality and scaling. 
Under the equality problem, we want to test whether $P_1 = P_2$,
akin to the classical paired sample testing problem.
This problem arises in a wide range of scientific problems, such as 
functional neuroimaging \citep{ginestet2017hypothesis}, anatomical brain structures \citep{bassett2008hierarchical}, and gene regulatory networks \citep{zhang2009differential}.

{The \textit{scaling} problem generalizes the equality problem by testing whether $P_1 = cP_2$ for some (unknown) $c > 0$, with $c=1$ corresponding to equality. For instance, \cite{gemmetto2017reconstruction} observed scaling relationships in global trading networks due to shared production or consumption patterns. Scaled relationships between layers are also common in sociology \citep{nicosia2015measuring}.
In this paper, we analyzed the Aarhus network of five interaction types (coauthor, leisure, work, lunch, Facebook) among 61 researchers \citep{rossi2015towards}. While the test of equality was rejected for all pairs of interactions, the test of scaling was not rejected for a majority of the pairs of interactions.
This suggests a proportional relationship that might reflect differences in ease of access, cost, or user preferences for specific modes of interaction. Such findings can lead to deeper sociological insights into communication habits and technological adoption. Note that equality corresponds to $\tau(P) = P$, and scaling to $\tau(P) = P/||P||$, where $||\cdot||$ is a matrix norm (e.g., Frobenius norm).}

{This paper introduces a versatile framework for 
matched network inference based on a Frobenius norm test statistic.
The central idea is to carry out a \textit{null-restricted} transformation of the estimated models to generate {parametric} bootstrap resamples, to estimate the sampling distribution of the test statistic under the null.
The proposed methodology is highly flexible as it accommodates both equality and scaling problems while remaining applicable across a wide array of network models and estimators.
We establish theoretical results under three widely-used models --- the stochastic blockmodel, the Chung-Lu model, and the random dot product graph model --- that emphasize that the proposed method is consistent under general settings.
We also present numerical results under six network models --- the three aforementioned models plus the degree-corrected blockmodel, the popularity adjusted blockmodel, and the latent space model --- that demonstrate the versatality of the proposed method in practice.}

While matched network inference is a relatively new area, some exciting progress has been made in recent years \citep{tang2017semiparametric,tang2017nonparametric,ghoshdastidar2017two, ghoshdastidar2017,ghoshdastidar2018practical,li2018two,levin2017central,agterberg2020nonparametric,jin2024optimal,wu2024two}.
In \cite{tang2017semiparametric}, \cite{tang2017nonparametric}, \cite{levin2017central}, and \cite{agterberg2020nonparametric}, the authors studied the problem under the random dot product graph model and its generalizations.
In \cite{li2018two}, the authors studied the problem under the stochastic blockmodel.
In \cite{ghoshdastidar2018practical}, the authors proposed a test based on the spectral norm, and in \cite{ghoshdastidar2017two} and \cite{ghoshdastidar2017}, the authors studied the problem from an information theoretic perspective to derive minimax bounds.
The methods most closely related to our work are \cite{tang2017semiparametric}, \cite{levin2017central}, and \cite{ghoshdastidar2018practical}, and we provide a comparative analysis of these methods with our approach.

{In related work, recent years have seen substantial growth in theoretical results for bootstrapping network data, with notable contributions from \cite{bhattacharyya2015subsampling}, \cite{green2022bootstrapping}, \cite{lundesubsampling}, and \cite{levin2019bootstrapping}.
However, most of the existing work is for \textit{non-parametric} bootstrap - a notable recent exception being \cite{shao2024parametric} - while our methodology is based on \textit{parametric} bootstrap.
}

The rest of the paper is organized as follows.
% We propose a matched network inference framework based on a Frobenius norm test statistic and parametric bootstrap.
Section \ref{sec:setting} describes the methodology and Section \ref{sec:theory} contains theoretical results.
 Section \ref{sec-sims} describes simulation studies with comparisons to existing methods.
A case study on the Aarhus network in Section \ref{sec-data} illustrates the practical usefulness of the proposed framework, and
Section \ref{sec:discussion} concludes the paper with discussions and next steps.
{Open-source R code is publicly available on GitHub.}

\section{Problem statement and methodology}
\label{sec:setting}

% In what follows, we consider simple, unweighted, undirected networks with no self-loops.
% Let $A$ be the $n$-by-$n$ adjacency matrix of such a network with $n$ nodes, i.e., $A(i,j) = A(j,i) = 1$ if nodes $i$ and $j$ are connected, and   $A(i,j) = A(j,i) =  0$  otherwise.
% $A \sim P$ is shorthand for the statement that for $1 \le i < j \le n$,
% $
% A(i,j) \sim Bernoulli(P(i,j))
% $
% independently.
% Consider a paired network inference problem with $n$ nodes, where $A_1$ and $A_2$ are the adjacency matrices, and the $i^{th}$ node of $A_1$ is matched to the $i^{th}$ node of $A_2$ for $i=1,\ldots,n$.
% Note that the paired network framework holds only when the given networks are perfectly aligned, i.e., we know how to match the nodes in $A_1$ to the nodes in $A_2$.
% In practice, as discussed in Section 1, this may not be the case.
% \subsection{A general test of similarity for paired networks}

{We begin with a heuristic outline of the proposed method for a generic similarity function, $\tau$.
Then we describe detailed methodology for the two specific versions of $\tau$ corresponding to the equality and scaling problems in subsections \ref{subsec:eq} and \ref{subsec:scale}.
}

Let $A_1 \sim P_1$ and $A_2 \sim P_2$, where $P_1, P_2 \in \mathcal{P}$ and $\mathcal{P}$ is a known class of network models  (e.g., the class of stochastic blockmodels).
We want to test 
$
H_0: \tau(P_1) = \tau(P_2) \; \text{vs.} \; H_1: \tau(P_1) \neq \tau(P_2)
$.
Suppose that the model class has a consistent estimator, and we can 
{generate estimators} $\hat{P}_k$  and $\tau(\hat{P}_k)$ from $A_k$ for $k=1,2$.
% and $\hat{P}_2  \in \mathcal{P}$ from $A_2$.
% For some given network feature $\tau(\cdot)$, we are interested in testing 
% $
% H_0: \tau(P_1) = \tau(P_2) \; \text{vs.} \; H_1: \tau(P_1) \neq \tau(P_2)
% $.
% Let us assume that $\tau(\cdot)$ is a ``well-behaved'' function such that $\tau(\hat{P})$ is a consistent estimator of $\tau({P})$ for $P \in \mathcal{P}$.
% % Consider the equality problem first.
Then, a natural test statistic is given by
$$
T(A_1, A_2) = ||\tau(\hat{P}_1) - \tau(\hat{P}_2)||_F,
$$
where $||M||_F = \sqrt{ \sum_i \sum_j |M(i,j)|^2}$ denotes the Frobenius norm of a matrix;
if $\tau$ is a vector instead of a matrix, we use the $l_2$ norm instead.
If the null hypothesis is true, then 
% $\tau(P_1) = \tau(P_2)$, which implies (by the triangle inequality) that 
% $$T(A_1, A_2) \le ||\tau(\hat{P}_1) - \tau({P}_1)||_F + ||\tau(\hat{P}_2) - \tau({P}_2)||_F. $$
% Given a consistent estimator and a well-behaved function $\tau$, 
it is reasonable to expect that $T(A_1, A_2)$ is small.
On the other hand, when $H_1$ is true and $||\tau(P_1) - \tau(P_2)||_F$ is large, 
% we have 
% $$
% T(A_1, A_2) 
% \geq ||\tau(P_1) - \tau(P_2)||_F - \left(||\tau(\hat{P}_1) - \tau({P}_1)||_F + ||\tau(\hat{P}_2) - \tau({P}_2)||_F \right),
% $$
% where $||\tau(\hat{P}_1) - \tau({P}_1)||_F$ and $||\tau(\hat{P}_2) - \tau({P}_2)||_F$ are small, which means
the test statistic is likely to be larger.
Therefore, we should reject when $T(A_1, A_2)$ is greater than some threshold value.
% Therefore, one should use a rejection region of the form 
% % \begin{equation*}
%     $T(A_1, A_2) > K_{\alpha}$
%   % \end{equation*}
% where $\alpha$ is the target level of the test, i.e., we would like to ensure that $Prob[ T(A_1, A_2) > K_{\alpha}] \le \alpha$ under $H_0$ and $Prob[ T(A_1, A_2) > K_{\alpha}] \rightarrow 1$ under $H_1$ as $n \rightarrow \infty$.
% The key question is: how to determine the rejection threshold of the test?
% A suitable choice for $K_{\alpha}$ would be the upper $(1-\alpha)$ quantile of the sampling distribution of $T(A_1, A_2)$ under $H_0$.
One way to determine the rejection threshold would be to analytically formulate the (asymptotic) sampling distribution of $T(A_1, A_2)$ under $H_0$, and compute the upper $(1-\alpha)$ quantile, where $\alpha$ is the target level of the test.
% \ssg{Change the next bit}
In \cite{tang2017semiparametric} and \cite{ghoshdastidar2018practical}, the authors pursued this approach under their specific model settings to derive rejection thresholds for their test statistics.
% However, the asymptotic threshold may not work well in finite samples, as reported by \cite{tang2017semiparametric}.
{However, this approach can lead to complications for practitioners, as they must know the asymptotic distribution in order to implement the test.}
% requires the threshold will likely vary from one model class to another, leading to complications for practitioners.

We propose a parametric bootstrap strategy to estimate the sampling distribution of $T(A_1, A_2)$ under the null.
Bootstrapping is a versatile and well-studied technique for estimating the sampling distribution of a random variable by drawing resamples as a proxy for samples and constructing an empirical distribution \citep{efron1979,singh1981asymptotic,beran1991asympotic, davison1997bootstrap,efron1994introduction,hall1993edgeworth,shao1995jackknife,lahiri2003resampling,chatterjee2011bootstrapping,hall2013simple,senguptadrw}. 
{This approach is conceptually appealing as it is simple to implement, and is automatic in nature, such that
practitioners can apply it without advanced statistical knowledge.}
To estimate the sampling distribution of the test statistic under $H_0$, we first need to transform $\hat{P}_1$ and $\hat{P}_2$ to their \textit{null-restricted} counterparts, $\hat{P}^0_1$ and $\hat{P}^0_2$, which satisfy 
\begin{equation}
    \tau(\hat{P}^0_1) = \tau(\hat{P}^0_2),
\label{eq:nullrestrict}
\end{equation}
such that 
$\hat{P}^0_1$ and $\hat{P}^0_2$ are accurate estimates of $P_1$ and $P_2$ under $H_0$ but not under $H_1$.
{Next, we generate $B$ matched network pairs $A^{*i}_1 \sim \hat{P}^0_1, A^{*i}_2 \sim \hat{P}^0_2$ and compute $T^{*i} = T(A^{*i}_1, A^{*i}_2)$ for $i=1, \ldots, B$.
The p-value is obtained by comparing the {observed} value of the test statistic, $T(A_1, A_2)$, with the empirical distribution of $\{T^{*i}\}_{i=1}^B$, i.e.
 $p = \frac{1}{B} \sum_{i=1}^{B} \mathbb{I}[T \le T^{*i}]$, where $\mathbb{I}$ is the indicator function, and we reject if $p < \alpha$.}

{The key challenge is how to transform the estimates $\hat{P}_1$ and $\hat{P}_2$ to their {null-restricted} counterparts that satisfy \eqref{eq:nullrestrict}.
There is no universal technique for accomplishing this for a generic $\tau$.
However, this transformation can be easily accomplished for the equality and scaling problems, as described in the next two subsections.}

% The steps involved are outlined in Algorithm \ref{algo-gen}.
% A key challenge is carrying out Step 3, which consists of transforming the estimates $\hat{P}_1$ and $\hat{P}_2$ into their null-restricted counterparts $\hat{P}^0_1$ and $\hat{P}^0_2$.
% In addition, $\hat{P}^0_1$ and $\hat{P}^0_2$ should be accurate estimates of $P_1$ and $P_2$ under the null, but not under the alternative.
% % It is challenging to come up with a general recipe for doing this that works for any $\tau(\cdot)$.
% In the next two subsections, we describe the proposed strategy in detail for two specific and important cases of $\tau(\cdot)$.

% Another related question is, how do we compute $\hat{P}_1$ and $\hat{P}_2$?
% This is addressed in Section \ref{sec-models}, where we describe the estimation strategies under a wide variety of statistical network models.
% , and compare our methods to existing methods that have been proposed in related literature.

\subsection{Test of equality}
\label{subsec:eq}
Here $\tau(P) = P$ and
% Then, for the equality problem, 
 we want to test
$    H_0: P_1 = P_2 \; vs. \; H_1: P_1 \neq P_2.
$
% If the null hypothesis is true, then $\hat{P}_1$ and $\hat{P}_2$ are two independent and identically distributed estimates of $P_1 = P_2$, and they should be close to each other.
% On the other hand, if the alternative hypothesis is true and ${P}_1$ and ${P}_2$ are sufficiently well separated, then $\hat{P}_1$ and $\hat{P}_2$ should also be well separated from each other.
The test statistic is
\begin{equation}
    T_\text{frob}(A_1, A_2) = ||\hat{P}_1 - \hat{P}_2||_F.
      \label{eq-tfrob}
\end{equation}
 We can transform $\hat{P}_1$ and $\hat{P}_2$ to their {null-restricted} counterparts by using the pooled estimator, i.e.,
 \begin{equation}
   \hat{P}^{(0)}_1 =  \hat{P}^{(0)}_2 =\frac{1}{2}(\hat{P}_1+\hat{P}_2).
    \label{eq-pooled}
\end{equation}
Clearly this satisfies \eqref{eq:nullrestrict}.
Crucially, $\hat{P}^{(0)}_k$ is a good estimator of $P_k$ for $k=1,2$ under the null  but it is biased under the alternative.
Therefore, resampling from $\hat{P}^{(0)}_1$ and $\hat{P}^{(0)}_2$ ensures that the bootstrapped distribution mimics the sampling distribution of the test statistic under the null, but not under the alternative.
% it tautologically follows that $\tau(\hat{P}^{(0)}_1) = \tau(\hat{P}^{(0)}_2)$, and therefore this 
% gives us the null-restricted distribution for bootstrapping.
The steps are outlined in Algorithm \ref{algo-gen}. 
% in the supplementary file.

\subsection{Test of scaling}
\label{subsec:scale}
Here $\tau(P) = \frac{P}{||P||_F}$, and
the test statistic is 
given by
\begin{equation}
    T_\text{scale}(A_1, A_2) = ||\frac{\hat{P}_1}{\hat{\rho}_1} - \frac{\hat{P}_2}{\hat{\rho}_2}||_F,
      \label{eq-tscale}
\end{equation}
where $\hat{\rho}_1 = ||\hat{P_1}||_F$ and $\hat{\rho}_2 = ||\hat{P_2}||_F$.
 Under the null, there could be a scaling difference between $P_1$ and $P_2$, which means we cannot use a simple pooled estimator to transform $\hat{P}_1$ and $\hat{P}_2$ to their null-restricted counterparts.
 However, under the null, the \textit{scaled} probability matrices are equal, i.e., $\frac{P_1}{||P_1||_F} = \frac{P_2}{||P_2||_F}$, and let us denote this as $H = \frac{P_1}{||P_1||_F} = \frac{P_2}{||P_2||_F}$.
  Then, under the null, $A_1 \sim ||P_1||_F H$ and $A_2 \sim ||P_2||_F H$.
 Therefore, we can use the pooled version of the scaled estimates to estimate $H$, and rescale the pooled version to obtain the null-restricted counterparts $\hat{P}^{(0)}_1$ and $\hat{P}^{(0)}_2$.

 That is, we first compute $
     \hat{H} = \frac{1}{2}\left(\frac{\hat{P}_1}{\hat{\rho}_1}+\frac{\hat{P}_2}{\hat{\rho}_2}\right)
    $, and then  we rescale it as 
  \begin{equation}
     \hat{P}^{(0)}_1 =\hat{\rho}_1\hat{H}, \; \;  
     \hat{P}^{(0)}_2 =\hat{\rho}_2\hat{H}.
     \label{eq-nullscaled}
 \end{equation}
As before, $\hat{P}^{(0)}_1$ and $\hat{P}^{(0)}_2$ now satisfy \eqref{eq:nullrestrict}.
Furthermore, $\hat{P}^{(0)}_k$ is a good estimator of $P_k$ for $k=1,2$ under the null  but it is biased under the alternative, which ensures that the resampling distribution approximates the sampling distribution under the null but not under the alternative, as intended.
The steps are outlined in Algorithm \ref{algo-gen}.

\begin{remark}
    \textbf{Sparse Graph Case}: We now consider the scenario where the networks are sparse. In this case, both \( P_1 \) and \( P_2 \) asymptotically converge to a matrix of zeroes, which trivializes the null hypothesis \( H_0 : P_1 = P_2 \) for the test of equality. Therefore, the case for sparse networks needs to be treated separately. To define sparsity, we represent the probability matrices \( P_1 \) and \( P_2 \) as \( P_k = \rho_{kn} \tilde{P}_k \) for \( k = 1, 2 \), where \( \rho_{kn} \) are the sparsity factors of the networks, and the entries of \( \tilde{P}_k \) are $O(1)$. Without loss of generality, assume that \( \rho_{1n} / \rho_{2n} \) does not diverge to infinity. 
In practice, the sparsity factors are not known a priori, and the objective is to compare the \textit{inherent} probability matrices \( \tilde{P}_1 \) and \( \tilde{P}_2 \), rather than the sparsity factors themselves. Therefore, we treat the sparsity factors as unknown sequences and test whether \( \tilde{P}_2 = \frac{\rho_{1n}}{\rho_{2n}} \tilde{P}_1 \). Equivalently, this is expressed as testing \( H_0 : \frac{\tilde{P}_1}{\|\tilde{P}_1\|_F} = \frac{\tilde{P}_2}{\|\tilde{P}_2\|_F} \), which translates to testing the scaled hypothesis \( H_0 : \frac{P_1}{\|P_1\|_F} = \frac{P_2}{\|P_2\|_F} \).
\end{remark}
 
%  We independently sample a pair of networks $A_1^* \sim \hat{P}^{(0)}_1$,
%  $A_2^* \sim \hat{P}^{(0)}_2$, and estimate $\hat{P}_1^* $ from $A_1^*$ and $\hat{P}_2^* $ from $A_2^*$.
%  Then 
% \begin{equation}
%     T^*_{scale}(A^*_1, A^*_2) = ||\frac{\hat{P}_1^*}{\hat{\rho}^*_1} - \frac{\hat{P}_1^*}{\hat{\rho}^*_2}||_F,
%       \label{eq-tscaleboot}
% \end{equation}
% is the parametric bootstrap version of $T_\text{scale}$.
% As before, we obtain the empirical distribution of $T^*_{scale}$ from $B$ parametric bootstrap iterations, and compute the p-value of $T_\text{scale}(A_1, A_2)$ with respect to this empirical distribution function.
% The steps are outlined in Algorithm \ref{algo-scale} in the supplementary file.

{\spacingset{1.2}
\IncMargin{1em}
\begin{algorithm}[h]
\SetKwData{Left}{left}\SetKwData{This}{this}\SetKwData{Up}{up}
\SetKwFunction{Union}{Union}\SetKwFunction{FindCompress}{FindCompress}
\SetKwInOut{Input}{Input}\SetKwInOut{Output}{Output}
\Input{Data $A_1, A_2$;
Network feature $\tau(\cdot)$;
Number of bootstraps $B$}
\Output{p-value for the test of similarity}
\BlankLine
\begin{enumerate}
    \item[1:] Compute $\hat{P}_1 $ from $A_1$ and $\hat{P}_2 $ from $A_2$
    \item[2:] Compute the test statistic $T = T(A_1, A_2) = ||\tau(\hat{P}_1) - \tau(\hat{P}_2)||_F$ using \eqref{eq-tfrob} or \eqref{eq-tscale}
    \item[3:] Transform $\hat{P}_1$ and $\hat{P}_2$ to their ``null-restricted'' counterparts, $\hat{P}^0_1$ and $\hat{P}^0_2$, \\ 
    using \eqref{eq-pooled} or \eqref{eq-nullscaled}.
    \item[4:] Parametric bootstrap: 
	\For{$i\leftarrow 1$ \KwTo $B$}{
\begin{enumerate}
		\item Generate $A^{*i}_1 \sim \hat{P}^{(0)}_1$,  $A^{*i}_2 \sim \hat{P}^{(0)}_2$
		\item Compute $\hat{P}^{*i}_1$ from $A^{*i}_1$ and $\hat{P}^{*i}_2$ from $A^{*i}_2$ 
		\item Compute $T^{*i} \leftarrow T(A^{*i}_1, A^{*i}_2) = ||\tau(\hat{P}^{*i}_1) - \tau(\hat{P}^{*i}_2)||_F$
	\end{enumerate}
}
	
    \item[5:] The p-value is $p \leftarrow \frac{1}{B} \sum_{i=1}^{B} \mathbb{I}[T \le T^{*i}]$, where $\mathbb{I}$ is the indicator function.
    % and we reject if $p < \alpha$.
\end{enumerate}
\caption{Test of equality and scaling for matched networks
\label{algo-gen}}
\end{algorithm}\DecMargin{1em}
}

\subsection{Models and estimators}
\label{sec-models}
% A strength of the proposed methodology is its generalizability to various statistical models.
So far, we have used a generic notation, $\mathcal{P}$, for the class of models.
We now describe six well-known model classes and the corresponding estimators.
When implementing the proposed tests, these estimators should be used for steps 1 and 4(b) of Algorithm \ref{algo-gen}.

\begin{itemize}[leftmargin=*]

	\item \textbf{Stochastic Blockmodel (SBM):}
 % The stochastic blockmodel \citep{holland1983stochastic,fienberg1985statistical} is probably the most well-studied network model in the statistics literature.
	Under an SBM with $K$ communities,
	% \[
	$P(i,j) = \omega_{c_ic_j},$
	% \]
	where $\omega$ is a $K$-by-$K$ symmetric matrix of community interaction probabilities, and ${\bf c} = \{c_i\}_{i=1}^n$ are the communities of the nodes, with $c_i$ taking its value in $1, \ldots, K$ \citep{holland1983stochastic}.
 A number of community detection methods \citep{rohe2011spectral,sengupta2015spectral,zhao2012consistency,gao2017achieving} can be used for estimating the communities $\{\hat{c}_i\}_{i=1}^n$.
 Once we have $\{\hat{c}_i\}_{i=1}^n$,
 the model parameters $\omega_{rs}$ are estimated as
	$
	\hat{\omega}_{rs} = \frac{\sum_{i,j: \hat{c}_i = \hat{c}_j = r} A(i,j)}{{\hat{n}_r(\hat{n}_r-1)}}${ when } $r=s$, { and } 
	$\hat{\omega}_{rs} = \frac{\sum_{i,j: \hat{c}_i, \hat{c}_j = r} A(i,j)}{\hat{n}_r \hat{n}_s}$ { when } $r \neq s$.
	% \]
	Here $\hat{n}_r$ is the size of the estimated $r^{th}$ community.
		We estimate $P$ as
		$\hat{P}(i,j) = \hat{\omega}_{\hat{c}_i\hat{c}_j}.$
	
	\item \textbf{Chung-Lu (CL) model:}
	Here
	% \[
$	P(i,j) = \theta_i \theta_j$,
	% \]
	where $\{\theta_i\}_{i=1}^n$ are the degree parameters \citep{chung2002average, dasgupta2022scalable}.
	We estimate $P$ as
	% \[
	$\hat{P}(i,j) = \frac{d_id_j}{2m},$
	% \]
	where $d_i$ is the degree of the $i^{th}$ node and $m=\sum_{i>j} A(i,j)$. 
 % is the total degree of the network.

 \item \textbf{Random dot product graph (RDPG) model:}
	Under the RDPG \citep{young2007random} with dimension $d$, we have
% \[
	${P} = XX',$
	% \]
	where $X_{n \times d}$ is a matrix of rank $d$ such that $[XX'](i,j) \in (0,1)$ for all pairs $(i,j)$.
	For estimation, we use the adjacency spectral embedding (ASE) of $A$, given by 
	% \[
	$\hat{X} = U_A S_A^{1/2}, $
		% \]
	where $S_A$ is the diagonal matrix of the $d$ largest eigenvalues of $(A'A)^{1/2}$ and $U_A$ is the 
 	$n$-by-$d$ 
	matrix 
 	whose columns consist 
	of the corresponding eigenvectors \citep{sussman2012consistent}.
	We estimate $P$ as
	% \[
$	\hat{P} = \hat{X}\hat{X}'.$
	% \]
		
	\item \textbf{Latent Space Model (LSM):}
	% Under the latent distance model of \cite{hoff2002latent}, 
 Here each node is assumed to have a latent position in $\mathbb{R}^d$  and
edge probabilities are determined by pairwise $l_2$ distances between the latent positions, i.e.,
% The probabilities are given by 
% \[
$\text{logit}({P}({i,j}))=\alpha - |{z}_i-{z}_j|$,
% \]
for $1\leq i<j\leq n$, where ${z}_i \in \mathbb{R}^d$ is the latent position of the $i^{th}$ and the parameter $\alpha$ controls overall sparsity \citep{hoff2002latent}.
We estimate $P$ by using the maximum likelihood estimation strategy described in  \cite{hoff2002latent} as implemented in the R package \textit{latentnet} \citep{krivitsky2008fitting}.
Note that we consider the null hypothesis to be conditional on latent positions, which is equivalent to the ``semiparametric'' framework considered in \cite{tang2017semiparametric} but different from the ``nonparametric'' framework considered in \cite{tang2017nonparametric}.

		\item \textbf{Degree Corrected Blockmodel (DCBM):} 
  % The degree-corrected stochastic blockmodel \citep{karrer2011stochastic} is a generalization of the stochastic blockmodel that allows for flexible degree distributions.
	Under the DCBM with $K$ communities,
	% \[
	$P(i,j) = \theta_i\omega_{c_i} \omega_{c_j}\theta_j,$
	% \]
	where $\omega$ is a $K$-by-$K$ symmetric matrix of community-community interaction probabilities, and $\{c_i\}_{i=1}^n$ are the communities of the nodes, and $\{\theta_i\}_{i=1}^n$ are degree parameters \citep{karrer2011stochastic}.
	Several community detection methods \citep{qin2013regularized,sengupta2015spectral, zhao2012consistency,gao2018community} can be used for estimating the communities $\{\hat{c}_i\}_{i=1}^n$.
 The remaining parameters are estimated as
	% \[
	$\hat{\omega}_{rs} = \sum_{i,j: \hat{c}_i, \hat{c}_j = r} A(i,j)$, 
 { and } $\hat{\theta}_i = \frac{d_i}{\delta_r},$
	{ where } $\delta_r = \sum_{i: \hat{c}_i=r} d_i$
	% \]
	 is the degree of the estimated $r^{th}$ community.
	We estimate $P$ as
	% \[
	$\hat{P}(i,j) = \hat{\theta}_i\hat{\omega}_{\hat{c}_i \hat{c}_j} \hat{\theta}_j.$
	% \]
	\item \textbf{Popularity Adjusted Blockmodel (PABM):} 
 % The popularity adjusted blockmodel was proposed by \cite{senguptapabm} for flexible modeling of node popularities in the presence of community structure. 
	Under the PABM with $K$ communities,
	$
	P(i,j) = \theta_{ic_j} \theta_{jc_i},
	$
	where $\theta_{ir}$ represents the popularity of the $i^{th}$ node in the $r^{th}$ community, and $\{c_i\}_{i=1}^n$ are the node communities \citep{senguptapabm}.
	We use the extreme points method of \cite{le2016optimization} to estimate communities, and estimate the popularity parameters as
	% \[
	$\hat{\theta}_{ir} = \frac{\sum_{j: \hat{c}_j=r} A(i,j)}{\sqrt{\sum_{i,j: \hat{c}_i, \hat{c}_j = r} A(i,j)}}.$
	% \]
	We estimate $P$ as
	% \[
	$\hat{P}(i,j) = \hat{\theta}_{i\hat{c}_j} \hat{\theta}_{j\hat{c}_i}.$
	% \]
\end{itemize}

\begin{remark}
We omitted the classical Erd\"{o}s-Ren\'{y}i model \citep{erdHos1959random}, where $P(i,j) = p$ for all pairs $(i,j)$,
because the test of equality reduces to the well-known two-sample test of equality of proportions, 
and the test of scaling is not meaningful since any two Erd\"{o}s-Ren\'{y}i models are scaled versions of each other.
\end{remark}

\section{Theoretical results}
\label{sec:theory}
We now demonstrate the consistency of the proposed methods under three models described in Section \ref{sec-models}.
Similar theoretical results can be proved under some of the other models, but we postpone this to future research.
We present the results in two subsections:
distributional consistency for the bootstrap methods under the SBM,
and testing consistency under the CL and RDPG models.
In both subsections, we analyze the equality and scaling problems under the specified models.
All technical proofs are in the supplementary file.

\subsection{Bootstrap distributional consistency under the SBM}
To prove distributional consistency of the bootstrap,
we first derive the limiting null distributions of
the relevant test statistics and then use these to
show that the proposed bootstrap procedures provide valid approximations in both testing problems.  
Recall the definitions from Section \ref{sec-models}.
In particular, the vectors ${\bf c}_{jn}$ and $\hat{{\bf c}}_{jn}$ are the true and estimated communities
({after necessary permutation  of the community labels}) for the $j$th network,  $j=1,2$
and ${n}_r$ and $\hat{n}_r$ are the sizes of the true and estimated $r^{th}$ community for $r=1, \ldots, K$,
where we assume that the number of communities $K$ is known.
{Note that under the matched network framework null hypotheses,
$K$ and $n_r$ is the same for both the networks given by $P_j = \big(\big(\omega_{j, rs}\big)\big)$, $j=1,2$}
We will make use of the following conditions:
\begin{itemize}
\item[(C.1)] 
For $r=1,\ldots,K$, 
$\lim_{\nti} \frac{n_r}{n}= \pi_r\sa >0$.

\item[(C.2)] There exists $\{a_n\}_{n\geq 1} \subset  (0,1]$
such that $a_n +n^{-2}a_n^{-1}  = o(1)$
and $\om_{j,rs} = a_n \om_{j,rs}\sa $ for all $1\leq r,s\leq K$, $j=1,2$,
where  $\om_{j,rs}\sa $ does not depend on $n$.

\item[(C.3)] 
%$P(\hK=K) \raw 1$ and 
$EH(\hat{\bf c}_{jn}, \bfc_{jn}) = o \big(n^{-1}\big)$ as $\nti$, $j=1,2$. 
%o\big(\min\{\sqrt{a_n}, \,
%n^{-1}\}\big)$, as $\nti$.

\end{itemize}

Condition (C.1) says that the $K$ 
underlying communities are of comparable size.
Condition (C.2) ensures that 
the network has some degree of sparsity.
Without this condition, we get a dense network
unsuitable for most  practical applications \citep{zhao2012consistency}
and,  therefore, we decided to focus on the sparse case. However, we also indicate the
limit distribution in the dense case in Remark 3.1 (which is slightly different from the sparse
case). 
Finally, 
Condition (C.3) gives the requirements for the 
community finding algorithm. 
{It requires  
the expected number of the misclassified
nodes to decay at a sufficiently fast rate.
While this condition is not exactly captured 
by the popular notions of weak or strong consistency of
a community detection algorithm, it can be proved 
using known bounds for different algorithms proposed 
in the literature when $a_n$ goes to zero slowly
 or is bounded away from zero (as in a dense network).
For example, for the two-stage algorithm proposed in \cite{gao2017achieving, gao2018community}, this condition holds for sparse assortative SBMs
when $a_n \gg n^{-1}\log n$,  as implied by   the exponential minimax rate
of misclassification proved therein. See Remark S1.2 of the Supplementary Materials
for more details. 
}

For analyzing the asymptotic properties of the two-sample test, it is pedagogically and notationally
simpler to consider the one-sample case first.
A straightforward extension of the arguments will be used to prove the limit distributions 
for the two-sample case.
To that end, for the first  testing problem $H_0: P_1=P_2 $ against $H_1:  P_1\neq P_2$,
write 
$$
T_{1n} = \frac{1}{\sqrt{a_n}} T_\text{frob} (A_1,A_2)
\equiv \frac{1}{\sqrt{a_n}}\|\hat{P}_1 - \hat{P}_2\|_F.
$$
Let $\{Z\sa_{rs} : 1\leq r\leq s\leq K\}$ be a collection of iid 
N(0,1) random variables and set $Z\sa_{rs}= Z\sa_{sr}$ if $r>s$. 
Write $\bfZ\sa=(( Z\sa_{rs}))_{K\times K}$.
Also, let $\bfZ_1\sa,\bfZ_2\sa$ be two independent
copies of $\bfZ\sa$. 
Then, we have the following result.

\begin{theorem}
    Suppose that Conditions (C.1)-(C.3) hold.
Then, $$T_{1n}\raw^d T_{1,\infty} $$
where $T_{1,\infty}=\|D_1\sa\odot \bfZ_1\sa -  
 D_2\sa\odot \bfZ_2\sa\|_F$, $\odot$ denotes the 
 Hadamard product and \\
 $D_k= \big(
 \big(\, \sqrt{[1+\ind(r=s)]\om_{k,rs}\sa}
 ~ \big)\big)$, $k=1,2$.
 \label{thm:sbmdist1}
\end{theorem}

\begin{remark}
    If $a_n\raw a\not = 0$ (i.e., $\{a_n\}_{n\geq 1}$  remains bounded away from zero), then we have a dense network 
and a version of Theorem \ref{thm:sbmdist1} still holds. In this case,
the test statistic $\|\hat{P}_1 -  \hat{P}_2  \|
\raw^d \tilde{\Gamma}_1$ where, with 
$\tilde{\tau}\sa_{k,rs} = a \omega_{k, rs}\sa (1-a \om_{k,rs}\sa)
$ for $r\neq s$ and 
$\tilde{\tau}\sa_{k,rs} = 2 a \omega_{k, rs}\sa
(1-a \om_{k,rs}\sa)$ for $r=s$, we have
$
\tilde{\Gamma}_1 
=
 \sum_{r=1}^K\sum_{s=1}^K
\Big[ \ttau\sa_{1, rs}
[Z_{1,rs}\sa]^2 + \ttau\sa_{2, rs}
[Z_{2,rs}\sa]^2
 - 2 \sqrt{ \ttau\sa_{1, rs} \ttau\sa_{2, rs}}
Z_{1,rs}\sa Z_{2,rs}\sa
\Big]. 
$
\end{remark}

Next, consider the test of scaling and define
$$
T_{2n}^2 = n^2 a_n [T_\text{scale} (A_1,A_2)]^2
\equiv n^2 a_n \Big\|\frac{\hat{P}_1}{\hat{\rho}_1} -
\frac{\hat{P}_2}{\hat{\rho}_2}\Big\|_F ^2.
$$
The following result establishes the distributional convergence of $T_{2n}$.

\begin{theorem}
Suppose that 
 %$c=c_n$ are such that with $\ta_n=\tilde{a}_n$,
Conditions (C.1)-(C.3) hold. 
%with $a_n$ replaced by $\ta_n$. 
Then, under the null hypotheses,
$$T_{2n}\raw^d T_{2,\infty}$$
where $T_{2,\infty}$ is a nondegenerate random variable whose exact definition is complicated and is given in Equation \eqref{t2-lim} of the supplementary files.  
\label{thm:sbmdist2}
\end{theorem}

Theorems \ref{thm:sbmdist1} and \ref{thm:sbmdist2} establish the limiting distributions of the (scaled) test statistics.
Next, we establish the validity of the bootstrap procedures. {Let $P_*$ and $E_*$  
respectively denote the bootstrap probability and the bootstrap expectation.}
In addition to Conditions (C.1)-(C.3), we will need the following condition for establishing validity of the bootstrap approximation to the null distribution of $T_{1n}$:

\noindent
{\bf (C.4)} 
$E_*H(\bfc^*_{kn}, \hat{\bfc}_{kn}) = 
o_p \big(n^{-1}\big)$ as $\nti$, $k=1,2$. \\[.in]

Condition (C.4)  requires that an analog of Condition (C.3)
be valid in the bootstrap setup. It is a conditional consistency 
condition on the community detection method that is 
assumed to have the same level of accuracy when applied to the 
bootstrap data set, in the weak (in probability) sense. 
{Again, this condition holds for the community detection algorithm of
\cite{gao2017achieving, gao2018community} for $a_n \gg n^{-1}\log n$, as the estimated parameters of the 
SBM lies in the minimaxity class with high probability, as implied by  Lemma S1.2 
and Remark S1.1 of the Supplementary Materials  below.
}
With this additional condition, we have the following result.
% In view of Lemma S.3 on concentration of the community-connection
% probabilities $\hat{\om}_{k, rs}$ in the SBM, a community detection algorithm
% that can accurately determine the community memberships
% for the original networks can typically attain a similar level
% of discrimination at the bootstrap level, with high probability. 

\begin{theorem}
    Suppose that Conditions (C.1)-(C.4) hold. 
    Also, suppose that the null hypotheses
hold for all
$n\geq 1$. Then, 
$$
\sup_{x\geq 0} \Big| P_*(T_{1n}^*\leq x) - P(T_{1n}\leq x |H_{0,n})  \Big| = o_p(1).
$$
\label{thm:sbmboot1}
\end{theorem}
Thus, the proposed bootstrap method provides a valid approximation to the null distribution of the test statistic.
Next, define the bootstrap version of the test statistic
$T_{2n}$ by replacing $(A_1, A_2)$ by $(A_1^*, A_2^*)$, i.e.,
$$
T_{2n}^* = n^2a_n\Big\|\frac{P_1^*}{\rho_1^*} - \frac{P_2^*}{\rho_2^*}      \Big\| 
$$
where $P^*_k$ is obtained by estimating the SBM model parameters
based on $A_k^*$ as in the original problem and where 
${\rho}^*_k=\|P^*_k\|$, $k=1,2$.

\begin{theorem}
Suppose that Conditions (C.1)-(C.4) hold. Also, suppose that the null hypotheses
hold for all
$n\geq 1$. Then, 
$$
\sup_{x\geq 0} \Big| P_*(T_{2n}^*\leq x) - P(T_{2n}\leq x |H_{0,n})  \Big| = o_p(1).
$$
\label{thm:sbmboot2}
\end{theorem}

\begin{remark}
As indicated in the statement of Theorem 3.2,
the limit distribution of the test statistic in the scale-hypothesis case is rather complicated, involving 
a nonlinear function of the population parameters 
and independent  standard Gaussian variables. For the
validity of the bootstrap, we show that the bootstrap
version of the test statistic can be closely {\it approximated}
by the same nonlinear function  but with population parameters 
replaced by their estimators and the standard Gaussian
random variables by some conditionally weakly convergent sequences  of bootstrap random variables.  
To show weak convergence of this bootstrap stochastic approximation to the same limit, we develop 
a technical result (a version of
the continuous mapping theorem for 
conditional weak convergence)
in Lemma S.4 of the Supplementary materials file 
that may be of independent interest. Further, as it is 
evident from the statement of Theorem 3.1 and 
Equation \eqref{t2-lim}, the limiting random variables for both $T_{1n}$ and $T_{2n}$ have a continuous 
distribution on the real line and therefore, 
both the tests can be calibrated using the bootstrap quantiles to ensure that they attain any 
prespecified size (i.e., the probability of Type I error), asymptotically.  See Remark S1.3 in the Supplementary materials file for more details
on using bootstrap calibration and its
theoretical justification. 
\end{remark}

\subsection{Consistency under the CL and RDPG models}
Consider a test statistic $T_n$ and a rejection region of the form $T_n > K$ for testing $H_0$ against $H_1$ at level $\alpha$.
Following \cite{tang2017semiparametric},
we define a testing method to be consistent if there exists $K>0$ such that for any $\eta > 0$,
\begin{enumerate}
    \item If $H_0$ is true, then $ \lim_{n \rightarrow \infty} P(T_n >K) \leq \alpha + \eta$, and
    \item If $H_1$ is true, then $\lim_{n \rightarrow \infty} P(T_n >K) > 1 - \eta$
    \end{enumerate}
The following theorems show that the proposed testing methods are consistent as per this definition.
% We first consider the equality case and then the scaling case.
% For each case, we are going to list some fairly mild conditions under which the corresponding inferential method is consistent.
% All technical proofs are in the supplement.

\subsubsection{Test of equality}
% Recall that in the equality case we are given a pair of networks $A_1 \sim P_1$ and $A_2 \sim P_2$, and we want to test
% \[
% H_0: P_1 = P_2, \hspace{0.2cm} vs. \hspace{0.2cm} \; H_1: P_1 \neq P_2.
% \]
% The test statistic is $T_\text{frob}(A_1, A_2) = ||\hat{P}_1 - \hat{P}_2||_F$ and we reject when $T_\text{frob} > K$ for some suitable $K$.

% \subsubsection*{Chung-Lu model}
First, we consider the CL modeling framework,
% case where both $P_1$ and $P_2$ belong to the Chung-Lu model class, 
i.e., we can write $P_1(i,j) = \theta_i \theta_j$ and $P_2(i,j) = \beta_i \beta_j$ for $1 \le i<j \le n$.
Let $\theta_{(1)} = \min_{1 \le i \le n} \theta_i$ and 
$\beta_{(1)} = \min_{1 \le i \le n} \beta_i$.

\begin{theorem}
Under the CL model, the test of equality is consistent under the following assumptions:
\begin{enumerate}
    \item For any $ \alpha > 0$,
    $\min \left(\frac{n^\alpha\theta_{(1)}\bar{\theta}}{\sqrt{\log(n)}},
        \frac{n^\alpha\beta_{(1)}\bar{\beta}}{\sqrt{\log(n)}}
        \right) \rightarrow \infty$.
        
     \item The quantity 
     $\max\left(\frac{\sum_i\theta_i^2}{n\Bar{\theta}^2},
         \frac{\sum_i\beta_i^2}{n\Bar{\beta}^2}\right)$
         does not diverge to infinity.
    \item Under $H_1$, $P_1\neq P_2$ is in the sense that $
        \frac{||P_1-P_2||_F}{n^{1/2+\alpha}}\to \infty~\text{for any}~\alpha>0$.
\end{enumerate}
\label{thm:cleq}
\end{theorem}

\begin{remark}
    The first two assumptions enforce regularity conditions.
The first assumption ensures that the networks are not too sparse and the expected degrees are not too small.
In particular, assumption 1 holds as long as the smallest degree parameter $\theta_{(1)}$ is of the order ${1}/{(\log(n))^k}$ for any $k < \infty$.
The second assumption ensures that the network parameters do not vary too much.
When $H_1$ is true, the third assumption provides a lower limit on the difference of the two models, given by $||P_1-P_2||_F$, such that the they can be told apart.
\end{remark}

Next, consider the case where both $P_1$ and $P_2$ belong to the RDPG model class, i.e., $P_1 = X_1 X_1'$ and $P_2 = X_2 X_2'$ where $X_1$ and $X_2$ are $n$-by-$d$ matrices.
For a matrix $M$ with singular values $\sigma_1(M) \geq \sigma_2(M) \geq ...$ and for a fixed $d$, we define the following quantities:
\begin{equation*}
    \delta(M)=\max_{1\leq i\leq n}\sum_{j=1}^n M_{ji}; \hspace{0.5cm} \hspace{0.5cm} \gamma(M)= \dfrac{\sigma_d(M)-\sigma_{d+1}(M)}{\delta(M)} \leq 1.
\end{equation*}
The definition of $\gamma$ depends implicitly on a parameter $d \in \mathbb{N}$. 
For a matrix $P = XX^T$ of rank $d$, $\delta(P)$ is simply the maximum expected degree of a graph $A \sim Bernoulli(P)$, $\gamma(P)$ is the $d$th largest eigenvalue normalized by $\delta(P)$.

\begin{theorem}
Under the RDPG model, the test of equality is consistent under the following assumptions:
\begin{enumerate}
    \item $\exists \; c_0 > 0$ such that $\min(\gamma(P_1),\gamma(P_2))>c_0$.
    \item $\exists \; \epsilon > 0$ such that $\min(\delta(P_1),\delta(P_2))>(\log n)^{2+\epsilon}$.
    \item Under $H_1$, $P_1\neq P_2$ is in the sense that $
        \frac{\|P_{1}-P_{2}\|_F}{\Gamma}\to \infty
    $ 
    where $\Gamma = \sum_{i=1}^{2} \big[3\sqrt{r_i}+(d\gamma^{-1}(P_{i}))^{1/2}\big]$.
\end{enumerate}
\label{thm:rdpgeq}
\end{theorem}

\begin{remark}
    These assumptions are similar to those used in \cite{tang2017semiparametric}.
    The first assumption is an eigengap type condition that ensures that the singular values are well separated.
The second assumption ensures that the expected degrees are not too small.
The third assumption ensures that under the alternative hypothesis, the two models are sufficiently well separated.

\end{remark}

\subsubsection{Test of scaling}
% For the scaling case, we are given a pair of networks $A_1 \sim P_1$ and $A_2 \sim P_2$.
% Let $\mathbb{C} = \mathbb{C}(P_2)$ denote the class of all positive constants $c$ for which all the entries of $cP_2$ belongs to $[0,1]$.
% Then, we want to test
% \[
% H_0: P_1 = cP_2 \text{ for some } c \in \mathbb{C}, \hspace{0.2cm} vs. \hspace{0.2cm} \; H_1: P_1 \neq cP_2 \text{ for any } c \in \mathbb{C}.
% \]
% In what follows, we will only write $c > 0$, but will always assume that $c \in \mathbb{C}$, since the
% problem is ill-posed otherwise.
% The test statistic is $T_\text{scale}(A_1, A_2) = ||\frac{\hat{P}_1}{\hat{\rho}_1} - \frac{\hat{P}_2}{\hat{\rho}_2}||_F$ where $\hat{\rho}_1 = ||\hat{P_1}||_F$ and $\hat{\rho}_2 = ||\hat{P_2}||_F$, and we reject when $T_\text{scale} > K$ for some suitable $K$.

% \subsubsection*{Chung-Lu model}
As before, we start with the CL model.
Recall the definitions before Theorem \ref{thm:cleq}.
% , which we do not repeat here in the interest of space.
\begin{theorem}
Under the Chung-Lu model, the test of scaling is consistent if the following assumptions hold:
\begin{enumerate}
    \item For any $ \alpha > 0$,
       $ \min \left(\frac{n^\alpha\theta_{(1)}\bar{\theta}}{\sqrt{\log(n)}},
        \frac{n^\alpha\beta_{(1)}\bar{\beta}}{\sqrt{\log(n)}}
        \right) \rightarrow \infty$ \label{eq:assm1}.
     \item The quantity 
         $\max\left(\frac{\sum_i\theta_i^2}{n\Bar{\theta}^2},
         \frac{\sum_i\beta_i^2}{n\Bar{\beta}^2}\right)$
 does not diverge to infinity.
    \item Under $H_1$, $P_1 \neq cP_2 \text{ for any } c$ is in the sense that 
        $\frac{||\frac{P_1}{\rho_1}-\frac{P_2}{\rho_2}||_F}{n^{1/2+\alpha}}\to \infty~\text{for any}~\alpha>0$,
        where $\rho_1 = ||P_1||_F$ and $\rho_2 = ||P_2||_F$.
\end{enumerate}
\label{thm:clscale}
\end{theorem}
These assumptions are similar to Theorem \ref{thm:cleq} and carry the same interpretations.

% \subsubsection*{RDPG Model}
Next, under the RDPG model, recall the definitions before Theorem \ref{thm:rdpgeq}, which we do not repeat here in the interest of space.
\begin{theorem}
Under the RDPG model, the test of scaling is consistent if the following assumptions hold:
\begin{enumerate}
    \item $\exists \; c_0 > 0$ such that $\min(\gamma(P_1),\gamma(P_2))>c_0$.
    \item $\exists \; \epsilon > 0$ such that $\min(\delta(P_1),\delta(P_2))>(\log n)^{2+\epsilon}$.
    \item Under $H_1$, $P_1\neq P_2$ is in the sense that
    % \footnote{Kaustav: please check, the last assumption here does not involve $\Gamma$ although the last condition in Theorem 3.2 did}
        $\|\frac{1}{\rho_{1}}P_{1}- \frac{1}{\rho_{2}}P_{2}\|_F\to \infty$
    where $\rho_1 = ||P_1||_F$ and $\rho_2 = ||P_2||_F$.
\end{enumerate}
\label{thm:rdpgscale}
\end{theorem}
The assumptions and their interpretations are similar to those for Theorem \ref{thm:rdpgeq}.

\begin{remark}
    We note that under the RDPG model, the adjacency spectral embedding (ASE) has better concentration properties for sparse graphs with
respect to the $2 \rightarrow \infty$ norm rather than the Frobenius norm \citep{xie2024entrywise}.
In this work, we have used the Frobenius norm as the test statistic, and using the $2 \rightarrow \infty$ norm instead could lead to sharper consistency results for the test.
We consider this to be an important avenue for future research.   
\end{remark}

% \begin{remark}
%     Theorems \ref{thm:cleq} and \ref{thm:rdpgeq} are analogous to Theorem 2 of \cite{tang2017semiparametric}, while Theorems \ref{thm:clscale} and \ref{thm:rdpgscale} are analogous to Theorem 3 of \cite{tang2017semiparametric}.
%     Thus, the above theorems establish the consistency of the proposed tests under more general model settings than current methods.
% \end{remark}

% To conclude this section,
% here we have provided theoretical justification for the inferential methods under three of the six models outlined in Section 2.3.
% It is also important to theoretically study the inferential methods under the four remaining models.
% We consider this endeavour an important next step beyond the scope of this paper. 

% \begin{remark}
% In these theorems, we established that there is some $K$ such that a rejection region of the form $T_n >K$ makes the test asymptotically valid.
% In practice one needs to obtain the value of this suitable threshold $K$ and we propose to do so by employing parametric bootstrap.
% The accuracy of the inferential method therefore depends on the accuracy of this parametric bootstrap strategy towards estimating the relevant quantile of the sampling distribution, and it is important to theoretically investigate the accuracy of the bootstrap strategy.
% We consider this endeavour as an important next step beyond the scope of this paper. 
% \end{remark}

\section{Simulation Study}
\label{sec-sims}
% \ssg{SELECT SOME AND MOVE OTHERS TO SUPP}.
We now study the numerical performance of the proposed $T_\text{frob}$ (for the test of equality) and $T_\text{scale}$ (for the test of scaling) methods across a range of network models and parametric scenarios.
We report two performance metrics: Type I error, i.e., probability of false rejection when $H_0$ is true, which should be close to $\alpha = 5\%$, and power, i.e., probability of true rejection when $H_1$ is true, which should be close to $100\%$.
We used $n=100, 200, 300, 400$, along with 2000 Monte Carlo simulations and $B=200$ bootstrap iterations in each experiment unless specified otherwise.
Comparisons with three existing methods:
$T_\text{ase}$ proposed by \cite{tang2017semiparametric}, 
$T_\text{omni}$ proposed by \cite{levin2017central}, and $T_\text{eig}$ proposed by \cite{ghoshdastidar2018practical} are included wherever appropriate;
see Section \ref{sec-compare} in the supplementary materials file for more details on these methods.
Note that $T_\text{ase}$ applies to both equality and scaling problems, whereas $T_\text{omni}$ and $T_\text{eig}$ apply only to the equality problem.

Since most of the existing work has focused on the RDPG model, we dedicate a full subsection, Section \ref{sec:RDPG}, to study various scenarios under this model.
Next,
Section \ref{sec:nonrdpg} covers three other models from Section \ref{sec-models}.
{Due to page limits, the following scenarios were moved to Section \ref{sec:simsupp} of the supplementary materials:
the sparse RDPG, the SBM, the DCBM, and a model misspecification case study.}

\subsection{RDPG model}
\label{sec:RDPG}

\subsubsection{Test of equality}
We consider three cases for the test of equality under the RDPG model.
Case 1 considers the scenario where, under $H_1$, all latent positions are different.
In Case 2, only a small fraction of the latent positions are different under $H_1$.
Case 3 (see Section \ref{sec:simsupp} of supplementary materials) considers a sparse RDPG model with higher sample sizes.

\noindent \textbf{Case 1 (all latent positions vary):}
We generated $A_1 \sim P_1 = X_1X_1^T$ and $A_2 \sim P_2 = X_2 X_2^T$, where $X_1$ and $X_2$ are two latent matrices of dimension $n \times 2$ (i.e., $d=2$). The rows of $X_1$ and $X_2$ are generated by sampling with replacement from the rows of $M_1$ and $M_2$, respectively, with probability vector $\pi$, where 
\begin{equation}
    M_1 = \begin{pmatrix} 0.6 & -0.4\\ 0.6 & 0.4 \end{pmatrix}; \; M_2 = \begin{pmatrix} 0.6 + \epsilon & -0.4 - \epsilon\\ 0.6 + \epsilon & 0.4 + \epsilon \end{pmatrix}; \; \pi  = \begin{pmatrix} 0.4\\ 0.6\end{pmatrix}.
    \label{eq:rdpgmodel}
\end{equation}
We first generate random variables $z_1, \ldots, z_n$ that take values 1 and 2 with probabilities 0.4 and 0.6, and construct the latent matrices $X_1$ and $X_2$ such that the $j^{th}$ row of $X_k$ is $z_j^{th}$ row of $M_k$, for  $k=1,2$ and $j=1,\ldots,n$.
We studied the performance of the methods $T_\text{frob}, T_\text{omni}, T_\text{ase}$ and $T_\text{eig}$ over values of $\epsilon = 0, 0.02, 0.04, 0.06, 0.08, 0.12$, and this experiment is performed for network sizes $n = 100, 200, 300, 400$.
The null hypothesis $H_0 : P_1 = P_2$ holds true when $\epsilon = 0$, and the alternative $H_1 : P_1 \neq P_2$ holds true when $\epsilon>0$. 

Figure \ref{Fin1} displays the mean rejection rates (with standard error bands) as a function of $\epsilon$. Note that the rejection rate for $\epsilon = 0$  is the type-I error and should be close to 5\%. 
We denote this by a solid black square in Figure \ref{Fin1} and all subsequent plots of rejection rates. We observe that $T_\text{omni}$ and $T_\text{frob}$ performs about the same, with $T_\text{omni}$ being slightly better. Both of them perform much better than $T_\text{ase}$ and $T_\text{eig}$.
In particular, $T_\text{ase}$ has very high type-I error as well as low power compared to $T_\text{omni}$ and $T_\text{frob}$.

\noindent \textbf{Case 2 (fraction of latent positions vary):}
Similar to Case 1, we use the model settings from \eqref{eq:rdpgmodel}.
However, under $H_1$, only $10\%$ of the rows of $X_2$ are sampled from $M_2$, and the remaining rows of $X_2$ are set to be the same as the corresponding rows of $X_1$. 
We consider the performance of the methods $T_\text{frob}, T_\text{omni}, T_\text{ase}$ and $T_\text{eig}$ over values of $\epsilon = 0, 0.02, 0.04, 0.06, 0.08, 0.12$ for network sizes $n = 100, 200, 300, 400$. The results are displayed in Figure \ref{Fin2}. We can see that the proposed test $T_\text{frob}$ performs better than $T_\text{ase}$, $T_\text{omni}$ and $T_\text{eig}$, although $T_\text{omni}$ catches up with $T_\text{frob}$ for $n=400$.

{\spacingset{1.4}{
\begin{figure}[h]
\centering
\includegraphics[width=\textwidth]{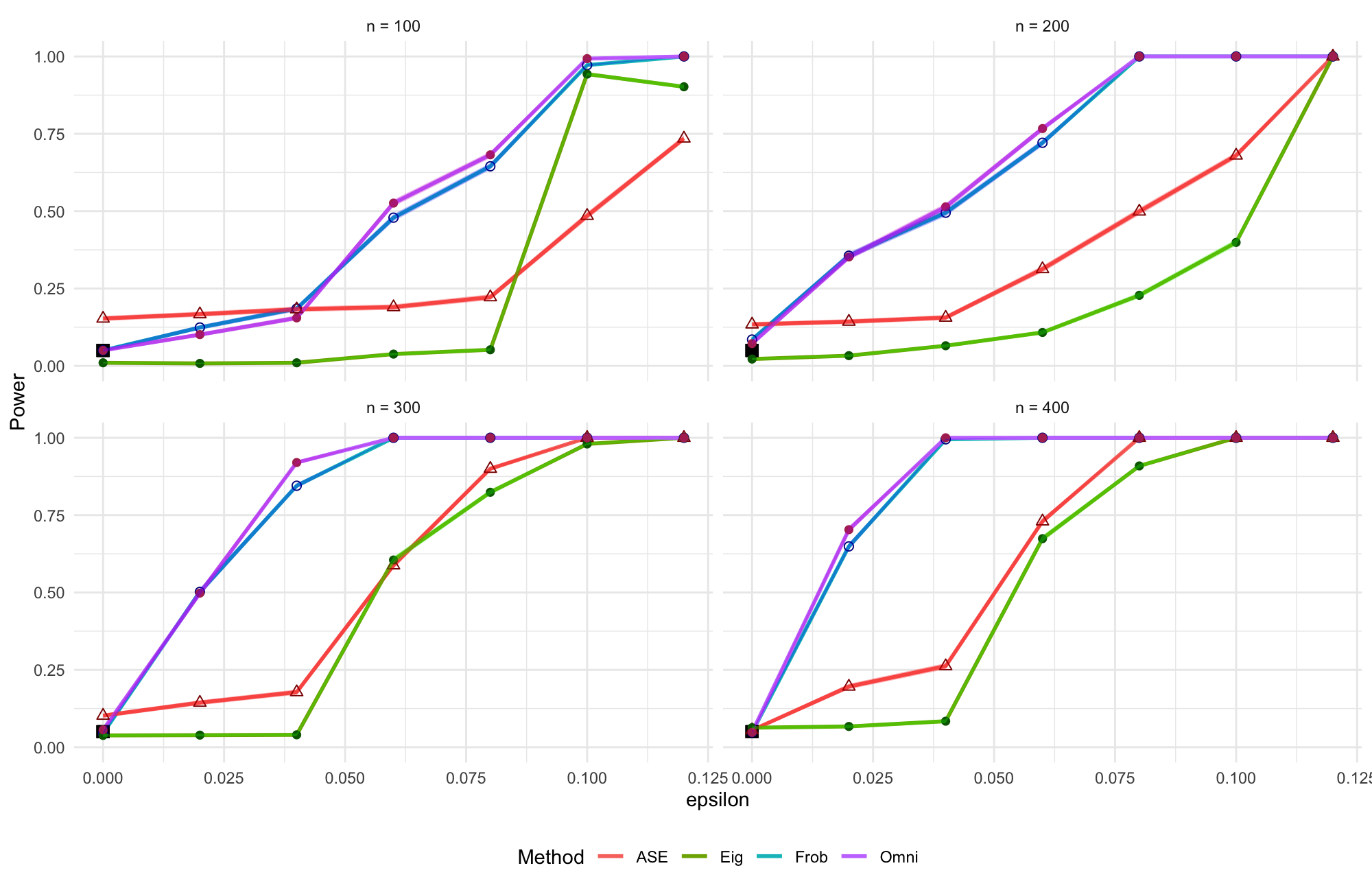}
\caption{\small Rejection rates for RDPG test of equality (Case 1).
The solid black square represents the 5\% significance level for under $H_0$ ($\epsilon = 0$).
\label{Fin1}}
\end{figure}
}

\spacingset{1.4}{
\begin{figure}[h]
\centering
\includegraphics[width=\textwidth]{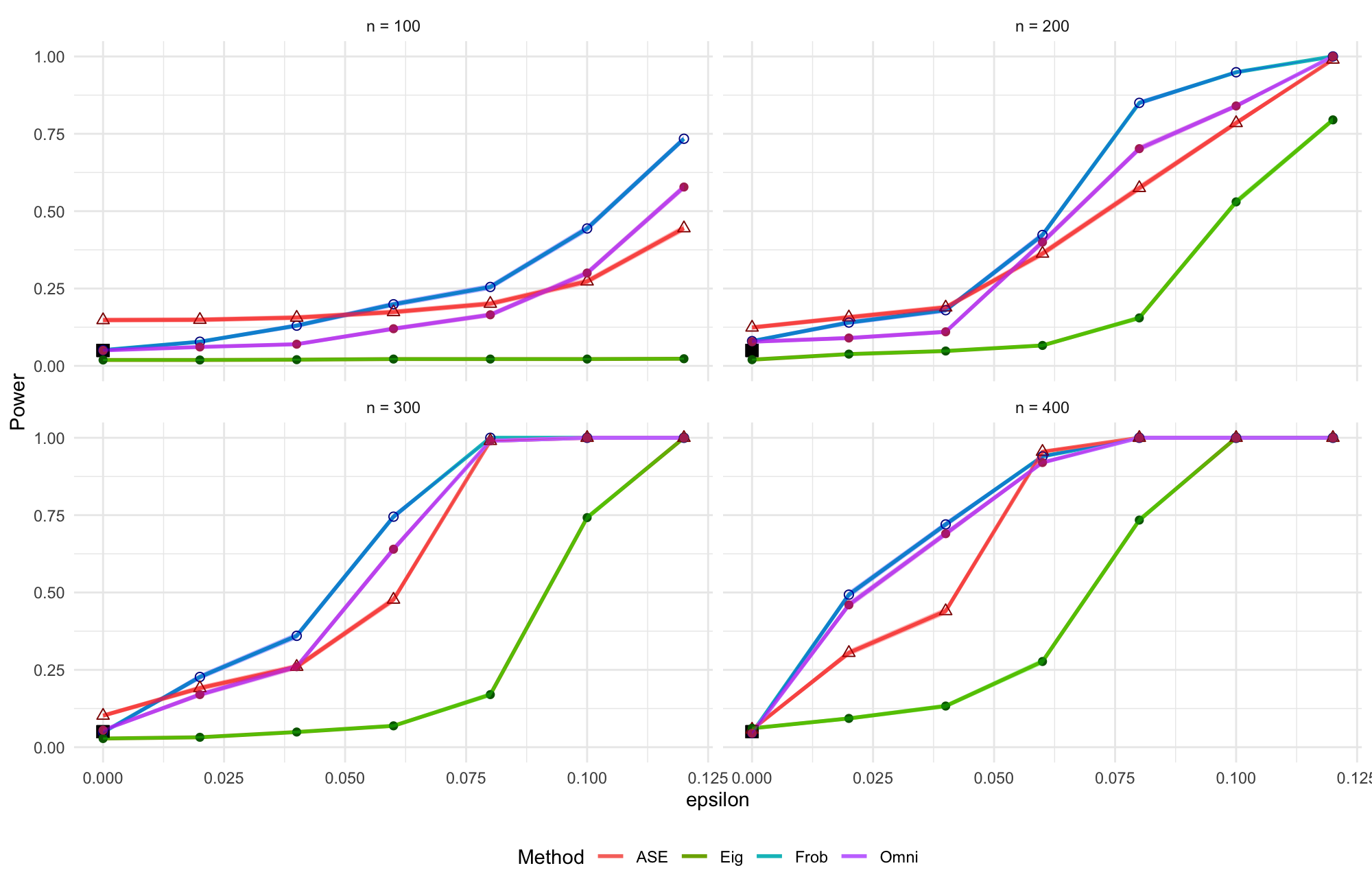}
\caption{\small Rejection rates for RDPG test of equality (Case 2).
The solid black square represents the 5\% significance level for under $H_0$ ($\epsilon = 0$).
\label{Fin2}}
\end{figure}
}}

To summarize, we observe that $T_\text{frob}$ and $T_\text{omni}$ consistently outperform $T_\text{ase}$ and $T_\text{eig}$ across all cases.
The proposed $T_\text{frob}$ has accuracy similar to $T_\text{omni}$, with $T_\text{frob}$ being more accurate in Case 2 and $T_\text{omni}$ being slightly more accurate in Case 1.
However, the difference between $T_\text{frob}$ and $T_\text{omni}$ becomes negligible when the sample size is increased to $n=400$.

Beyond statistical accuracy, it is also important to consider computational cost. 
We report the runtimes of the four methods in Figure \ref{Time} as a function of $n$.
We observe that $T_\text{frob}$ is approximately four times faster than $T_\text{omni}$ and twice as fast as $T_\text{ase}$, as $T_\text{omni}$ involves spectral decomposition of a $2n \times 2n$ matrix \citep{levin2017central} and $T_\text{ase}$ requires Procrustes transformation.
Thus, $T_\text{frob}$ and $T_\text{omni}$ have comparable accuracy, and both of them are much more accurate than $T_\text{ase}$ and $T_\text{eig}$.
Given its much lower computational cost than $T_\text{omni}$, the proposed $T_\text{frob}$ is therefore the preferred method.

\spacingset{1.4}{
\begin{figure}[h!]
\centering
\includegraphics[width=0.75\textwidth]{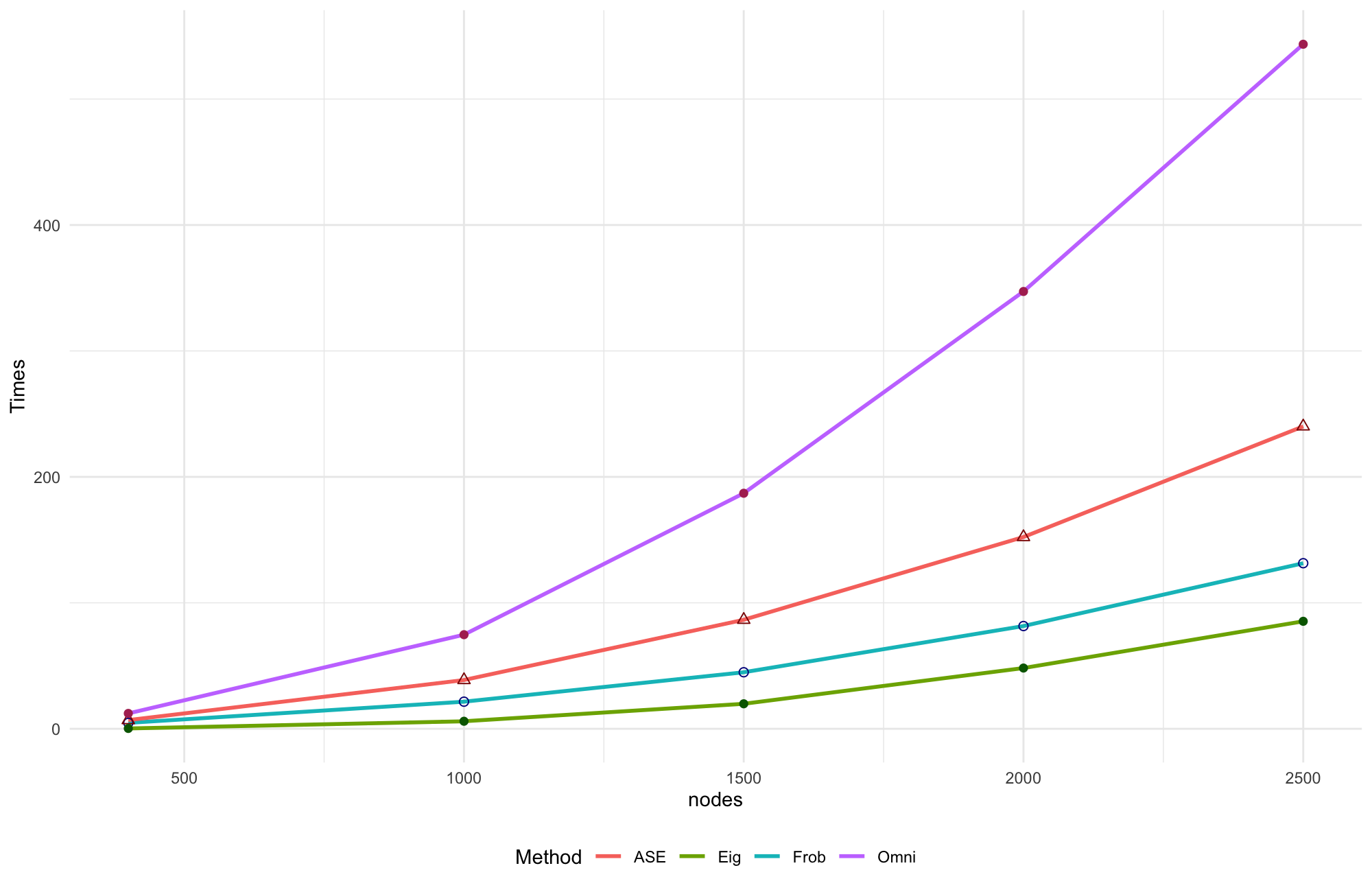}
\caption{\small Runtime comparison between various methods
\label{Time}}
\end{figure}
}

\subsubsection{Test of scaling}
As before, we first generate random variables $z_1, \ldots, z_n$ that take values 1 and 2 with probabilities 0.4 and 0.6, and construct the latent matrices $X_1$ and $X_2$ such that the $j^{th}$ row of $X_k$ is $z_j^{th}$ row of $M_k$, for  $k=1,2$ and $j=1,\ldots,n$, where
\begin{equation}
    M_1 = \begin{pmatrix} 0.6 & -0.4\\ 0.6 & 0.4 \end{pmatrix}; \; M_2 = \begin{pmatrix} 0.3 & -0.1\\ 0.3 & 0.1 \end{pmatrix}.
\end{equation}
 Next, we generate three network adjacency matrices:
 $A_1 \sim P_1 = X_1X_1^T$,
 $A_2 \sim P_2 = (\sqrt{c}X_1) (\sqrt{c}X_1)^T$ for some constant $c>0$, and
 $A_3 \sim P_3 = X_2 X_2^T$.
 Here,  the pair $(A_1, A_2)$ satisfies $H_0$ and we performed the test of scaling with $c=0.5, 0.7, 0.75, 0.8, 0.9$ to obtain Type-1 error rates. 
 We also performed the test using the pair $(A_1, A_3)$ to obtain the power under $H_1$.
  % Finally, for finding the rejection rate under the alternative hypothesis $H_0: P_1 \neq cP_2$, we perform the test between $A_1$ and $A_3$, since by construction $P_3$ is not a scaled version of $P_1$. The tests are repeated over 2000 Monte Carlo iterations to obtain the rejection rates.
Table \ref{tab_rdpg2} shows that $T_\text{ase}$ has low power for $n=100$ and $n=200$, and Type-1 error close to zero in most cases.
The proposed method, $T_\text{scale}$, performed much better with Type I error rates closer to the target value of $\alpha = 5\%$ under $H_0$ and higher power under $H_1$.
The only exception is $n=200, c = 0.9$ where the type-I error from the $T_\text{ase}$ test is closer to  $\alpha = 5\%$.
Note that $T_\text{eig}$ and $T_\text{omni}$ do not apply to the test of scaling.

\spacingset{1.4}{
\begin{table}[h]
\centering
\begin{tabular}{|c||cccccccccc|cc|}
\hline
\multirow{2}{*}{} & \multicolumn{10}{c|}{$H_0$ is true}                                                                                                                                                                                                                                                                                          & \multicolumn{2}{c|}{$H_1$ is true}           \\ \cline{2-13} 
                  & \multicolumn{2}{c|}{$P_2 = 0.5P_1$}                               & \multicolumn{2}{c|}{$P_2 = 0.7P_1$}                               & \multicolumn{2}{c|}{$P_2 = 0.75P_1$}                              & \multicolumn{2}{c|}{$P_2 = 0.8P_1$}                               & \multicolumn{2}{c|}{$P_2 = 0.9P_1$}          & \multicolumn{2}{c|}{$P_2 \neq cP_1$}         \\ \hline
$n$               & \multicolumn{1}{c|}{$T_\text{scale}$} & \multicolumn{1}{c|}{$T_\text{ase}$} & \multicolumn{1}{c|}{$T_\text{scale}$} & \multicolumn{1}{c|}{$T_\text{ase}$} & \multicolumn{1}{c|}{$T_\text{scale}$} & \multicolumn{1}{c|}{$T_\text{ase}$} & \multicolumn{1}{c|}{$T_\text{scale}$} & \multicolumn{1}{c|}{$T_\text{ase}$} & \multicolumn{1}{c|}{$T_\text{scale}$} & $T_\text{ase}$ & \multicolumn{1}{c|}{$T_\text{scale}$} & $T_\text{ase}$ \\ \hline \hline
100               & \multicolumn{1}{c|}{9.2}        & \multicolumn{1}{c|}{1.2}       & \multicolumn{1}{c|}{7.2}         & \multicolumn{1}{c|}{1.6}       & \multicolumn{1}{c|}{6.6}         & \multicolumn{1}{c|}{1.4}       & \multicolumn{1}{c|}{6.8}         & \multicolumn{1}{c|}{2.2}       & \multicolumn{1}{c|}{4.7}         & 6.6       & \multicolumn{1}{c|}{85.4}        & 25.5     \\ \hline
200               & \multicolumn{1}{c|}{10.0}         & \multicolumn{1}{c|}{0.0}       & \multicolumn{1}{c|}{6.1}         & \multicolumn{1}{c|}{0.0}       & \multicolumn{1}{c|}{7.4}         & \multicolumn{1}{c|}{0.0}       & \multicolumn{1}{c|}{5.6}         & \multicolumn{1}{c|}{0.0}       & \multicolumn{1}{c|}{8.4}         & 3.9       & \multicolumn{1}{c|}{100}         & 98.3      \\ \hline
300               & \multicolumn{1}{c|}{8.4}         & \multicolumn{1}{c|}{0.0}       & \multicolumn{1}{c|}{5.6}         & \multicolumn{1}{c|}{0.0}       & \multicolumn{1}{c|}{6.1}         & \multicolumn{1}{c|}{0.0}       & \multicolumn{1}{c|}{5.9}         & \multicolumn{1}{c|}{0.0}       & \multicolumn{1}{c|}{8.4}         & 1.8       & \multicolumn{1}{c|}{100}         & 100      \\ \hline
400               & \multicolumn{1}{c|}{6.5}         & \multicolumn{1}{c|}{0.0}       & \multicolumn{1}{c|}{5.8}         & \multicolumn{1}{c|}{0.0}       & \multicolumn{1}{c|}{6.4}         & \multicolumn{1}{c|}{0.0}       & \multicolumn{1}{c|}{6.1}         & \multicolumn{1}{c|}{0.0}       & \multicolumn{1}{c|}{4.5}         & 0.4       & \multicolumn{1}{c|}{100}         & 100      \\ \hline
\end{tabular}
\caption{\small RDPG scaling case: Rejection rates (in percentage) from $T_\text{scale}$ (the proposed method) and $T_\text{ase}$ for the scaling case  using $B=200$ bootstrap iterations and averaged over 2000 Monte Carlo simulations.
The first five scenarios refer to Type I error rates, and the third scenario refers to the power of the test.
Our method performed much better than $T_\text{ase}$, with Type I error rates closer to the target value of $\alpha = 5\%$ and higher power.
% \ssg{Add more scaling values like the other scaling tables}
\label{tab_rdpg2}}
\end{table}
}

% \clearpage

\subsection{Results for non-RDPG models}
\label{sec:nonrdpg}
\subsubsection{Chung-Lu model}
\begin{figure}[h]
\centering
\includegraphics[width=\textwidth]{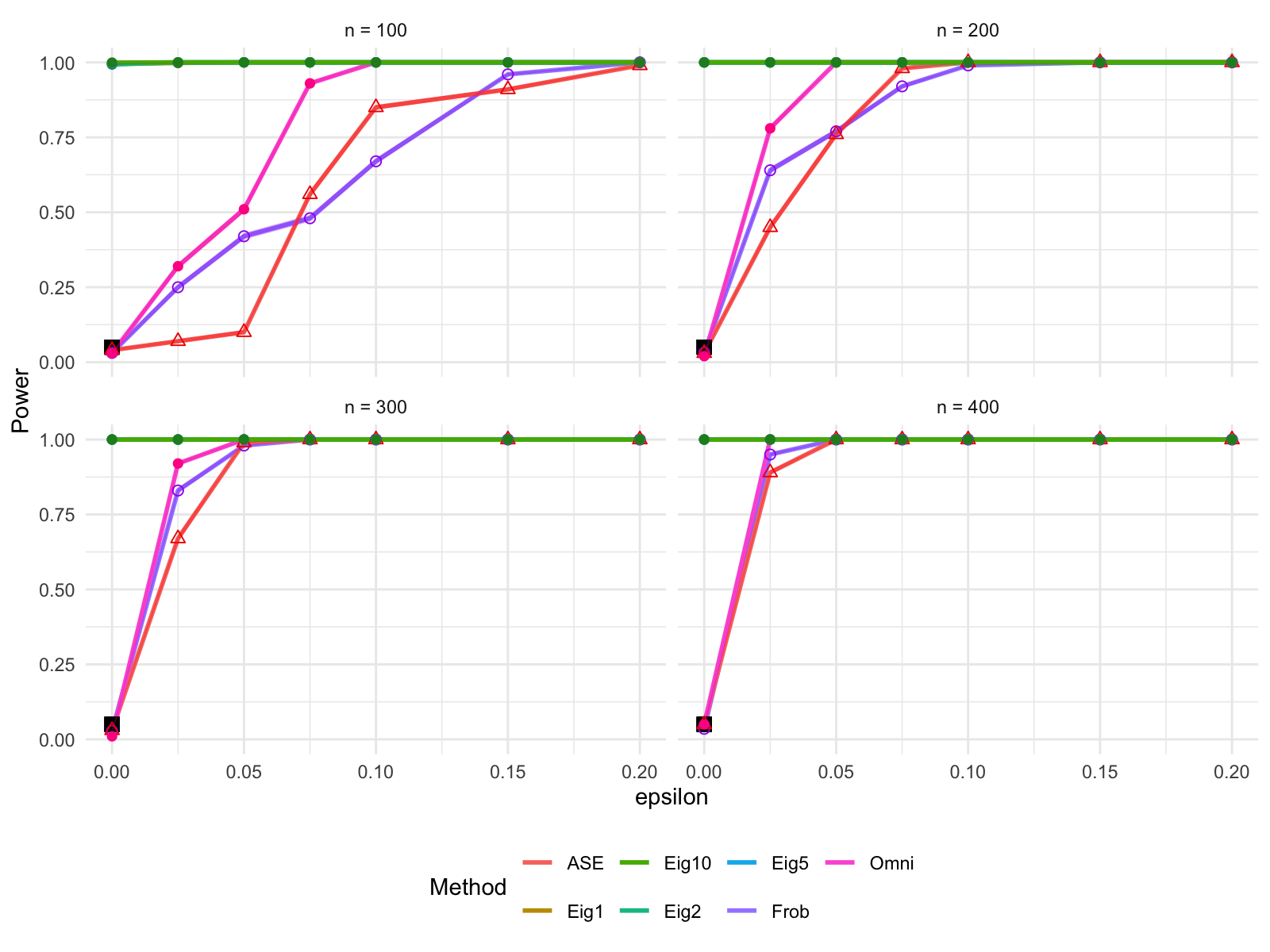}
\caption{\small Chung-Lu Equality test: Rejection rates for deviating alternatives. $T_\text{frob}$, $T_{\text{ase}}$, and $T_{\text{omni}}$ performed well with Type I error rates close to the target value of $\alpha = 5\%$ and power close  to 1.
Type I error rates from $T_\text{eig}$ were very high for all values of $r=1,2,5,10$.
\label{plot_chunglu}}
\end{figure}

% \textcolor{red}{Writing updated by Kaustav}\\
For the Chung-Lu model, we sampled one set of parameters $\mathbf{\theta}_i\sim \text{Beta}(a=1,b=5)$ for $i=1, \ldots, n$, and used $P_1(i,j) = \theta_i \theta_j$, and used $P_2 = P_1$ to configure the null scenario under the equality case.
To configure the alternative scenario, we set $P_2(i,j)=(\theta_i+\epsilon)(\theta_j+\epsilon)$.
For comparison, we have implemented the $T_{\text{ase}}$ method of \cite{tang2017semiparametric}, the $T_{\text{omni}}$ method of \cite{levin2017central} and the $T_\text{eig}$ method of \cite{ghoshdastidar2018practical}.
As noted earlier, the number of blocks $r$, has to be provided as an input to the $T_\text{eig}$ method and the authors did not provide a strategy for obtaining $r$.
We used four ad-hoc values, $r=1,2,5,10$, to study the performance of $T_\text{eig}$.
Figure \ref{plot_chunglu} shows the increase in the rejection rate for the equality case in the range  $\epsilon=0$, $0.025$, $0.05$, $0.075$, $0.1$, $0.15$, $0.2$. We can observe that the three methods $T_{\text{ase}}$, $T_{\text{omni}}$ and $T_{\text{frob}}$ perform well.
$T_{\text{omni}}$ performs the best for smaller values of $n$
 but $T_{\text{frob}}$ and $T_{\text{ase}}$ catch up to $T_{\text{omni}}$ for $n=400$. $T_\text{eig}$ has a high rejection rate even when null is true for different $r$ values, making it the worst performing method in this case.

For the scaling case, we used the same $P_1$ as the equality case, and set $P_2 = cP_1$ with $c=0.5,0.7, 0.75, 0.8, 0.9,$ to configure a range of scenarios satisfying $H_0$.
We generated another separate set of parameters $\eta_i\sim\text{Beta}(a=4,b=3)$ and used $P_2(i,j)=\eta_i\eta_j$ under $H_1$.
Note that no existing method applies to the test of scaling under the Chung-Lu model.
Table \ref{tab_cl2} shows that the proposed method, $T_\text{scale}$,
performed reasonably well, with Type I error rates somewhat higher than $\alpha = 5\%$, and power equal to 1.

\begin{table}[ht]
{
  \begin{center}
    \begin{tabular}{|c|c|c|c|c|c|c|}
    \hline
    & \multicolumn{5}{c|}{$H_0$ is true} & 
    {$H_1$ is true}\\
    \hline
    $n$ &  $P_2 = 0.5 P_1$ & $ P_2 = 0.7P_1$ & $P_2 = 0.75 P_1$ & 
    $ P_2 = 0.8P_1$ & $ P_2 = 0.9P_1$ & $ P_2 \neq cP_1$ \\
    \hline     \hline
     100 & $11.4$ & $8.8$ & 8.2 & $9.0$ & $8.0$ & $100$ \\
    \hline
    200 & $9.1$ & $9.8$ & 8.6 & $8.3$ & $8.6$ & $100$ \\
    \hline
    300 & $8.3$ & $8.5$ & 7.8 & $7.4$ & $8.2$ & $100$\\
    \hline
    400 & $7.6$ & $7.8$ & 7.4 & $8.1$ & $7.0$ & $100$\\
    \hline \hline
\end{tabular}
\end{center}
\caption{\small Chung-Lu scaling case: Rejection rates (in percentage) from $T_\text{scale}$ using $B=200$ bootstrap iterations and averaged over 2000 Monte Carlo simulations.
Scenarios $1-5$ satisfy $H_0$ and the sixth scenario satisfies the alternative.
Our method performed reasonably well with Type I error rates somewhat higher than $\alpha = 5\%$ and power equal to 1.
\label{tab_cl2}}}
 \end{table}%

\subsubsection{Popularity Adjusted Block Model}
Here we used parameter configurations from the simulation study of \cite{senguptapabm}. 
We consider networks with $K=2$ equally sized communities. 
Model parameters are set as $\lambda_{ir} = \alpha \sqrt{\frac{h}{1+h}}$ when $r=c_i$, and $\lambda_{ir} = \beta \sqrt{\frac{1}{1+h}}$ when $r \neq c_i$, where $h$ is the homophily factor.
In each community, we designated $50\%$ of the nodes as Category 1 and $50\%$ as Category 2.
We set $\alpha= 0.8, \beta = 0.2$ for category 1 nodes and $\alpha= 0.2, \beta = 0.8$ for category 2 nodes.
For the equality case, we used $h=4$ for $P_1$, set $P_1 = P_2$ under the null, and used $h-\epsilon$ for $P_2$ under the alternative, for $\epsilon=0$, $0.025$, $0.05$, $0.075$, $0.1$, $0.15$, $0.2$, similar to the Chung-Lu case.
Since $T_\text{eig}$ had very high rejection rates under the Chung-Lu model (Figure \ref{plot_chunglu}), we skipped it for the PABM and subsequent models.
For the test of scaling, we used $h=4$ for $P_1$, set $P_2 = cP_1$ under $H_0$ with $c=0.5,0.7,0.75,0.8,0.9$, and used $h=2$ for $P_2$ under $H_1$.

\begin{figure}[h]
\centering
\includegraphics[width=\textwidth]{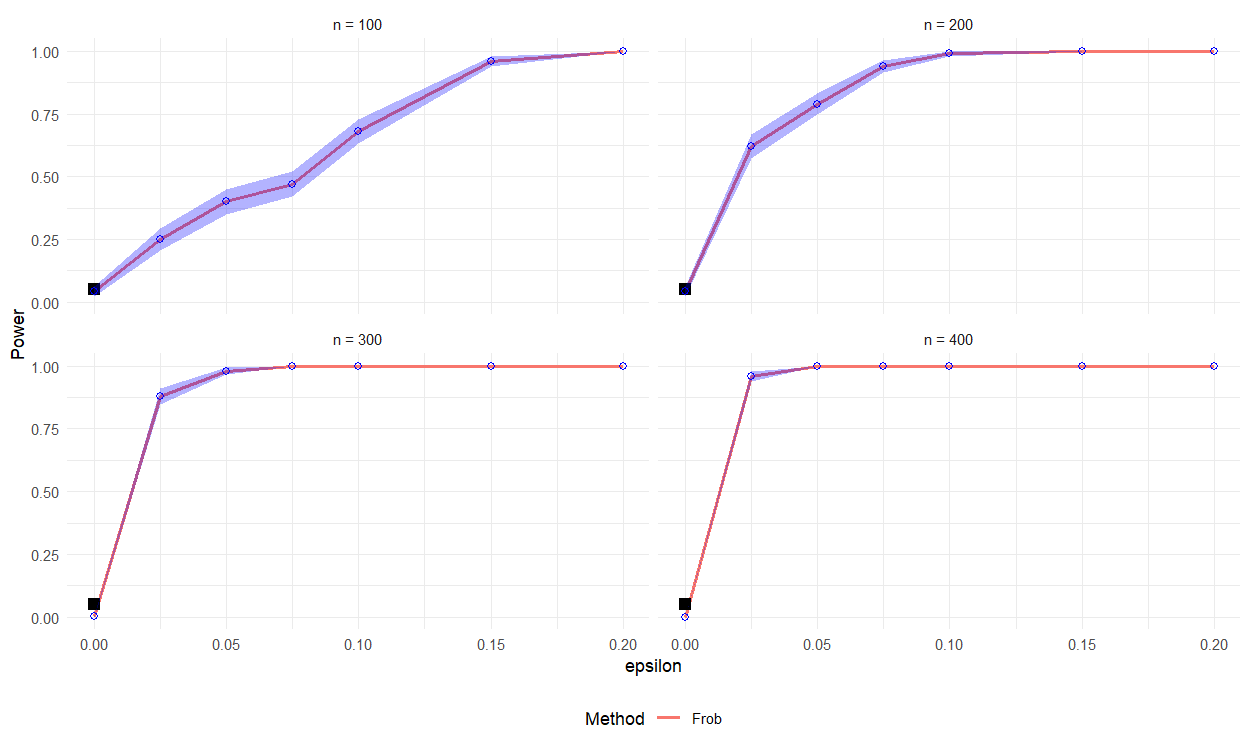}
\caption{\small PABM equality case: Rejection rates for deviating alternatives using $2000$  Monte Carlo simulations.
\label{plot_pabm}}
\end{figure}

The results for the test of equality case are plotted in Figure \ref{plot_pabm} and the same for the test of scaling case are reported in Table \ref{tab_pabm}.
% that makes parameter estimation computationally expensive, 
We note that the proposed tests performed quite well.
The type I errors were somewhat conservative both for the test of equality and the test of scaling.
The power in both cases improved from $n=100$ to $n=400$, as expected.

\begin{table}[ht]
    \begin{center}
    \begin{tabular}{|c||c|c|c|c|c|c|}
    \hline
    & \multicolumn{5}{c|}{$H_0$ is true} & 
    {$H_1$ is true}\\\hline
     & {$ P_2 = 0.5  P_1$} & {$ P_2 = 0.7  P_1$} & {$ P_2 = 0.75 P_1$} &  {$ P_2 = 0.8  P_1$} & {$ P_2 = 0.9  P_1$} & {$ P_2 \neq c P_1$}\\
    \hline
     $n$ & $T_\text{scale}$ & $T_\text{scale}$ & $T_\text{scale}$ & $T_\text{scale}$ & $T_\text{scale}$ & $T_\text{scale}$ \\
     \hline
     100 & $1.1$ & $1.2$ & $1.5$ & $ 2.1$ & $2.9$ &  $34.9$ \\
    \hline
    200 & $0.9$ & $1.1$ & $1.6$ & $1.7$ & $2.3$ &  $99.0$ \\
    \hline
    300 & $0.6$ & $0.9$ & $0.8$ & $1.1$ & $0.8$ &  $100$\\
    \hline
    400 & $0.0$ & $0.2$ & $0.4$ & $0.6$ & $0.8$ & $100$\\
    \hline
   \end{tabular}
\end{center}

    \caption{\small PABM scaling case: Rejection rates (in percentage) from $T_\text{scale}$ using $B=200$ bootstrap iterations.
    Results are averaged over 2000 Monte Carlo simulations. 
       \label{tab_pabm}}
\end{table}

\subsubsection{Latent Space Model} 
Under the latent distance model of \cite{hoff2002latent},
 we used $d=3$, $\alpha = 3$, and sampled the latent positions $z_1, \ldots, z_n \sim N_3(\mathbf{0}, \mathbf{I})$ independently for $P_1$.
 For the test of equality, we set $P_2=P_1$ under $H_0$.
 To configure the alternative scenario for the equality case, we used $\alpha=3-\epsilon$ for $\epsilon=0$, $0.025$, $0.05$, $0.075$, $0.1$, $0.15$, $0.2$, and sampled a different set of $z_1, \ldots, z_n \sim N_3(\mathbf{0}, \mathbf{I})$ independently  for $P_2$.  For the test of scaling, we set $P_2 = cP_1$ under $H_0$ with $c=0.5,0.7,0.75,0.8,0.9$, and used the same $P_2$ used in the test of equality under $H_1$.
 For the scaling case, we used $P_2 = c \times P_1$ under $H_0$, and used $\alpha=3$ and sampled another different set of $z_1, \ldots, z_n \sim N_3(\mathbf{0}, \mathbf{I})$ independently to create $P_2$  under the alternative model.
  It has been well documented 
 that estimation under the latent space model is computationally expensive \citep{raftery2012,salter2013variational}.
 The computational expense for our inferential method is further exacerbated due to bootstrap resampling.
 Therefore, we used smaller sample sizes, $n=30, 40, 50$, and carried out $500$ Monte Carlo iterations for each sample size, and we used $B=200$ bootstrap iterations as before.
  We note that the computational issue can potentially be resolved by using approximation techniques \citep{raftery2012} or variational inference \citep{salter2013variational}; however, we did not pursue this direction in this work, and we consider this as an important future direction.

\begin{figure}[h]
\centering
\includegraphics[width=\textwidth]{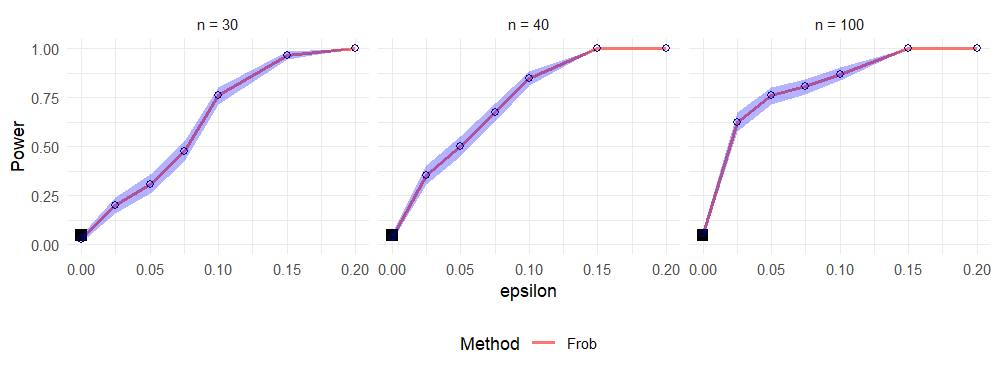}
\caption{\small Rejection rates for the test of equality under the latent space model.
\label{plot_LSM}}
\end{figure}

 The results for the test of equality and the test of scaling are reported in Figure \ref{plot_LSM} and Table \ref{tab_lsm}, respectively.
 Our methods work quite well in both cases, with Type I error rates close to the nominal value of $5\%$ and power equal to $100\%$.
 
 \begin{table}[ht]
    \begin{center}
    \begin{tabular}{|c||c|c|c|c|c|c||}
    \hline
    & \multicolumn{5}{c|}{$H_0$ is true} & 
    {$H_1$ is true}\\\hline
     & {$ P_2 = 0.5  P_1$} & {$ P_2 = 0.7  P_1$} & {$ P_2 = 0.75 P_1$} &  {$ P_2 = 0.8  P_1$} & {$ P_2 = 0.9  P_1$} & {$ P_2 \neq c P_1$}\\
    \hline
     n & $T_\text{scale}$ & $T_\text{scale}$ & $T_\text{scale}$ & $T_\text{scale}$ & $T_\text{scale}$ & $T_\text{scale}$ \\
     \hline
     30 & $9.2$ & $7.2$ & $6.0$ & $6.8$ & $5.8$ & $100$ \\
    \hline
    40 & $6.2$ & $5.6$ & $3.6$ & $4.2$ & $5.0$ & $100$ \\
    \hline
    50 & $5.4$ & $5.8$ & $5.2$ & $4.6$ & $5.0$ & $100$ \\
    \hline
\end{tabular}
\end{center}

    \caption{\small Latent space model scaling case: Rejection rates (in percentage) from $T_\text{scale}$ using $B=200$ bootstrap iterations and averaged over 500 Monte Carlo simulations.
    \label{tab_lsm}}
\end{table}

\subsection{Computational considerations}
    {The computational costs of implementing the proposed $T_\text{frob}$ and $T_\text{scale}$ methods are primarily driven by two factors: the cost of computing the estimators $\hat{P}_1$ and $\hat{P}_2$, and the cost of generating bootstrap resamples.
    While the proposed methods are shown to be faster than existing alternatives with comparable statistical accuracy (as demonstrated in Figure \ref{Time}), their efficiency can be further enhanced by improving the scalability of these two components.
    Recent years have witnessed promising advances towards improving the computational efficiency of both components.
    In the network inference literature,
    computationally efficient methods for estimating model parameters have been developed for blockmodels \citep{mukherjee2021two,chakrabarty2023sonnet}, the RDPG model \citep{chakraborty2025scalable}, and the LSM \citep{raftery2012,salter2013variational}.
    Similarly, in the bootstrap literature, computationally efficient alternatives to bootstrap resampling have been introduced by \cite{kleiner2014scalable}, \cite{sengupta2016subsampled}, \cite{politis2024scalable}, and \cite{ganguly2023scalable}.
    Integrating these advances into the proposed methods represents a promising future research direction to further enhance computational efficiency.}

\section{Aarhus Computer Science Department Network}
\label{sec-data}
This anonymized network dataset was collected by \cite{rossi2015towards} at the Department of Computer Science at Aarhus University.
It includes five kinds of interaction (coauthor, leisure, work, lunch, and Facebook) between  $n=61$ researchers, including professors, postdocs, and Ph.D. students.
Importantly, these interactions span both social (e.g., leisure and Facebook) and professional domains (e.g., coauthor and work).
Each kind of interaction is represented as an undirected network. 

We carried out tests for 
equality (using $T_\text{frob}$) and scaling (using $T_\text{scale}$) for each pair of interactions with {$\alpha = 5\%$} and $B=10000$ bootstrap iterations.
Since this involves 20 simultaneous tests of hypothesis (${5 \choose 2}$ pairs each for equality and scaling), we use Bonferroni's correction for multiple testing and set the significance level for each pairwise test at ${\alpha}/{20} = 0.0025$.
% \ssg{Multiple testing?}
Following \cite{han2015consistent}, we used an RDPG model with $d=4$.
The equality hypothesis ($H_0: P_1=P_2$) was rejected for all interaction pairs.

From Table \ref{tab_aucs}, we observe that the scaling hypothesis ($H_0: P_1=cP_2$ for some $c>0$)
was rejected for three pairs of interaction types: \textit{leisure-facebook}, \textit{work-facebook}, and \textit{lunch-facebook}.
The test of scaling was \textit{not} rejected for the seven remaining pairs of interactions.
Notably, the co-author network displayed scaling similarity with every other network, indicating that co-authors are likely to have other kinds of interactions like work, lunch, leisure, and social media, in a \textit{proportional} manner.
This aligns well with our understanding of how academic interactions often intermingle, and how a co-authorship relation can be associated with other forms of professional and social interactions. 
We also observed that the highest p-value in Table \ref{tab_aucs} is for the \textit{leisure} and \textit{lunch} pair, underscoring the strength in their scaling similarity, which makes intuitive sense as these two activities are similar.
Thus, the test results reveal interesting similarities across social and professional modes of interaction while also reflecting differences in communication \textit{intensity} across these modes.
Such findings are valuable as they simplify the analysis of multilayer networks by reducing the effective dimensionality of interactions, enabling domain scientists to focus on the underlying patterns that govern similar behavior across multiple interaction types. 

{\spacingset{1.4}
\begin{table}[h]
    \centering
    \begin{tabular}{|c|c|c|c|c|c|}
    \hline
     \textbf{Interaction type} & coauthor & leisure & work & lunch & facebook\\
        \hline
        coauthor & - & 0.1325 & 0.1104 & 0.1393 & 0.0538 \\
          \hline
         leisure  & - & - & 0.0270 & 0.1956 & 0.0008 \\
          \hline
          work & - & - & - & 0.0144 & 0.0014 \\
          \hline
          lunch & - & - & - & - & 0.0000 \\
          \hline
                \end{tabular}  
    \caption{\small Aarhus network: p-values for the test of scaling (using $T_\text{scale}$) for all interaction pairs.
    % The RDPG model with $d=4$ was used to carry out the tests with $B=1000$ bootstrap iterations.
    % The scaling hypothesis ($P_1=cP_2$ for some $c>0$) was \textit{not rejected} for five interaction pairs: the \textit{co-author} network with all other interactions (\textit{leisure}, \textit{work}, \textit{lunch}, and \textit{facebook}), as well as \textit{leisure} and \textit{lunch}.
    }
    \label{tab_aucs}
\end{table}
} % end new spacing

% \subsubsection{British MP Twitter network}
% \ssg{Drop this?}
% This dataset was curated by \cite{greene2013producing}  and
% consists of nodes as user accounts of 419 British members of Parliament (MPs) on the social media platform \emph{twitter.com} and three kinds of interaction between them, namely \emph{mentions}, \emph{follows}, and \emph{retweets}.
% %The true community assignments are given by the political party affiliations of the MPs.
% % consists of 419 nodes belonging to five political parties,  and three `layers' of edges between them, namely \emph{mentions}, \emph{follows}, and \emph{retweets}.
% %, which are three kinds of interactions that can happen between Twitter users.
% Following \cite{senguptapabm}, we analyze the largest connected component which has 329 nodes, and use the PABM model with $K=2$, using $B=1000$ bootstrap iterations.
% As before, we consider each pair of interactions and carry out matched network inference for both equality and scaling.
% We observed that all p-values were very close to zero, which means both hypotheses of equality and scaling were rejected for all the network pairs.
% % This results makes intuitive sense since professional politicians in the modern era use Twitter actively, and the actions or 

\section{Discussion}
\label{sec:discussion}
This paper studies the matched network inference problem, where the statistician is given two independent networks on the same set of entities, and the goal is to determine whether the two networks are similar.
We propose a bootstrap-based testing framework for two key problems: the equality problem, testing if the networks originate from the same random graph model, and the scaling problem, testing whether their underlying probability matrices are scaled versions of each other.
The proposed methodology works well on a wide range of random graph models, as demonstrated by our theoretical and empirical results, and outperforms existing approaches in terms of flexibility, computational efficiency, and statistical accuracy across a wide range of scenarios.
% We applied the proposed tests on two well-studied network datasets and obtained some interesting results.

For future research, an important next step will be to move beyond equality and scaling and develop testing methods for more general and intricate notions of similarity, $\tau(P)$.
Examples of network features of interest include expected subgraph counts \citep{bhattacharyya2015subsampling}, clique numbers \citep{sengupta2018anomaly}, graph spectra \citep{van2010graph,jovanovic2012spectral,dasgupta2022scalable}, path lengths \citep{watts1998collective,lovekar2021testing}, and so on.
There are two challenges involved in extending the proposed framework to more general $\tau(P)$.
First, it needs to be established that the convergence of $\hat{P}$ to $P$ leads to the  convergence of $\tau(\hat{P})$ to $\tau(P)$.
This hinges on the smoothness properties of $\tau(\cdot)$, and the approximations and expansions analyzed in \cite{zhang2022edgeworth} and \cite{levin2019bootstrapping} could be excellent starting points towards such results.
The second challenge is how to transform $\tau(\hat{P}_1)$ and $\tau(\hat{P}_2)$ into their null-restricted counterparts which can be used to generate parametric bootstrap resamples to estimate the sampling distribution of the test statistic under the null.
We look forward to 
methodological innovation from the research community that will address these challenges.

Another impactful direction of future research would be extending the proposed framework to dynamic or time-varying networks.
Network monitoring has emerged as a highly active area of research in recent years \citep{woodall2017overview,jeske2018statistical,sengupta2018discussion,stevens2021broader,stevens2021foundations,stevens2021interdisciplinary,stevens2021research}.
In this context, the proposed framework can be used for changepoint analysis and anomaly detection for identifying sudden shifts in connectivity patterns as the network evolves over time, by testing for $H_0: \tau(P_t) = \tau(P_{t+1})$ where $t$ denotes time.

\section*{Supplementary Material}
The supplementary material contains algorithms and proofs of technical results

\bibliographystyle{apalike}
\bibliography{ref}

\clearpage
\setcounter{page}{1}
\setcounter{equation}{0} 
\setcounter{section}{0} 
\renewcommand{\theequation}{S\arabic{equation}}
\renewcommand{\thesection}{S\arabic{section}}

\begin{center}
		{\Large \bf Supplementary materials for \\
  ``A Bootstrap-based Method for Testing Similarity of Matched Networks''}
	\end{center}

\section{Proofs of bootstrap distributional results (Theorems \ref{thm:sbmdist1} - \ref{thm:sbmboot2})}
For  vectors $\bfc, \bfd\in \Sigma^n$ for some finite alphabet $\Sigma$, define the Hamming distance between 
$\bfc=(c_1,\ldots,c_n)'$ and $\bfd
=(d_1,\ldots,d_n)'$ as 
$$
H(\bfc, \bfd) = \sum_{i=1}^n \ind(c_i\neq d_i),
$$
where $\ind(\cdot)$ denotes the indicator function. Thus, $H$ counts the number of mismatches among the components 
of $\bfc$ and $\bfd$. For $n\geq 1$, 
write $\hat{\bfc}_n=(\hc_1,\ldots,\hc_n)'$ 
and $\bfc_n=(c_1,\ldots,c_n)'$
for the estimated and the true community assignments (after relabeling as 
necessary).\\

\begin{lemma}
For any $1\leq r,s\leq K$,
$$ \big|\hn_{rs}-n_{rs}\big| \leq 2 n H(\hat{\bfc}_n, \bfc_n)
\qmq{for all} n\geq 1.
$$
\label{lemmas1}
\end{lemma}

\noindent
{\bf Proof:}~Clearly,
\begin{align*}
   \big|   \hn_{rs}-n_{rs}   \big|  & = \Big|  \sum_{i\neq j} \ind(\hc_i=r, \hc_j=s)
    -    \sum_{i\neq j} \ind(c_i=r, c_j=s)
   \Big| \\
      &=  \Big|  \sum_{i\neq j} \Big\{  \ind(\hc_i=r)
    -    \ind(c_i=r)\Big\} \ind( \hc_j=s)\Big\} 
   \Big| \\
    & \hspace{.5in} + \Big|  \sum_{i\neq j} \Big\{ \ind(\hc_j=s)
    -    \ind(c_j=s)\Big\} \ind( c_i=r)\Big\} 
   \Big| \\
    &\leq  \sum_{i\neq j} \ind(\hc_i\neq c_i) + \sum_{i\neq j} \ind(\hc_j\neq c_j),
 %   &\leq 2nH(\hat{\bfc}_n, \bfc_n). 
\end{align*}
which proves the lemma. \qb

\begin{corollary} Under conditions (C.1) and (C.3), 
$$\hn_{rs}
= n_{rs}[1+o_p(1)] \qmq{ for all}  1\leq r,s\leq K. 
$$
\label{cors2}
\end{corollary}

Next let $\{Z\sa_{rs} : 1\leq r\leq s\leq K\}$ be a collection of iid 
N(0,1) random variables and set $Z\sa_{rs}= Z\sa_{sr}$ if $r>s$. 
Write $\bfZ\sa=(( Z\sa_{rs}))_{K\times K}$. 
Also, let 
$\{Z_{n,rs}: 1\leq r, s\leq K\}_{n\geq 1}$ 
be a collection of (generic) random variables 
such that 
$\{Z_{n,rs}: 1\leq r \leq s\leq K\}$ 
are 
independent for each $ {n\geq 1}$,
and $Z_{n,rs}= Z_{n,sr}$ if $r>s$,
and $Z_{n,rs} \raw^d Z\sa_{rs} $
for all $r,s$.  Also, write $\C_r=\{ i: 1\leq i\leq n, c_i=r\}$ for the index set for 
the $r$th 
community. \\

\begin{lemma}
For any  $ 1\leq r,s\leq K$, we have the representation: 
\begin{align*}
\Big(\hat{\om}_{rs} - \om_{rs}  \Big)
& = \sqrt{\frac{[1+\ind(r=s)]\om_{rs}(1-\om_{rs})}{n_{rs}}} \cdot Z_{n,rs} +R_{n,rs}
\end{align*}
where $R_{n,rs}\leq 
4nH(\hat{\bfc}_n, \bfc_n)/\hn_{rs}$
for all 
$n\geq 1$.
\label{lemmas3}
\end{lemma}

\noindent
{\bf Proof:}~
Write $\tilde{\om}_{rs} = \sum_{i\neq j}  A(i,j) \ind(c_i=r, c_j=s)/ n_{rs}    $. Then, it is easy to check that for $r\neq s$, 
\begin{align*}
    \big(\hat{\om}_{rs} - \om_{rs}  \big)
     & = \big(\tilde{\om}_{rs} - \om_{rs}  \big) + \big(\hat{\om}_{rs} - \tilde{\om}_{rs}  \big)\\
    & = n_{rs}^{-1}\sum_{i\in \C_r}\sum_{j\in \C_s} [A(i,j) - \om_{rs}] 
    +       
   \sum_{i\in \C_r}\sum_{j\in \C_s} A(i,j)\frac{[n_{rs}-\hn_{rs}]}{ \hn_{rs}  n_{rs}}
    \\
     & + \hn_{rs}^{-1}\sum_{(i,j): i\neq j} A(i,j)\big[\ind(\hc_i=r, \hc_j=s)
              - \ind(c_i=r, c_j=s) \big] \\
          & \equiv I_{1n} + I_{2n} +I_{3n}, \qmq{say.}
\end{align*}
By Lemma \ref{lemmas1} and the fact that $|A(i,j)|\leq 1$, 
it follows that 
$I_{2n}\leq 2nH(\hat{\bfc}_n, \bfc_n)/\hn_{rs}$. Also, using arguments similar to the proof of 
Lemma \ref{lemmas1}, one can show that 
$I_{3n} \leq 2n H(\hat{\bfc}_n, \bfc_n)/\hn_{rs}$. Now
 set $R_{n,rs}= I_{2n}+I_{3n}$ and 
 $Z_{n, rs}=  [\om_{rs}(1-\om_{rs})n_{rs}]^{-1/2}\sum_{i\in \C_r}\sum_{j\in \C_s} [A(i,j) - \om_{rs}] $. Then,
 by the independence of $\{A(i,j) : 1\leq i<j\leq n\}$ and the Central Limit Theorem (CLT), the conclusions of the lemma follow for $r\neq s$. 
 The proof of the case $r=s$ is similar. The only notable change is due to the fact for
 $r=s$, $A(i,j)=A(j,i)$ for all $i\neq j, i,j\in \C_r$ and therefore, only half as many variables in the sum are independent.
 As a result, we write 
 the first term (i.e., $I_{1n}$) as
 \begin{align*}
 I_{1n}&= n_{rr}^{-1} \sum_{(i,j): i\neq j} \ind(c_i=r=c_j) [A(i,j) - \om_{rr}]\\
 & = 2 n_{rr}^{-1} \sum_{(i,j): i< j} \ind(c_i=r=c_j) [A(i,j) - \om_{rr}]\\
 &\equiv  [2\om_{rs}(1-\om_{rs})/n_{rs}]^{1/2} Z_{n, rr},
\end{align*}
where, it is now easy to check that  $Z_{n,rr}$ satisfies the independence and the asymptotic normality requirements of the lemma. 
\qb

\begin{remark}
    Note that for 
the expected number of edges to grow with $n$,
we must have 
$a_n. n^2\raw \infty$ which we tacitly assumed in (C.2). 
%Thus,  Condition (C.3) implies that $EH(\hat{\bfc}_n, \bfc_n) = o (\sqrt{n})$. 
This, in particular,  implies
that 
$$
\Big(\hat{\om}_{rs} - \om_{rs}  \Big)
= \sqrt{\frac{[1+\ind(r=s)]\om_{rs}(1-\om_{rs})}{n_{rs}}} \cdot Z_{n,rs} \big[1+o_p(1)\big]. 
$$
\end{remark}

\begin{remark}
Here we provide further clarification about the sufficient condition on the 
sparsity factor $a_n$ for the validity of Condition (C.3) for the algorthm in
\cite{gao2017achieving,  gao2018community}. Let $q^0 \equiv \max_{1\leq r\neq s\leq K} \om_{rs}^0 <  
p^0
\equiv \min_{1\leq r\leq K} \om_{rr}^0 $. Then, the minimax bound on the misclassification 
rate in Section 2.2 of
\cite{gao2018community}  yields (in our notation) 
\begin{equation*}
EH(\hat{\bfc}_n, \bfc) = O\Big( n \exp\big( - C(K) na_n \big[\sqrt{p^0} - \sqrt{q^0}\, \big]^2\big)\Big)
\qmq{as $\nti$,}
\end{equation*}
      for some constant $C(K)\in (0,\infty)$, depending only on 
    $K$ and $\{\pi_r : 1\leq r \leq K\}$ (and not on $n$). Condition (C.3) 
    now easily follows from this when $na_n\gg \log n$.  
\end{remark}

% We are now ready to prove Theorem \ref{thm:sbmdist1}.

\subsection{Proof of Theorem \ref{thm:sbmdist1}}
Note that under $H_0: P_1=P_2$,
\begin{align}
T_{1n}^2 &= a_n^{-1}  \|\hat{P}_1 - \hat{P}_2\|_F^2
                 \nonumber\\
&= \sum_{(i,j) : i\neq j} 
                  a_n^{-1} [\hP_1(i,j) - P_1(i,j)]^2
+ \sum_{(i,j) : i\neq j} 
                  a_n^{-1}[\hP_2(i,j) - P_2(i,j)]^2 \nonumber\\
 & \hspace{.8in} -2
\sum_{(i,j) : i\neq j} 
                  a_n^{-1}[\hP_1(i,j) - P_1(i,j)]
                  [\hP_2(i,j) - P_2(i,j)]. 
                  \label{tstat1}
\end{align}

Consider the first term. Using the fact that for
any $i=1,\ldots,n$, 
$\sum_{r=1}^K \ind(c_i=r) =1$, we have 
\begin{align}
   & \sum_{(i,j) : i\neq j} 
                  a_n^{-1}[\hP_1(i,j) - P_1(i,j)]^2 
= \sum_{(i,j) : i\neq j} 
                  a_n^{-1}[\hat{\om}_{\hc_i, \hc_j} - 
                  \om_{c_i, c_j} ]^2 \nonumber\\
&
=\sum_{r=1}^K\sum_{s=1}^K 
     \sum_{(i,j) : i\neq j} 
                  a_n^{-1}[\hat{\om}_{\hc_i, \hc_j} - \om_{rs}]^2
                     \ind(c_i=r, c_j=s)\nonumber\\
&
=\sum_{r=1}^K\sum_{s=1}^K    \sum_{(i,j) : i\neq j}
a_n^{-1}[\hat{\om}_{rs} - \om_{rs}]^2
                     \ind(c_i=r, c_j=s)\nonumber\\
&+\sum_{r=1}^K\sum_{s=1}^K 
     \sum_{(i,j) : i\neq j} a_n^{-1}
\Big\{ [\hat{\om}_{rs} - \om_{rs}]^2 +
                  [\hat{\om}_{\hc_i, \hc_j} - \om_{rs}]^2
                  \Big\} 
                     \ind(c_i=r, c_j=s)
                     \big[1- \ind(\hc_i=r, \hc_j=s)\big]
                    \nonumber \\                 
&   \equiv 
   J_{1n}+J_{2n}, \qmq{(say).}
  \label{j12}
\end{align}
Then, by Corollary \ref{cors2}, Lemma \ref{lemmas3}, and Condition (C.3), 
\begin{align}
J_{1n} &= \sum_{r=1}^K\sum_{s=1}^K n_{rs} a_n^{-1} [\hat{\om}_{rs} - \om_{rs}]^2~ = \sum_{r=1}^K\sum_{s=1}^K \tau_{n,rs} Z_{n,rs}^2 +o_p(1)
 \label{j1}
\end{align}
where $\tau_{n, rs}= \om\sa_{rs} (1-a_n \om_{rs}\sa) $ if 
$r\neq s$ and $\tau_{n, rs}= 
2\om\sa_{rs} (1-a_n\om_{rs}\sa) $ if 
$r=s$. Further, by Corollary \ref{cors2} and Lemma \ref{lemmas3},  
\begin{align}
J_{2n} &\leq \sum_{r=1}^K\sum_{s=1}^K 
 \sum_{(i,j) : i\neq j} a_n^{-1}
2 \Big\{ \hat{\om}_{rs}^2+2\om_{rs}^2 +
                  \hat{\om}_{\hc_i, \hc_j}^2
                  \Big\} 
                     \ind(c_i=r, c_j=s)
                     \big[\ind(\hc_i\not=r)+
                     \ind(
                     \hc_j\not=s)\big]
                    \nonumber \\                 
& \leq  4a_n^{-1} \max_{1\leq r,s\leq K}\{\hat{\om}_{rs}^2+\om_{rs}^2\} \,\, \cdot \,\Big[
 2 \sum_{r=1}^K\sum_{s=1}^K 
 \sum_{(i,j) : i\neq j}
 \ind(c_i=r, c_j=s)
\ind(\hc_i\not=r)\Big]
  \nonumber \\                 
& =  8a_n^{-1} \max_{1\leq r,s\leq K}\{\hat{\om}_{rs}^2+\om_{rs}^2\} 
\,\, \cdot \,\Big[ (n-1)H(\hat{\bfc}_n, \bfc_n)\Big] 
=o_p(1),\qmq{by (C.3).}
 \label{j2}
\end{align}

Now using the identity \eqref{tstat1}, the independence of 
the adjacency matrix $A_1$ and $A_2$, and 
relations \eqref{j1} and \eqref{j2} (and their analogs for
$(\hat{P}_2-P_2)$, by the continuous mapping theorem, we conclude
that 
$$
T_{1n}^2\raw^d ~  \sum_{r=1}^K\sum_{s=1}^K
\Big( \tau\sa_{1, rs}
[Z_{1,rs}\sa]^2 + \tau\sa_{2, rs}
[Z_{2,rs}\sa]^2
 - 2 \sqrt{ \tau\sa_{1, rs} \tau\sa_{2, rs}}
Z_{1,rs}\sa Z_{2,rs}\sa
\Big), 
$$
where $\tau\sa_{1, rs} = \om_{rs}\sa$ 
for $r\neq s$ and $\tau\sa_{1, r2} = 2 \om_{rr}\sa$,  
and $Z_{1,rs}\sa$
is the $(r,s)$th component of $\bfZ_1\sa$, and 
$\tau\sa_{2, rs}$,  $Z_{2,rs}\sa$ are similarly defined for 
Population 2. This proves the theorem. \qb

\subsection{Proof of Theorem \ref{thm:sbmdist2}}

To prove Theorem \ref{thm:sbmdist2}, we will need the following lemma.

\noindent
\begin{lemma}
    Under Conditions (C.1)-(C.3), 
$$
\hat{\rho}_k - \rho_k = \sqrt{a_n} \cdot V_k\sa \cdot \big(1+o_p(1)\big)
$$
where $ V_k\sa=  \frac{\sum_{r=1}^K\sum_{s=1}^K \tau_{k, rs}\sa \pi_{rs}\sa \om_{k,rs}\sa Z_{k,rs}\sa}{
\sqrt{\sum_{r=1}^K\sum_{s=1}^K \pi_{rs}\sa (\om_{k,rs}\sa)^2 } 
}, 
$ $k=1,2$. 
\label{lemmas6}
\end{lemma}

\noindent
{\bf Proof:} 
We only consider the case $k=1$. Also, 
for notational simplicity,  we drop the subscript $1$
from $\tau_{1,rs}, \om_{1,rs}, ...$ etc. 
Note that using Lemma \ref{lemmas3}, Condition (C.2) and 
the arguments  in the derivation of \eqref{j2}, one can show that 
\begin{eqnarray*}
\hat{\rho}_1^2 - \rho_1^2 &=& \|\hat{P}_1\|_F^2 - \|P_1\|^2_F
\\
&=& \sum_{r=1}^K\sum_{s=1}^K \sum_{(i,j): i\neq j} [\hat{P}(i,j)^2 -\om_{rs}^2]
                   \ind(c_i=r, c_j=s)\\
&=& \sum_{r=1}^K\sum_{s=1}^K n_{rs}
      [\tilde{\om}_{rs}-\om_{rs}] [\tilde{\om}_{rs}+\om_{rs}]
                \big[1+o_p(1)\big]\\       
&=& \sum_{r=1}^K\sum_{s=1}^K n_{rs}
      \Big\{\frac{\sqrt{a_n}}{n}(\tau_{rs}\sa Z_{rs}\sa\Big\} \Big[ 2a_n \om_{rs}\sa 
      +O_p\Big(\frac{
      \sqrt{a_n}}{n}\Big) \Big ] \cdot 
              \big [1+o_p(1)\big]\\                   
    &=&2 n a_n^{3/2}  \sum_{r=1}^K\sum_{s=1}^K \pi_{rs}\sa \tau_{rs}\sa Z_{rs}\sa\om_{rs}\sa \cdot 
              \big [1+o_p(1)\big].
\end{eqnarray*}
Since $\rho_1^2 = \sum_{(i,j): i\neq j} P_1(i,j)^2 
 =\sum_{r=1}^K\sum_{s=1}^K \pi_{rs}\sa  [\om_{rs}\sa]^2 n^2 a_n^2  (1+o(1)) ~
\equiv  [\rho_1\sa]^2 n^2a_n^2 [1+o(1)]$,
from Condition (C.2) and the above, it follows that 
$$
\hat{\rho}_1^2 = \rho_1^2 +O_p(n a_n^{3/2} )= \rho_1^2 \big[ 1+o_p(1)\big]. 
$$
Hence, it follows that 
\begin{eqnarray*}
    \hat{\rho}_1 - \rho_1 &=& \frac{\hat{(\rho}_1 - \rho_1)(\hat{\rho}_1 + \rho_1)}{\hat{\rho}_1 + \rho_1} \\
     &=& \frac{2 na_n^{3/2}\Big(  
     \sum_{r=1}^K\sum_{s=1}^K \pi_{rs}\sa \tau_{rs}\sa Z_{rs}\sa\om_{rs}\sa \Big) \cdot 
              \big [1+o_p(1)\big] }{ 2\rho_1 (1+o_p(1))} \\
 &=& \frac{\sqrt{a_n}\Big(  
     \sum_{r=1}^K\sum_{s=1}^K \pi_{rs}\sa \tau_{rs}\sa Z_{rs}\sa\om_{rs}\sa \Big) }{ 
            \sqrt{\sum_{r=1}^K\sum_{s=1}^K \pi_{rs}\sa  [\om_{rs}\sa]^2 } }
 \cdot 
              \big [1+o_p(1)\big].   
\end{eqnarray*}
This proves Lemma \ref{lemmas6}.\qb 

\noindent {\bf Proof of Theorem \ref{thm:sbmdist2}:}
Note that under $H_0$, 
\begin{align}
&\Big\| \frac{\hP_1}{\hat{\rho}_1}  -  
      \frac{\hP_2}{\hat{\rho}_2}    \Big\|^2_F\nonumber\\
=  ~ & \Big\| \frac{\hP_1 -P_1}{\hat{\rho}_1} 
+ P_2\Big( \Big[\frac{c}{\hat{\rho}_1} - 
                            \frac{c}{{\rho}_1}\Big] 
    -    \Big[\frac{1}{\hat{\rho}_2} - 
                            \frac{1}{{\rho}_2}\Big]      
\Big)
-  
      \frac{\hP_2 -P_2}{\hat{\rho}_2}    \Big\|^2_F\nonumber\\
= ~ &  {\hat{\rho}_1}^{-2} \|{\hP_1 -P_1}\|_F^2 
        +{\hat{\rho}_2}^{-2} \|{\hP_2 -P_2}\|_F^2 
+ \|P_2\|^2\Big( \Big[\frac{c}{\hat{\rho}_1} - 
                            \frac{c}{{\rho}_1}\Big]
 -    \Big[\frac{1}{\hat{\rho}_2} - 
                            \frac{1}{{\rho}_2}\Big]      
                                \Big)^2\nonumber\\
& \hspace{.5in} + 2 {\hat{\rho}_1}^{-1}\sum_{(i,j) : i\neq j} 
                  [\hP_1(i,j) - P_1(i,j)]   P_2(i,j)
                  \Big( \Big[\frac{c}{\hat{\rho}_1} - 
                            \frac{c}{{\rho}_1}\Big]
 -    \Big[\frac{1}{\hat{\rho}_2} - 
                            \frac{1}{{\rho}_2}\Big]      
                                \Big)
\nonumber\\
& \hspace{.5in} - 2 {\hat{\rho}_2}^{-1}\sum_{(i,j) : i\neq j} 
                  [\hP_2(i,j) - P_2(i,j)]   P_2(i,j)
                  \Big( \Big[\frac{c}{\hat{\rho}_1} - 
                            \frac{c}{{\rho}_1}\Big]
 -    \Big[\frac{1}{\hat{\rho}_2} - 
                            \frac{1}{{\rho}_2}\Big]      
                                \Big)\nonumber\\
& \hspace{.5in} 
-2 [ {\hat{\rho}_1}{\hat{\rho}_2}]^{-1}\sum_{(i,j) : i\neq j} 
                  [\hP_1(i,j) - P_1(i,j)] [\hP_2(i,j) - P_2(i,j)]\nonumber\\
\equiv & ~ I_{1n}+\ldots+I_{6n}, \qmq{(say).}
\label{t2-expn}
\end{align}
Using arguments in the proof of Theorem \ref{thm:sbmdist1} and Lemma \ref{lemmas6},
one can show that for $k=1,2$,
\begin{align}
I_{kn}  & = \frac{a_n \sum_r\sum_s  \tau_{k,rs}\sa [Z_{k,rs}^0]^2 (1+o_p(1))}{\rho_k^2 (1+o_p(1))}\nonumber\\
    & = \frac{1}{n^2a_n} \cdot 
    \frac{  \sum_r\sum_s  \tau_{k,rs}\sa [Z_{k,rs}^0]^2}
        { [\rho_k\sa]^2    } \cdot \big(1+o_p(1)\big)\nonumber\\
    &  \equiv  \frac{1}{n^2a_n} \cdot W_{k}\sa\cdot \big(1+o_p(1)\big), \qmq{(say).} 
        \label{i12-lim}
\end{align}
By similar arguments,
\begin{equation}
    I_{6n}= \frac{1}{n^2a_n} \cdot   
    \frac{2 \sum_r\sum_s 
    \sqrt{\tau_{1,rs}\sa   \tau_{2,rs}\sa}
                      Z_{1,rs}^0  Z_{2,rs}^0}
                        {\rho_1\sa \rho_2\sa } \cdot \big(1+o_p(1)\big)
    \equiv  \frac{1}{n^2a_n} \cdot W_{6}\sa\cdot \big(1+o_p(1)\big), \qmq{(say).}              
\label{i6-lim}
\end{equation}

Next consider $I_{3n}$. Then, using Lemma \ref{lemmas6}
and defining $V_2$ analogously for $P_2$, it follows that 
\begin{align}
I_{3n} & ~ = \rho_2^2 \cdot \Big( \Big[\frac{c}{\hat{\rho}_1} - 
                            \frac{c}{{\rho}_1}\Big]
 -    \Big[\frac{1}{\hat{\rho}_2} - 
                            \frac{1}{{\rho}_2}\Big]      
                                \Big)^2\nonumber\\    
&~ = \rho_2^2 \cdot \Big( \frac{c \big(\rho_1-\hat{\rho}_1\big)}  
                    {{\rho}_1^2 (1+o_p(1))}
- 
\frac{\big(\rho_2-\hat{\rho}_2\big)}  
                    {{\rho}_2^2 (1+o_p(1))}    
                                \Big)^2\nonumber\\ 
&~= \frac{1}{n^2a_n} \cdot [\rho_2\sa]^2 \cdot 
\Big( \frac{c V_1\sa}{[{\rho}_1\sa]^2}
- \frac{ V_2\sa}{[{\rho}_2\sa]^2}
                    \Big)^2
                    \cdot \big(1+o_p(1)\big)    
                                \nonumber\\ 
        &  \equiv  \frac{1}{n^2a_n} \cdot W_{3}\sa\cdot \big(1+o_p(1)\big), \qmq{(say).}   
        \label{i3-lim}
\end{align}
Also, by Lemmas \ref{lemmas3} and \ref{lemmas6}, we have for $k=1,2$,
\begin{align}
&(-1)^{k+1}I_{(k+3)n} \nonumber\\
&~= 
2 {\hat{\rho}_k}^{-1}\sum_{(i,j) : i\neq j} 
                  [\hP_k(i,j) - P_k(i,j)]   P_2(i,j)
                  \Big( \Big[\frac{c}{\hat{\rho}_1} - 
                            \frac{c}{{\rho}_1}\Big]
 -    \Big[\frac{1}{\hat{\rho}_2} - 
                            \frac{1}{{\rho}_2}\Big]      
                                \Big)
\nonumber\\
 & ~= 
2 {{\rho}_k}^{-1} \sum_{r=1}^K\sum_{s=1}^K \pi_{rs}\sa n^2
    \Big[ \sqrt{\frac{{(1+\ind(r=s))a_n \om_{k,rs}\sa} 
   }{\pi_{rs}\sa n^2} }  Z_{k,rs}\sa a_n \om_{k, rs}\sa
    \Big] (1+o_p(1))\nonumber\\
 & \hspace{1in} 
    \times \frac{1}{\rho_k\sa n a_n} 
\Big( \Big[\frac{c (\rho_1 -\hat{\rho}_1)}{\rho_1^2 (1+o_p(1))}\Big]
 -   \Big[\frac{\rho_2 -\hat{\rho}_2}{\rho_2^2 (1+o_p(1))}\Big]  \Big)
\nonumber\\
& ~= - \frac{2}{n^2a_n\rho_k\sa}   \sum_{r=1}^K\sum_{s=1}^K 
      \sqrt{\pi_{rs}\sa \big(1+\ind(r=s) \om_{k, rs}\sa    \big)}   
       \om_{k, rs}\sa Z_{k,rs}\sa  
 \cdot
\Big[\frac{c V_1}{[\rho_1\sa]^2 }
-
\frac{V_2}{[\rho_2\sa]^2 }
\Big](1+o_p(1))\nonumber\\
 &  \equiv  \frac{1}{n^2a_n} \cdot W_{k+3}\sa\cdot \big(1+o_p(1)\big), \qmq{(say).}  
 \label{i56-lim}
\end{align}
Then, from \eqref{t2-expn}-\eqref{i56-lim}, it follows that 
\begin{equation}
T_{2n} \raw^d  \sqrt{W_1\sa+\ldots+W_6\sa}.
\label{t2-lim}
\end{equation}

\subsection{Proof of Theorem \ref{thm:sbmboot1}}
 It is enough to show that 
$$
\sup_{x\geq 0} \Big| P_*(T_{1n}^*\leq x) - P(T_{1,\infty}\leq x)  \Big| = o_p(1).
$$
We proceed as in the proof of Theorem \ref{thm:sbmdist1}. Note that 
$$
T_{1n}^{* 2} = a_n^{-1}\|P_1^* - P_2^*\|^2 = 
a_n^{-1}\|(P_1^* - \hP)  - (P_2^*-\hP)\|^2
$$
where $\hP= [\hP_1+\hP_2]/2$. Restricting attention to $(P_1^* - \hP)/\sqrt{a_n}$ first, we note that the bootstrap random variables driving the 
distribution of  $(P_1^* - \hP)/\sqrt{a_n}$ are given by
$$\{\om_{1,rs}^* - \tilde{\om}_{1,rs} : 1\leq r,s\leq K\}$$
where $\om_{1,rs}^* =
\frac{\sum_{(i,j): c_i^*=r, c^*_j=s, i\neq j}
A_1^*(i,j)}{ \sum_{(i,j): c_i^*=r, c^*_j=s, i\neq j} 1  }$
and where $\tilde{\om}_{1,rs}= [\hat{\om}_{1,rs}+ \hat{\om}_{2,rs}]/2$,
Consider the set $B_n=\{\hat{\bfc}_{jn} = 
{\bfc}_{jn}, j=1,2\}$.
Note that $P(B_n^c) \leq 
P(H(  \hat{\bfc}_{1n},
{\bfc}_{1n}) \geq 1) + P(H(  \hat{\bfc}_{2n},
{\bfc}_{2n}) \geq 1) = o(n^{-1})$, by 
(C.3). Under $H_0$,  the estimated number of communities
as well as estimated community memberships 
coincide on $B_n$. 
Now conditional on the set $B_n$,  
 repeating the arguments in the proofs of Lemmas \ref{lemmas1} and \ref{lemmas3}, one can show that on $\{K_1^*=K\}$,
\begin{align}
\om_{1,rs}^* - \tilde{\om}_{1,rs} 
& = \sqrt{ 
\frac{[1+\ind(r=s)] \tilde{\om}_{1,rs} (1-\tilde{\om}_{1,rs})}{\hat{n}_{1,rs}}
}
Z_{1n,rs}^* +R^*_{1n,rs}
\label{om1-st}
\end{align}
where $Z_{1n,rs}^* = \big[\tilde{\om}_{1,rs} (1-\tilde{\om}_{1,rs}) \hat{n}_{1,rs}\big]^{-1/2} \sum_{i\in \hat{\C}_r, j\in  \hat{\C}_s}
[A^*_{1}(i,j) - \tilde{\om}_{1,rs}]
$
for $r\neq s$, and 
$Z_{1n,rr}^* = \big[\tilde{\om}_{1,rr} (1-\tilde{\om}_{1,rr}) \hat{n}_{1,rr}/2\big]^{-1/2} \sum_{i\in \hat{\C}_r, j\in  \hat{\C}_r
i<j}
[A^*_{1}(i,j) - \tilde{\om}_{1,rr}]
$
for
$r=s$ case, and where $|R^*_{1,rs}|\leq  
4 n H(\bfc^*_{1n},
\hat{\bfc}_{1n})/n^*_{1,rs}$ for all $r,s$. 
Note that under $H_0: P_1=P_2$, $a_n^{-1}\tilde{\om}_{rs}
\raw_p \om_{1,rs}\sa$ for all $r,s$. 
Now using a subsequence argument and the multivariate Lindeberg CLT
(cf. Chapter 11 of \citet{athreya2006measure}), one can show that 
$$
d_P\Big(  
~[Z_{1,rs}\sa]_{1\leq r,s\leq K}, ~~ [Z_{1,rs}^*]_{1\leq r,s\leq K}\Big| A_1, A_2\Big) =o_p(1)
$$
where, for random vectors $X$, $Y$ and a $\sigma$-field ${\mathcal G}$,  $d_P(Y, X|{\mathcal G})$ denotes the Prohorov distance between
the distribution of $Y$
and
the 
conditional distribution of $X$ given ${\mathcal G}$. Also, using Condition
(C.4), it is easy to show that for any $\epsilon>0$, 
$$
P_*\Big( \max_{1\leq r,s\leq K}~  |R_{1n,rs}^*|>\epsilon \sqrt{a_n}/n \Big) =o_p(1).
$$
The proof of Theorem \ref{thm:sbmboot1} can now be completed by retracing the 
arguments in the proof of Theorem \ref{thm:sbmdist1} and 
using the following lemma. We omit the routine details.
\qb

\begin{lemma}
    (An extended  conditional continuous mapping theorem).
For each $n\geq 1$, 
let $V_n^*\in \bbr^p, U_n^*\in \bbr^q$ and $W_n\in \bbr^r$
be random vectors defined on a probability space $(\Omega,
{\mathcal F}, P)$ and let $\G_n$ be a $\sigma$-field 
such that $W_n$ is $\G_n$-measurable where $p,q,r\geq 1$.
Suppose that
\begin{enumerate}
    \item[(i)] $d_P(V\sa, V_n^*|\G_n)=o_p(1)$,
\item[(ii)] there exists a nonrandom 
$u\sa\in \bbr^q$ such that 
$P( \|U_n^* - u\sa\| >\epsilon |\G_n) =o_p(1)$ 
for any $\epsilon>0$, and 
 \item[(iii)]  $W_n\raw_p w\sa$
for some (nonrandom) $w\sa\in \bbr^r$.
\end{enumerate}
Let $g:\bbr^{p+q+r}
\raw \bbr^s$ be a measurable function such that  
(a) for any $M>0$,
$$
\sup_{\|x\|\leq M} ~ \| g(x,y,w) - g(x, u\sa, w\sa)\| \raw 0
$$
as $y\raw u\sa$ and $w\raw w\sa$, and 
(b) $P(V\sa\in D_g)=0$, where $ D_g$ is the set of discontinuity
points of $g(\cdot, u\sa,w\sa)$. Then, 
\begin{align}
d_P\Big( g\big(V_n^*, U_n^*, W_n\big),
    ~ g\big(V_n\sa, u\sa, w\sa\big) \Big) =o_p(1). 
    \label{ccmt}
\end{align}
\label{lemmas8}
\end{lemma}

\noindent
{\bf Proof:} Note that (ii) is equivalent to the following
seemingly stronger statement
(Chapter 3 of \citet{lahiri2003resampling}): \\[.1in] 
(ii)':  there exists a sequence $\{\ep_n\}$ with $\ep_n\downarrow 0$
such that $P( \|U_n^* - u\sa\| >\ep_n |\G_n) =o_p(1)$.
We will use (ii)' in place of (ii) in the proof. 
Using the characterization of convergence in probability 
in terms of subsequential almost sure convergence,
given any subsequence $\{n_i\}$, it is enough to extract a 
further subsequence $\{n_{i_k}\}\subset \{n_i\}$ such that 
the left side of \eqref{ccmt} converges to zero almost surely
along $\{n_{i_k}\}$. For notational simplicity, we will 
write  $\{m\}$ for the subsequence  $\{n_{i_k}\}$. Indeed, 
using (i), (ii)' and (iii), we can extract 
the subsequence  $\{m\}\subset \{n_i\}$ is 
such that on a set $B\in \F$ with $P(B)=1$,
\begin{equation}
W_m\raw w\sa, 
P( \|U_m^* - u\sa\| >\ep_m |\G_m) =o(1),\qmq{and} 
d_P(V\sa, V_m)=o(1)
\label{L8-bd1}
\end{equation}
as $m\raw \infty$. 
Given $\ep_0\in (0,1)$, there exists $M_0\in (1,\infty)$
and $N_0\geq 1$ such that 
\begin{equation}
    P(\|V\sa\|> M_0-1) < \ep_0 \qmq{and} d_P(V\sa, V_m^*|\G_m) < \ep_0
    \label{L8-bd2}
\end{equation}
for all $m>n_0$, on $B$. In particular, by the definition of the 
Prohorov distance, this implies that on $B$, 
$$
P\big( \|V_m^*\|> M_0\big| \G_n\big) \leq 
   P\big( \|V\sa\|> M_0 - \ep_0\big) +\ep_0 \leq 2\ep_0
$$
for all $m>N_0$. Let $B_m^*=\{ \|V^*_m\| \leq M_0, \|U_m^* -u\sa\|
\leq \ep_m\}$. Also, let $\ci=\sqrt{-1}$. 
Then, on $B$ and for  $m>N_0$, for any $t\in\bbr^s$,
\begin{align*}
  &   E\big(\exp(\ci t'g(V_n^*, U_n^*,W_n)|\G_n\big)\\
  = & E\Big(\exp(\ci t'g(V_n^*, U_n^*,W_n)\ind(B_m^*)\Big|\G_n\Big)
      +  E\Big(\exp(\ci t'g(V_n^*, U_n^*,W_n)[1-\ind(B_m^*)]\Big|\G_n\Big)
                \\
             =   &
   E\Big(\Big[\exp(\ci t'g(V_n^*, U_n^*,W_n)  - \exp(\ci t'g(V_n^*, u\sa, w\sa)\Big]\ind(B_m^*)\Big|\G_n\Big) \\
     & \hspace{.5in}
     + E\Big(\exp(\ci t'g(V_n^*, u\sa, w\sa)\ind(B_m^*)\Big|\G_n\Big)
      +  E\Big(\exp(\ci t'g(V_n^*, U_n^*,W_n)[1-\ind(B_m^*)]\Big|\G_n\Big)\\
     = & 
    E\big(\exp(\ci t'g(V_n^*, u\sa, w\sa)|\G_n\big) +R_m,\qmq{(say)}
\end{align*}
where, using  \eqref{L8-bd1} and  \eqref{L8-bd2}, on $B$, one gets 
\begin{align*}
|R_m| \leq & 2 P(B_m^{*c}|\G_m)+ |t|
E \Big(\big| g(V^*_m, U_m^*, W_m) - g(V_m^*, u\sa,w\sa)\big|
\ind(B_m) 
\Big|\G_m\Big)\\
\leq & 2 \Big[P(\|V_m^*\|>M_0|\G_n)+
P( \|U_m^* - u\sa\| >\ep_m |\G_m) 
\Big]\\
& \hspace{.6in}  + |t|\times ~ \sup_{\|x\|\leq M_0, \|y\|\leq \ep_m, \|z\|
\leq \|W_m-w\sa\|}~ \big| g(x, u\sa+y, w\sa+z) - g(x, u\sa,w\sa)\big|  
\\
\leq & 4\ep_0 + o(1).
\end{align*}
Since $\ep_0\in (0,1)$ is arbitrary, this implies
$$
 E\big(\exp(\ci t'g(V_n^*, U_n^*,W_n)|\G_n\big)
 =  E\big(\exp(\ci t'g(V_n^*, u\sa, w\sa)|\G_n\big) + o(1)
$$
as $m\raw\infty$ for all $t\in \bbr^s$, on $B$. 
Now using condition (b) on $g$, \eqref{L8-bd1} and the standard 
version of the Continuous Mapping Theorem (cf. Chapter 9, \citet{athreya2006measure}), one can easily conclude that, 
 on the set $B$, 
 $$E\big(\exp(\ci t'g(V_n^*, u\sa, w\sa)|\G_n\big)
 =  E\big(\exp(\ci t'g(V\sa, u\sa, w\sa)\big)+o(1)
 \qmq{for all $t\in \bbr^s$.} 
 $$
This completes the proof of Lemma \ref{lemmas8}. \qb

\begin{remark}
    Lemma \ref{lemmas8} can be used to prove the convergence of functions that 
are polynomials in $V_n^*$ and $U_n^*$
with coefficients that are rational functions of $W_n$ with a finite limit.
In particular, a mixed version of the (Conditional) Slutsky's Theorem holds:\\
 \hspace*{1in} $W_n \odot V_n^* + U_n^* \raw ^d 
w\sa \odot V\sa + u\sa 
$, in probability. 
\end{remark} 

\begin{remark}
Theorem \ref{thm:sbmboot1} establishes that the proposed bootstrap method provides a valid approximation to the null distribution of the test statistic $T_{1n}$. 
 We can use the 
bootstrap quantiles to calibrate the test statistic $T_{1n}$. 
Specifically, let $\hat{q}_{1n}(u)$ denote the
$u$-quantile of the conditional distribution of $T^*_{1n}$ given $A_1,
A_2$ for
 $u\in (0,1)$. Then, we would reject $H_{0}: P_1=P_2$ at level $\al_0\in (0,1)$ if $T_{1n} > \hat{q}_{1n}(1-\al_0)$. Since the limit distribution 
 of $T_{1n}$ under $H_{0,n}$ is continuous with a positive density
 on $(0,\infty)$, it follows that 
$$
\hat{q}_{1n}(1-\al_0) \raw_p q_{1,\infty}(1-\al_0)
$$ 
where $q_{1,\infty}(u)$ is the $u$-quantile of $T_{1,\infty}$.
As a result,
\begin{align}
& 
P\big(T_{1n}> \hat{q}_{1n}(1-\al_0) | H_{0,n}\big)\nonumber\\
 = &
P\big(T_{1n}> {q}_{1,\infty}(1-\al_0) | H_{0,n}\big)+o(1)\nonumber\\
= &
  P\big(T_{1,\infty} >  {q}_{1,\infty}(1-\al_0)\big)+o(1)\nonumber\\
= & \al_0+o(1),
\label{Boot-CI-cons}
\end{align}
and the proposed test attains the desired  size 
$\al_0$ asymptotically. 
\end{remark}

\subsection{Proof of Theorem \ref{thm:sbmboot2}}
Let $\rho^0_k$ be as defined in the proof of 
Lemma \ref{lemmas6}, $k=1,2$ and let 
$\aal_0= \rho^0_1/ \rho^0_2$.  
Using the approximations for 
$\hat{\rho}_k$ from Lemma \ref{lemmas6}, it is easy to check that 
\begin{align*}
    \hat{\aal}_n& =\frac{\hat{\rho}_{1n}}{\hat{\rho}_{2n}} = 
    \frac{\rho_{1}}{\rho_{2}}(1+o_p(1)) 
    %= \aal_n (1+o_p(1)) 
    \equiv \aal_0 + o_p(1). 
\end{align*}
Also, note that using Corollary \ref{cors2}, Lemma \ref{lemmas3}, the relation above and the identities
`$\om_{1,rs}\sa = \aal_0 \om_{2,rs}\sa$
for all $r,s$' and 
`$\hat{\rho}_{1n}= \hat{\aal}_n \hat{\rho}_{2n}$
for all $n\geq 1$', 
one gets 
\begin{align}
    \tilde{\rho}^2_{1n} \equiv ~ & ~ \|\tilde{P}_{1,n}\|^2
          =   \| \hP_{1,n}+ \hat{\aal}_n \hP_{2,n}\|^2/4 \nonumber \\
= & ~ \sum_{(i,j): i\neq j} \Big[ \hP_1(i,j)^2 + \hat{\aal}^2 \hP_2(i,j)^2 + 
2\hat{\aal} \hP_1(i,j)\hP_2(i,j)  \Big]/4 \qmq{(Dropping the subscript '$n$')}
\nonumber
\\
= & ~    \Big[ \hat{\rho}_1^2 + \hat{\aal}^2 \hat{\rho}_2^2
+ 2\hat{\aal} \sum_{(i,j): i\neq j} \hP_1(i,j)\hP_2(i,j)  \Big]/4 \\
& ~   =  \Big[ 2\hat{\rho}_1^2 
+ 2\hat{\aal} \sum_r \sum_s n_{rs} \om_{1,rs}\om_{2,rs}   \Big(1+o_p(1)\Big) \Big]/4 \nonumber\\
& ~   =  \Big[ 2 (\rho_1\sa)^2 n^2a_n^2 +  
+ 2{\aal}_0 n^2a_n^2  \sum_r \sum_s n_{rs} 
\pi_{rs}\sa \om_{1,rs}\sa \om_{2,rs}\sa \Big(1+o_p(1)\Big) \Big]/4 \nonumber\\
& ~   =  (\rho_1\sa)^2 n^2a_n^2 \Big(1+o_p(1)\Big).
\label{rho-til1}
\end{align}
Similarly,   $\tilde{\rho}^2_{2n} \equiv \|\tilde{P}_{2,n}\|^2=  (\rho_2\sa)^2 n^2a_n^2 \big(1+o_p(1)\big)$. 
Now retracing the steps in the proof of Theorem \ref{thm:sbmdist2} and using \eqref{rho-til1}
(and its analog for $k=2$) and 
Lemma \ref{lemmas8}, one 
can complete the proof of Theorem S.9. \qb

\section{Proof of Theorem \ref{thm:cleq}}
First note that, for the Chung-Lu estimate $\hat{P}^n$ of underlying probability matrix $P$
\begin{equation}
\begin{split}
0\leq||\hat{P}^n-P||^2=& \sum_{i,j}(\hat{P}^n(i,j)-P(i,j))^2\\
& =\sum_{i,j}(\hat{\theta}_i\hat{\theta}_j-\theta_i\theta_j)^2\\
& =\sum_{i,j}(\hat{\theta}_i\hat{\theta}_j-\theta_i\hat{\theta}_j+\theta_i\hat{\theta}_j-\theta_i\theta_j)^2\\
& \leq 2\Big[\sum_{i,j}\hat{\theta}_j^2(\hat{\theta}_i-\theta_i)^2+\sum_{i,j}\theta_i^2(\hat{\theta}_j^2-\theta_j)^2\Big]\\
& \leq 4n\sum_i(\hat{\theta}_i-\theta_i)^2 \hspace{1cm} since~ \hat{\theta}_i, \theta_i <1\hspace{0.1cm}\forall\hspace{0.1cm}i=1(1)n
\end{split}
\end{equation}

Again note that $$E(d_i)=n\theta_i\Bar{\theta}-\theta_i^2\hspace{0.1cm}=\delta_i\hspace{0.05cm}(say)\hspace{0.1cm}\forall \hspace{0.1cm} i=1(1)n$$ where $d_i$ is the degree of the $i^{th}$ vertex and $\Bar{\theta}=\frac{1}{n}\theta_i$ 

So $\sum_i\delta_i=(n\Bar{\theta})^2-\sum_i\theta_i^2\geq (n\Bar{\theta})^2-n\Bar{\theta}$~(since $\theta_i \leq 1~\forall~i$).
Since $n\Bar{\theta}\ll(n\bar{\theta})^2$, $\exists~ C~(\text{a constant})>0~\text{with}~\sum_i\delta_i>C(n\bar{\theta})^2$. 

For fixed $n$, choose $\epsilon_n=3\frac{\sqrt{n\log (n)}}{\gamma_n}$

Taking $d_j=\sum_{i\neq j}A(i,j)$ where $A(i,j)\sim Ber(p(i,j))$ and using Hoeffding concentration inequality, we have
\begin{equation}
\begin{split}
\mathbb{P}\Big[\bigcup_i\{|d_i-\delta_i|\geq\epsilon_n\}\Big]\leq&\sum_i\mathbb{P}\Big[|d_i-\delta_i|\geq\epsilon_n\Big]\\
&\leq \sum_ie^{-(n/5)(\delta_i/n)(\epsilon_n/n)^2}\\
& = \sum_ie^{-\delta_i\epsilon_n^2/5n^2}\\
& = \sum_ie^{-\theta_i\Bar{\theta}\epsilon_n^2/5n}\\
&\leq ne^{-\gamma_n^2\epsilon_n^2/5n}\hspace{1cm}[since~\theta_i\geq\gamma_n~\forall~i=1(1)n]\\
& = \exp\Big\{-\frac{\gamma_n^2\epsilon_n^2}{5n}+\log(n)\Big\}\\
&= \exp\Big\{-\frac{4}{5}\log(n)\Big\}\to 0\\
\end{split}
\end{equation}

Hence under the assumptions, $$\mathbb{P}\Big[\bigcup_i\{|d_i-\delta_i|\geq\epsilon_n\}\Big]\to 0~as~n\to\infty$$

Observe that
\begin{equation}
\begin{split}
\frac{\epsilon_n}{n^{1/2+\alpha}\Bar{\theta}} =&
3\frac{\sqrt{n\log(n)}}{n^{1/2+\alpha}\Bar{\theta}\gamma_n}\\
& = 3\frac{\sqrt{\log(n)}}{n^{\alpha}\gamma_n\Bar{\theta}}\\
& \to 0\hspace{2cm}\text{by assumption $(\ref{eq:assm1}$})\\
\end{split}
\label{eq16}
\end{equation}

Also observe that
\begin{equation}
\begin{split}
\frac{\epsilon_n^2\sum_i\theta_i^2}{n^{2+2\alpha}\Bar{\theta}^4} =&
9\frac{n\log(n)}{\gamma_n^2n^{2+2\alpha}\Bar{\theta}^4}\sum_i\theta_i^2\\
& =9\Big(\frac{\sum_i\theta_i^2}{n\Bar{\theta}^2}\Big)\Big(\frac{n\log(n)}{n^{1+2\alpha}\gamma_n^2\Bar{\theta}^2}\Big)\\
& = 9\Big(\frac{\sum_i\theta_i^2}{n\Bar{\theta}^2}\Big)\Big(\frac{\sqrt{n\log(n)}}{n^{1/2+\alpha}\gamma_n\Bar{\theta}}\Big)^2\\
\end{split}
\label{eq17}
\end{equation}

Hence from assumption $1,2$ $$\frac{\epsilon_n^2\sum_i\theta_i^2}{n^{2+2\alpha}\Bar{\theta}^4}=o(1)$$

Hence, on the asymptotically $\mathbb{P}$robability-1 set $\bigcap_i\Big\{|d_i-\delta_i|<\epsilon_n\Big\}$, observe that

$\sum_id_i=(n\Bar{\theta})^2-\sum_i\theta_i^2+nk_n$ and hence $\sqrt{\sum_id_i}=O\Big(n\Bar{\theta}+\sqrt{nk_n}\Big)$ where $k_n\leq O(\epsilon_n)$ and so $\sum_id_i=O((n\bar{\theta})^2)$ since $\theta_i\leq1$ and so $\sum_i\theta_i^2\leq \sum_i\theta_i=n\Bar{\theta}$.

Hence, on the above mentioned set,

\begin{equation}
\begin{split}
|\hat{\theta}_i-\theta|^2\leq& \Big|\frac{d_i}{\sqrt{\sum_id_i}}-\theta_i\Big|^2\\
&\leq3 \Big[\Big(\frac{d_i}{\sqrt{\sum_id_i}}-\frac{\delta_i}{\sqrt{\sum_id_i}}\Big)^2+\Big(\frac{\delta_i}{\sqrt{\sum_id_i}}-\frac{\delta_i}{\sqrt{\sum_i\delta_i}})^2+\Big(\frac{\delta_i}{\sqrt{\sum_i\delta_i}}-\theta_i\Big)^2\Big]\\
& \leq 3\frac{\epsilon_n^2}{\sum_id_i}+3\delta_i^2\Big(\frac{1}{\sqrt{\sum_id_i}}-\frac{1}{\sqrt{\sum_i\delta_i}}\Big)^2+\frac{3}{\sqrt{C}}\Big(\frac{n\theta_i\Bar{\theta}-\theta_i^2}{n\Bar{\theta}}-\theta_i\Big)^2\hspace{0.5cm}[\text{since}~\sum_i\delta_i\geq C(n\bar{\theta})^2]\\ 
& \leq 3\frac{\epsilon_n^2}{\sum_id_i}+3\frac{(n\theta_i\Bar{\theta})^2}{\sum_id_i\sum_i\delta_i}\Big(\sqrt{\sum_i\delta_i}-\sqrt{\sum_id_i}\Big)^2+\frac{3}{\sqrt{C}}\frac{\theta_i^4}{(n\Bar{\theta})^2}\hspace{1cm}[\text{since $n\theta_i\Bar{\theta}\geq\theta_i^2$}]\\
& \leq 3\frac{\epsilon_n^2}{\sum_id_i}+\frac{3}{C}\frac{\theta_i^2}{\sum_id_i}\Big(\sqrt{\sum_i\delta_i}-\sqrt{\sum_id_i}\Big)^2+\frac{3}{\sqrt{C}}\frac{\theta_i^4}{(n\Bar{\theta})^2}\hspace{1cm}[\text{since}~\sum_i\delta_i\geq C(n\bar{\theta})^2]\\
\end{split}    
\end{equation}

Hence
\begin{equation}
\begin{split}
T^2_n=||\hat{P}^n-P||_F^2\leq&
4n\sum_i(\hat{\theta}_i-\theta_i)^2\\
&\leq 12n^2\frac{\epsilon_n^2}{\sum_id_i}+\frac{12n}{C}\frac{(\sqrt{\sum_i\delta_i}-\sqrt{\sum_id_i})^2}{\sum_id_i}\sum_i\theta_i^2+\frac{12n}{\sqrt{C}(n\Bar{\theta})^2}\sum_i\theta_i^4\\
\end{split}
\label{eq19}
\end{equation}

Again note that
\begin{equation*}
\begin{split}
\Big(\sqrt{\sum_i\delta_i}-\sqrt{\sum_id_i}\Big)^2 =&
\Big(\frac{\sum_i\delta_i-\sum_id_i}{\sqrt{\sum_i\delta_i}+\sqrt{\sum_id_i}}\Big)^2\\
& \leq \frac{(n\epsilon_n)^2}{\sum_i\delta_i+\sum_id_i}\hspace{1cm}[\text{since}~(a+b)^2\geq a^2+b^2]\\
& = O\Big(\frac{(n\epsilon)^2}{(n\Bar{\theta})^2}\Big)\hspace{1cm}[\text{since}~\sum_i\delta_i=O((n\bar{\theta})^2),~\sum_id_i=O((n\bar{\theta})^2)]\\
& = O\Big(\frac{\epsilon_n^2}{\Bar{\theta}^2}\Big)\\
\end{split}
\end{equation*}

Hence putting in \eqref{eq19}
\begin{equation}
\begin{split}
T^2_n=||\hat{P}^n-P||_F^2\leq&
O\Big(n^2\frac{\epsilon_n^2}{\sum_id_i}+n\frac{\epsilon_n^2\sum_i\theta_i^2}{\Bar{\theta}^2\sum_id_i}+\frac{n}{(n\Bar{\theta})^2}\sum_i\theta_i^4\Big)\\
& = O\Big(n^2\frac{\epsilon_n^2}{(n\Bar{\theta})^2}+n\frac{\epsilon_n^2\sum_i\theta_i^2}{\Bar{\theta}^2(n\Bar{\theta})^2}+\frac{n}{(n\Bar{\theta})^2}\sum_i\theta_i^4\Big)\\
\end{split}
\end{equation}

And hence, for any $\alpha>0$,
\begin{equation}
\begin{split}
\frac{1}{n^{1+2\alpha}}T^2_n=\frac{1}{n^{1+2\alpha}}||\hat{P}^n-P||_F^2\leq&
O\Big(n^{-2\alpha}\Big[\frac{\epsilon_n^2}{n\Bar{\theta}^2}+\frac{\epsilon_n^2}{n^2\Bar{\theta}^4}\sum_i\theta_i^2+\frac{1}{(n\Bar{\theta})^2}\sum_i\theta_i^4\Big]\Big)\\
& = o(1)\hspace{2cm}[\text{by \eqref{eq16} and \eqref{eq17}}]\\
\end{split}
\label{eq21}
\end{equation}

Also note that the rejection region of the test is of the form $$R=\{T_n> c_n\}$$

So, we can write
\begin{equation}
\begin{split}
\mathbb{P}(T_n\notin R)\leq& 
\mathbb{P}(||\hat{P}_1^n-\hat{P}^n_2||_F\leq c_n)\\
&\leq \mathbb{P}(||P_1-P_2||_F-||\hat{P}^n_1-P_1||_F-||\hat{P}^n_2-P_2||_F\leq c_n)\\
& = \mathbb{P}(||\hat{P}^n_1-P_1||_F+||\hat{P}^n_2-P_2||_F+c_n\geq||P_1-P_2||_F)\\
\end{split}
\label{eq22}
\end{equation}

For any $\alpha>0$, taking $c_n=O(n^{1/2+\alpha})$ the LHS inside the probability expression above is $o(1)$, but under alternative, $P_1$ and $P_2$ are significantly apart in the sense that $$\frac{||P_1-P_2||_F}{n^{1/2+\alpha}}\to \infty~\text{for any}~\alpha>0$$ (by assumption). Hence $$\mathbb{P}(T_n\notin R)\leq \beta_n$$ for small $\beta_n\to 0$ and so $$\mathbb{P}(T_n\in R)\geq 1-\beta_n\to 1.$$ Hence the test is consistent.  

\section{Proof of Theorem \ref{thm:rdpgeq}}
% \begin{proof}
Suppose that the null hypothesis $H_0$ is true, so $P_{1} = P_{2}$. Let $\alpha$ be given, and let $\eta < \alpha/4$. We consider the latent positions $X_1$ and $X_2$ corresponding to $P_{1}$ and $P_{2}$. From Theorem 1 in \cite{tang2017semiparametric}, for all $n$ sufficiently large, there exists orthogonal matrices $W_{1}$ and $W_{2}$ $\in \mathbb{O}(d)$ such that with probability at least $1-\eta$,
$$\big|\|\Hat{X}_i - \Tilde{X}_i\|_F - C(X_i) \big| \leq  \dfrac{Cdlog(n/\eta)}{C(X_i)\sqrt{\gamma^5(P_i)\delta (P_i)}} \triangleq g(X_i)\hspace{1cm} \text{for $i=1,2$},$$
where $\Tilde{X}_i = X_i W_{i}$ for $i= 1,2$. Additionally, there exists a constant $K_0>0$ such that $K_0<C(X_{i})<(d\gamma^{-1}(P_{i}))^{1/2}$ for $i= 1,2$. Now,
\begin{equation*}
\begin{split}
    g(X_i) & = \dfrac{Cdlog(n/\eta)}{C(Z_n)\sqrt{\gamma^5(P_i)\delta (P_i)}}\\
    & < \dfrac{Cdlog(n/\eta)}{K_0\sqrt{c_0^5(\log n)^{2+\epsilon}}} \rightarrow 0 \hspace{1cm} as \hspace{0.2cm} n \rightarrow \infty.
\end{split}
\end{equation*}

Let, the estimates of $P_{1}$ and $P_{2}$ be $\Hat{P}_1=\Hat{X}_1 \Hat{X}_1^T$ and $\Hat{P}_2=\Hat{X}_2 \Hat{X}_2^T$. Then for $i=1,2$,
\begin{equation}
    \begin{split}
    \|\hat{P}_i-P_i \|_F & = \|\Tilde{X}_i \Tilde{X}_i^T - \hat{X}_i \hat{X}_i^T\|_F\\
    & = \| \Tilde{X}_i \Tilde{X}_i^T - \Tilde{X}_i \hat{X}_i^T + \Tilde{X}_i \hat{X}_i^T - \hat{X}_i \hat{X}_i^T\|_F\\
    & \leq \|\Tilde{X}_i(\Tilde{X}_i - \hat{X}_i)^T\|_F + \|(\Tilde{X}_i - \hat{X}_i) \hat{X}_i^T\|_F\\
    & \leq \|\Tilde{X}_i\|_F \|\Tilde{X}_i - \hat{X}_i\|_F + \|\hat{X}_i\|_F \|\Tilde{X}_i - \hat{X}_i\|_F\\
    & = \sqrt{r_i} \|\Tilde{X}_i - \hat{X}_i\|_F + \|\hat{X}_i\|_F \|\Tilde{X}_i - \hat{X}_i\|_F\\
    \end{split}
\end{equation}
where $r_i = trace(P_i)$.\\
Now, by the Theorem 2.1. from \cite{tang2017semiparametric}, both $C(X_1)$ and $C(X_2)$ are bounded above by $(d\gamma^{-1}(P_{1}))^{1/2}$ and $(d\gamma^{-1}(P_{2}))^{1/2}$ respectively. Then,
\begin{equation*}
    \begin{split}
    \|\hat{X}_i\|_F & \leq \|\Tilde{X}_i\|_F + \|\hat{X}_i-\Tilde{X}_i\|_F\\
    & \leq \|\Tilde{X}_i\|_F + C(X_i) + g(X_i)\\
    & \leq \sqrt{r_i} + (d\gamma^{-1}(P_i))^{1/2} + g(X_i)\\
    & \leq 2\sqrt{r_i} + (d\gamma^{-1}(P_i))^{1/2} \hspace{3cm} (g(X_i) \leq \sqrt{r_i} \hspace{0.2cm} \text{for large n})\\
    \end{split}
\end{equation*}
Hence,
\begin{equation}
    \begin{split}
    \|\Hat{P}_i-P_i\|_F &= \|\Tilde{X}_i \Tilde{X}_i^T - \hat{X}_i \hat{X}_i^T\|_F\\
    & \leq (3\sqrt{r_i}+(d\gamma^{-1}(P_i))^{1/2})\|\hat{X}_i-\Tilde{X}_i\|_F\\
    & = \Gamma_i \|\hat{X}_i-\Tilde{X}_i\|_F \hspace{4cm} (say)\\
    \end{split}
\label{eq28}
\end{equation}
Let, $\Gamma = \Gamma_1 + \Gamma_2$. Then,
\begin{equation*}
\begin{split}
    \frac{1}{\Gamma}\|\Hat{P}_1-\Hat{P}_2\|_F &= \frac{1}{\Gamma}\|(\Hat{P}_1-P_{1})+(P_{2}-\Hat{P}_2)\|_F \hspace{2cm} (as \hspace{0.2cm} P_{1}=P_{2})\\
    & \leq \frac{1}{\Gamma}(\|\Hat{P}_1-P_{1}\|_F + \|\Hat{P}_2-P_{2}\|_F)\\
    & \leq \frac{1}{\Gamma_1}\|\Hat{P}_1-P_{1}\|_F + \frac{1}{\Gamma_2}\|\Hat{P}_2-P_{2}\|_F\\
    & \leq \|\Hat{X}_1 - \Tilde{X}_1\|_F + \|\Hat{X}_2 - \Tilde{X}_2\|_F \hspace{2cm} (\text{By Equation \eqref{eq28})}\\
    & \leq C(X_1)+C(X_2)+g(X_1)+g(X_2)\\
\end{split}
\end{equation*}
Hence, with probability at least $1 - \alpha$,

$$\dfrac{\|\hat{P_{1}} - \hat{P_{2}}\|_F}{\Gamma(\sqrt{d\gamma^{-1}(P_{1})}+\sqrt{d\gamma^{-1}(P_{2})})}
\leq 1 + r(\alpha,n)$$
where $r(\alpha,n) \rightarrow 0$ as $n \rightarrow \infty$ for a fixed $\alpha$.\\
Hence,
$$T_n = \|\hat{P}_1 - \hat{P}_2\|_F
\leq \Gamma(1 + r(\alpha,n))(\sqrt{d\gamma^{-1}(P_1)}+\sqrt{d\gamma^{-1}(P_{2})})$$

$$\implies T_n = \|\hat{P}_1 - \hat{P}_2\|_F
\leq 2\Gamma(\sqrt{d\gamma^{-1}(P_1)}+\sqrt{d\gamma^{-1}(P_{2})})$$
with probability at least $1-\alpha$.\\
Then for $P_{1}, P_{2}$ satisfying $P_{1} = P_{2}$, we conclude
$$P(T_n \in R) < \alpha$$
where $R = \{t : t > 2\Gamma(\sqrt{d\gamma^{-1}(P_1)}+\sqrt{d\gamma^{-1}(P_{2})})\}$
\\
Now suppose the alternative hypothesis is true.
We note that,
\begin{equation*}
\begin{split}
\|P_{1}-P_{2}\|_F &= \|(P_{1}-\hat{P}_1)+(\hat{P}_1-\hat{P}_2)+(\hat{P}_2-P_{2})\|_F\\
&\leq \|P_{1}-\hat{P}_1\|_F+\|\hat{P}_1-\hat{P}_2\|_F+\|\hat{P}_2-P_{2}\|_F\\
\end{split}
\end{equation*}

$$\implies \|\hat{P}_1-\hat{P}_2\|_F \geq \|P_{1}-P_{2}\|_F-\|P_{1}- \hat{P}_1\|_F-\|P_{2}-\hat{P}_2\|_F$$
Therefore, for all $n$,

\begin{equation*}
\begin{split}
P(T_n \notin R) & \leq P(\|\hat{P}_1 - \hat{P}_2\|_F
\leq C)\\
& \leq P(\|P_{1}- \hat{P}_1\|_F+\|P_{2}-\hat{P}_2\|_F+C \geq \|P_{1}-P_{2}\|_F)
\end{split}
\end{equation*}
\\
Now, let $\beta > 0$ be given. By the convergence of $\|\Hat{X}_1 - \Tilde{X}_1\|_F$ to $C(X_1)$ in Theorem 1 in \cite{tang2017semiparametric}, we deduce that there exists a constant $M_1(\beta)$ and a positive integer $n_0=n_0(\alpha,\beta)$ so that, for all $n \geq n_0(\alpha,\beta)$,

$$P\Big(\|\Hat{X}_1 - \Tilde{X}_1\|_F+k\sqrt{d\gamma^{-1}(P_1)}\geq M_1/2\Big) \leq \beta/2$$
$$P\Big(\|\Hat{X}_2 - \Tilde{X}_2\|_F+k\sqrt{d\gamma^{-1}(P_2)}\geq M_1/2\Big) \leq \beta/2$$
By Equation \eqref{eq28}
$$P\Big(\|\hat{P}_1 - P_1\|_F+k\Gamma\sqrt{d\gamma^{-1}(P_1)}\geq \Gamma M_1/2\Big) \leq \beta/2$$
$$P\Big(\|\hat{P}_2 - P_2\|_F+k\Gamma\sqrt{d\gamma^{-1}(P_2)}\geq \Gamma M_1/2\Big) \leq \beta/2$$
where $C = k\Gamma(\sqrt{d\gamma^{-1}(P_1)}+\sqrt{d\gamma^{-1}(P_2)})$.\\
As $d_n \rightarrow \infty$ there exists some $n_2=n_2(\alpha,\beta,C)$ such that for all $n \geq n_2$, $\|P_{1}-P_{2}\|_F \geq \Gamma M_1$. Hence, for all $n \geq n_2$, $P(T_n \notin R) \leq \beta$, i.e., our test statistic $T_n$ lies within the rejection region $R$ with probability at least $1 - \beta$, as
required.
% \end{proof}

\section*{Proof of Theorem \ref{thm:clscale}}
Let $\hat{\rho}$ be the analogue of $\rho$ corresponding to $\hat{P}$

Define $Q=\frac{P}{\rho}$, hence $\hat{Q}=\frac{\hat{P}}{\hat{\rho}}$, i.e.~
\begin{equation*}
\begin{split}
\hat{Q}(i,j)=\frac{\hat{P}(i,j)}{\hat{\rho}}=&
n\frac{\hat{P}(i,j)}{\sqrt{\sum_{i,j}(\hat{P}(i,j)})^2}\\
\end{split}    
\end{equation*}

So, for any $\alpha>0$, 
\begin{equation}
\begin{split}
\|\hat{Q}-Q\|_F=&
\|\frac{\hat{P}}{\hat{\rho}}-\frac{P}{\rho}\|_F\\
& = n\Big\|\frac{\hat{P}}{\|\hat{P}\|_F}-\frac{P}{\|P\|_F}\Big\|_F\\
& \leq n\Big[\Big\|\frac{\hat{P}-P}{\|\hat{P}\|_F}\Big\|_F+\|P\|_F\Big|\frac{1}{\|\hat{P}\|_F}-\frac{1}{\|P\|_F}\Big|\Big]\hspace{1cm}\text{By triangle inequality}\\
& = n\Big[\Big\|\frac{\hat{P}-P}{\|\hat{P}\|_F}\Big\|_F+\frac{|\|P\|_F-\|\hat{P}\|_F|}{\|\hat{P}\|_F}\Big]\\
& \leq 2n\frac{\|\hat{P}-P\|_F}{\|\hat{P}\|_F}\hspace{2cm}\text{By triangle inequality}\\
& \leq O\Big(\frac{n^{3/2+\alpha}}{\|\hat{P}\|_F}\Big)
\end{split}
\end{equation}

The last line of the above equation follows from the calculation of previous (equality) case.

Now observe
\begin{equation*}
\begin{split}
\|\hat{P}\|_F=&
\|P-(P-\hat{P})\|_F\\
& \geq \|P\|_F-\|\hat{P}-P\|_F\hspace{1cm}\text{By triangle inequality}\\
& \geq (\sum_i\theta_i^2)-o(n^{1/2+\alpha})\\
& \geq n\Bar{\theta}^2-o(n^{1/2+\alpha})\hspace{1cm}~\text{since}~RMS\geq AM\\
& = n\Bar{\theta}^2\\
\end{split}
\end{equation*}

Also note that $\Bar{\theta}\geq \gamma_n$ since $\theta_i\geq \gamma_n~\forall~i$. So from assumption $(1)$, we get $$o(1)=\frac{\sqrt{\log(n)}}{n^\alpha\gamma_n\Bar{\theta}}\geq\frac{\sqrt{\log(n)}}{n^\alpha\Bar{\theta}^2}$$. Hence $\frac{1}{\Bar{\theta}^2}\leq o(\frac{n^\alpha}{\sqrt{\log(n)}})$

So $\|\hat{Q}-Q\|_F\leq O\Big(\dfrac{n^{1/2+\alpha}}{\Bar{\theta}^2}\Big)\leq o\Big(\frac{n^{1/2+\alpha}\sqrt{\log(n)}}{n^\alpha}\Big)=o(n^{1/2}\sqrt{\log(n)})\leq o(n^{1/2+\alpha})$ for any $\alpha>0$

Hence proceeding for the probability calculation part as for the equality case, one can show that the test is consistent and hence the theorem follows.

\section*{Proof of Theorem \ref{thm:rdpgscale}}
% \begin{proof}
The proof of this result is almost identical to that of Theorem \ref{thm:rdpgeq}. We only describe here the necessary modifications.
Let $\alpha$ be given and let $\eta = \alpha/4$. From Theorem 2.1 of \cite{tang2017semiparametric}, for sufficiently large $n$, there exists some orthogonal matrices $W_1, W_2 \in \mathbb{O}(d)$ such that, with probability $1-\eta$,
$$\|\Hat{X}_1 - \Tilde{X}_1\|_F \leq C(X_1) + g(X_1)$$
$$\|\Hat{X}_2 - \Tilde{X}_2\|_F \leq C(X_2) + g(X_2)$$
where $\Tilde{X}_1 = X_1 W_{1}$ and $\Tilde{X}_2 = X_2 W_{2}$; $g(X_1) \rightarrow 0$ and $g(X_2) \rightarrow 0$.
Now, for $i = 1,2$
\begin{equation*}
\|\frac{1}{\|\hat{P}_i\|_F}\hat{P}_i - \frac{1}{\|P_1\|_F}P_1\|_F = \|\hat{Z}_i\hat{Z}_i^T - \Tilde{Z}_i\Tilde{Z}_i^T\|_F
\end{equation*}
where $\hat{Z}_i =\frac{\hat{X}_i}{\|\hat{X}_i\|_F}, \Tilde{Z}_i =\frac{\Tilde{X}_i}{\|\Tilde{X}_i\|_F}, \Tilde{X}_i = XW$ for any $W \in \mathbb{O}(d)$

\begin{equation*}
\begin{split}
\|\hat{Z}_i\hat{Z}_i^T - \Tilde{Z}_i\Tilde{Z}_i^T\|_F & \leq \|\hat{Z}_i\|_F\|\hat{Z}_i - \Tilde{Z}_i\|_F + \|\Tilde{Z}_i\|_F\|\hat{Z}_i - \Tilde{Z}_i\|_F\\
&= 2\|\hat{Z}_i - \Tilde{Z}_i\|_F \hspace{5cm} (\text{Since $\|\hat{Z}_i\|_F= \|\Tilde{Z}_i\|_F = 1$})\\
&= 2\|\frac{\hat{X}_i}{\|\hat{X}_i\|_F}-\frac{\Tilde{X}_i}{\|\Tilde{X}_i\|_F}\|_F\\
&\leq 2\frac{\|\Hat{X}_i-\Tilde{X}_i\|_F}{\|\Hat{X}_i\|_F}+2\|\Tilde{X}_i\|_F|\frac{1}{\|\hat{X}_i\|_F}-\frac{1}{\|\Tilde{X}_i\|_F}|\\
&\leq 2\frac{\|\Hat{X}_i-\Tilde{X}_i\|_F}{\|\Hat{X}_i\|_F}+2\frac{|\|\Hat{X}_i\|_F-\|\Tilde{X}_i\|_F|}{\|\Hat{X}_i\|_F}\\
&\leq 4\frac{\|\Hat{X}_i-\Tilde{X}_i\|_F}{\|\Hat{X}_i\|_F}\\
&\leq 4\frac{C(X_i)+g(X_i)}{\|\Hat{X}_i\|_F}
\end{split}
\end{equation*}
with probability at least $1-\eta$.

Under the null hypothesis, $P_1 = cP_2$ for some $c>0$, and by construction, $\rho_1 = c\rho_2$. Then,

$$\|\frac{1}{\rho_1}\hat{P}_1 - \frac{1}{\rho_2}\hat{P}_2\|_F \leq 4\frac{C(X_1)+g(X_1)}{\|\Hat{X}_1\|_F}+4\frac{C(X_2)+g(X_2)}{\|\Hat{X}_2\|_F}$$

Hence, for sufficiently large $n$,
$$\dfrac{\|\frac{1}{\rho_1}\hat{P}_1 - \frac{1}{\rho_2}\hat{P}_2\|_F}{4\sqrt{d\gamma^{-1}(A_1)}/\|\Hat{X}_1\|_F+4\sqrt{d\gamma^{-1}(A_2)}/\|\Hat{X}_2\|_F} \leq 1 + r(\alpha,n)$$
where $r(\alpha,n) \rightarrow 0$ as $n \rightarrow \infty$ for a fixed $\alpha$. Then,

\begin{equation*}
T_n = \|\frac{1}{\rho_1}\hat{P}_1 - \frac{1}{\rho_2}\hat{P}_2\|_F \leq 8(\sqrt{d\gamma^{-1}(A_1)/\rho_1}+\sqrt{d\gamma^{-1}(A_2)/\rho_2})
\end{equation*}

Then for $P_{1}, P_{2}$ satisfying $P_{1} = cP_{2}$, we conclude
$$P(T_n \in R) < \alpha$$
where $R = \{t : t > 8(\sqrt{d\gamma^{-1}(A_1)/\rho_1}+\sqrt{d\gamma^{-1}(A_2)/\rho_2})\}$.\\

The proof of consistency proceeds in an almost identical manner to that in Theorem \ref{thm:rdpgeq}, and we omit the details.

\section{Comparison with current methods}
\label{sec-compare}
% \ssg{SHORTEN AND MOVE TO SUPP}.
Here, we compare the proposed methodology with the current state-of-the-art methods from \cite{tang2017semiparametric} and \cite{ghoshdastidar2018practical}.
In \cite{tang2017semiparametric}, the authors proposed methods for both equality and scaling problems, whereas \cite{ghoshdastidar2018practical} and \cite{levin2017central} only considered the equality problem.
This section will focus on the equality problem for brevity, as the insights drawn therein have analogous implications for the scaling test. 
The ensuing comparison draws from theoretical and conceptual considerations, while a more in-depth simulation-based assessment is in Section \ref{sec-sims}.

\begin{itemize}[leftmargin=*]
\item In \cite{tang2017semiparametric}, the authors studied the problem under the RDPG model, proposing 
\begin{equation}
T_\text{ase}(A_1, A_2) = \min_{W \in \mathcal{O}_n}||\hat{X}_1-\hat{X}_2W||_F,
\label{eq-tase}
\end{equation}
as the test statistic.
Here $\hat{X}_1$ and $\hat{X}_2$ are estimated from $A_1$ and $A_2$ by using ASE and $\mathcal{O}_n$ is the set of all $n$-by-$n$ orthogonal matrices.
Two independent sets of parametric bootstrap iterations are carried out, using the generative models $\hat{X}_1\hat{X}_1^T$ and $\hat{X}_2\hat{X}_2^T$, and the p-value is defined as the larger of the two bootstrap p-values.

The proposed methodology offers several theoretical and computational advantages compared to the $T_\text{ase}$ method:
(i) taking the larger of the two p-values makes the $T_\text{ase}$ test overly conservative, with a nominal type-I error of $\alpha^2$ instead of $\alpha$. In contrast, the proposed test has the correct type-I error calibration;
(ii) while the $T_\text{ase}$ test is confined to the RDPG model, the proposed test is versatile as it applies to various statistical models; and
(iii) from a computational perspective, the proposed method is cheaper since it requires a single set of bootstrap iterations, compared to two sets for the $T_\text{ase}$ test.

\item In \cite{levin2017central}, the authors proposed the Omnibus embedding method.
Given two  adjacency matrices $A_1$ and $A_2$ , the \textit{omnibus} matrix is given by
$$
M = \begin{pmatrix}
    A_1 & \frac{A_1 + A_2}{2} \\
     \frac{A_1 + A_2}{2} & A_2.
\end{pmatrix}
$$
Next, let $S_M$ represent the diagonal matrix of the top $d$ eigenvalues of $M$, and let $U_M$ be the $(m+n) \times d$ matrix of their eigenvalues.
The omnibus embedding is defined as $\hat{X}_M U_M S_M^{1/2}$, and the test statistic is given by
$$
T_\text{omni}(A_1, A_2) = ||\hat{X}_{M1} - \hat{X}_{M2}||_F,
$$
where $\hat{X}_{M1}$ is the $n \times d$ matrix consisting of rows $1, \ldots, n$ of $\hat{X}_M$ and $\hat{X}_{M2}$ is the $n \times d$ matrix consisting of rows $n+1, \ldots, 2n$ of $\hat{X}_M$.

Note that $T_\text{omni}$ further exacerbates the computational issues described in the context of $T_\text{ase}$, since it requires spectral decomposition of a $2n \times 2n$ matrix.
Furthermore, \cite{levin2017central} did not provide any theoretical consistency results for the test. 

\item In \cite{ghoshdastidar2018practical}, the authors studied the problem under the generic inhomogeneous Erd\"{o}s-Ren\'{y}i model.
Given $A_1 \sim P_1$ and $A_2 \sim P_2$, 
they propose estimating $\hat{P}_1$ and $\hat{P}_2$ as stochastic blockmodel approximations with $r$ communities \citep{lovasz2012large, olhede2014network,zhang2017estimating}.
Then, they consider the differenced adjacency matrix, $C$, and construct a scaled version given by
\[
\Tilde{C}(i,j) = \frac{A_1(i,j) - A_2(i,j)}
{\sqrt{(n-1)\left(\hat{P}_1(i,j) (1-\hat{P}_1(i,j) +
\hat{P}_2(i,j) (1-\hat{P}_2(i,j) \right)}}.
\]
The test statistic is given by
\begin{equation}
T_\text{eig}(A_1, A_2) = n^{2/3}(||\Tilde{C}|| -2),
\label{eq-teig}
\end{equation}
where $||\cdot||$ denotes the spectral norm of a matrix.
The test is rejected for large values of $T_\text{eig}$ with p-values computed from the
standard Tracy-Widom distribution \citep{tracy1996orthogonal}.

The $T_\text{eig}$ test has two advantages: first, it is computationally efficient since no bootstrapping is needed, and second, it is highly versatile as it does not require model specification.
However, these advantages come with some associated drawbacks.
For the $T_\text{eig}$ test the number of communities, $r$, needs to be provided as an input to the algorithm.
The authors did not offer a strategy or recommendation for determining this crucial tuning parameter, and, in practice, the results vary quite a bit for different choices of $r$.
Furthermore, the asymptotic thresholds do not work very well for finite samples.
Finally, the largest singular value may not be sufficiently sensitive to the difference between the null model and the alternative model, making their test less effective.
These issues are empirically demonstrated in the rest of this section.

\end{itemize}

\subsection{Distributional comparison of test statistics}
% We now discuss an important point that relates to the comparison of $T_\text{frob}$ with both $T_\text{ase}$ and $T_\text{eig}$.
% The general strategy of statistical inference consists of computing the observed value of a test statistic from available data, determining (or estimating) a suitable threshold based on the sampling distribution of this test statistic under the null hypothesis, and rejecting the null hypothesis if the observed value of the test statistic exceeds this threshold.
Consider a point null and a point alternative, and let $F_0$ and $F_1$ be the corresponding distributions of some test statistic.
Then, the effectiveness of the test statistic is intrinsically linked to the separation between $F_1$ and $F_0$.
If the separation between $F_0$ and $F_1$ for one test statistic is greater than that for another test statistic, then the former is more likely to distinguish the null and the alternative.
This gives us a natural means to compare different test statistics, by fixing a null model and an alternative model and comparing the separation between $F_0$ and $F_1$.
Based on this notion, we carried out a comparative analysis of the test statistics $T_\text{frob}$ (the proposed method), $T_\text{ase}$, and $T_\text{eig}$ using simulated network data.
We deliberately used model configurations from \cite{tang2017semiparametric} and \cite{ghoshdastidar2018practical} for this analysis.
We leave out the $T_\text{omni}$ method from this analysis since \citep{levin2017central} did not provide theoretical consistency results for this test.
% to illustrate the results from their methods.
% It was observed that the null and alternative sampling distributions of $T_\text{frob}$ are much better separated than those of $T_\text{ase}$ and $T_\text{eig}$, which demonstrates the advantage of our method.
% Detailed results are reported in Section 4 together with other simulation experiments.
% We note that this notion of comparing test statistics is somewhat related to the formal and classical concepts of asymptotic relative efficiency named after Pitman and Bahadur \citep{wieand1976condition,kremer1979approximate,rothe1981some}.
% In this paper, we have  carried out an empirical analysis for comparing the test statistics instead of taking the more rigorous approach of theoretically deriving their asymptotic relative efficiency.
% We consider this theoretical endeavour as an important direction of future research.
% \subsection{Distributional comparison of $T_\text{frob}, T_\text{ase},$ and $T_\text{eig}$}
% We first study the null sampling distribution ($F_0$) and the alternative sampling distribution ($F_1$) of the three test statistics $T_\text{frob}, T_\text{ase},$ and $T_\text{eig}$. 
% Following the earlier discussion in Section \ref{sec-compare}, we would prefer the test statistic which, under the same null model and alternative model, has the largest separation between $F_0$ and $F_1$.
% To be fair to $T_\text{ase}$ and $T_\text{eig}$ we carried out this study under the modeling framework that was used in both \cite{tang2017semiparametric} and \cite{ghoshdastidar2018practical}.
Let $A_1 \sim P_1$ and $A_2 \sim P_2$ where $P_1, P_2$ are two-community stochastic blockmodels with 
$P_1(i,j) = B(c_i, c_j)$, and $P_2(i,j) = B_\epsilon(c_i, c_j)$,
where
$$B = \begin{pmatrix} 0.5  & 0.2\\ 0.2 & 0.5  \\ \end{pmatrix}, \; \text{ and } 
B_\epsilon = \begin{pmatrix} 0.5+\epsilon  & 0.2\\ 0.2 & 0.5+\epsilon  \\ \end{pmatrix}.$$
The community membership vector $(c_1, \ldots, c_n) \in \{1,2\}^n$ was sampled from the multinomial distribution with $\pi = (0.4, 0.6)$.
Thus, $P_1$ and $P_2$ share the same community assignment,  but have different block probability matrices when $\epsilon \neq 0$.
We used model parameters $\epsilon = 0, 0.5, 0.1$ along with $n=100, 200, 300.$
Note that the null hypothesis $H_0: P_1=P_2$ is satisfied when $\epsilon=0$, and $H_1$ is satisfied when $\epsilon = 0.05, 0.1$.
For each combination of $\epsilon$ and $n$, we carried out 100,000 Monte Carlo simulations to ensure that the empirical distributions are accurate proxies for the nominal distributions.
% Each simulation consists of randomly generating $A_1 \sim P_1$ and $A_2 \sim P_2$, and computing the test statistics $T_\text{frob}(A_1, A_2)$, $T_\text{ase}(A_1, A_2)$, and $T_\text{eig}(A_1, A_2)$.
% We used a large number of Monte Carlo simulations to ensure that the empirical distributions are accurate proxies for the nominal distributions.
% For each test statistic and for each value of $n$, we constructed the null sampling distribution or $F_0$ under $\epsilon = 0$, and the alternative sampling distribution or $F_1$ under $\epsilon = 0.05$ and $\epsilon = 0.1$.
% If $F_0$ and $F_1$ are well separated, then the test statistic can be effective in telling apart the alternative from the null.

\begin{figure}[ht]
	\centering
    \includegraphics[trim={0cm 1cm 1cm 1cm},clip,width=0.48\textwidth]{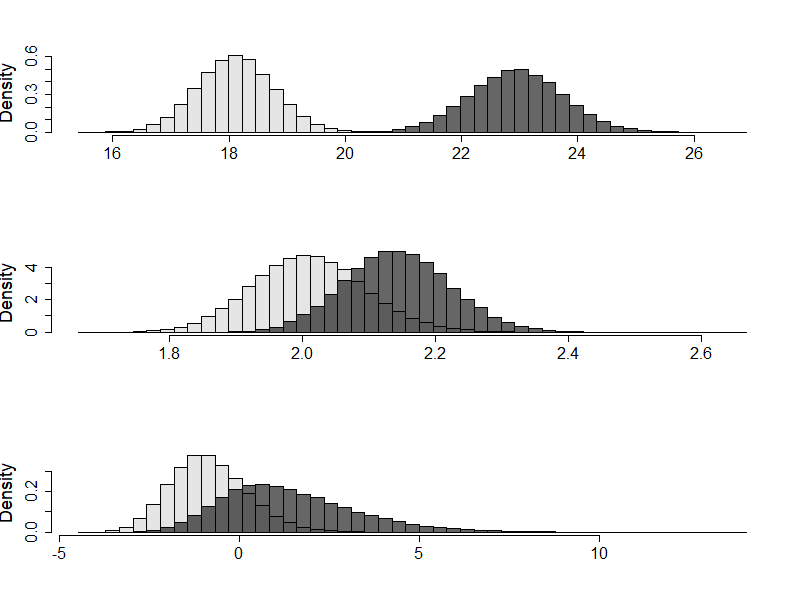}
	\includegraphics[trim={0cm 1cm 1cm 1cm},clip,width=0.48\textwidth]{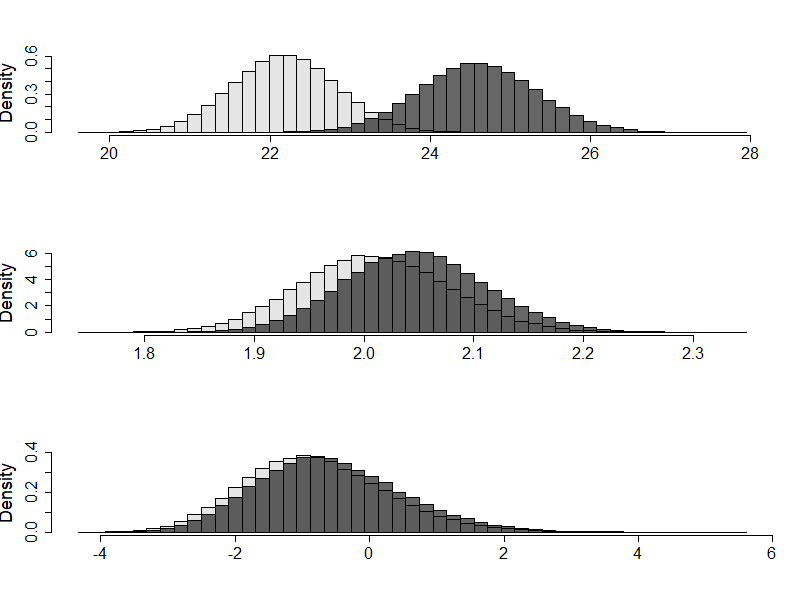}
	{ \caption{\small Histograms of $T_\text{frob}$ (top), $T_\text{ase}$ (middle), and $T_\text{eig}$ (bottom) for (left) $\epsilon = 0$ vs. $\epsilon = 0.05$ with $n=300$, and (right) $\epsilon = 0$ vs. $\epsilon = 0.1$ with $n=200$.
	The light-colored histogram shows the null sampling distribution or $F_0$ ($\epsilon = 0$) and the dark-colored histogram shows the alternative sampling distribution or $F_1$ ($\epsilon >0$).
	We observe that $F_1$ is well separated from $F_0$ for $T_\text{frob}$ but not for $T_\text{ase}$ or $T_\text{eig}$, which means the proposed test is more effective than the current tests.
	}\label{Scenab}}
\end{figure}

In the interest of space, out of the six scenarios studied, we report results from two scenarios in Figure \ref{Scenab}: (a) $\epsilon = 0$ vs. $\epsilon = 0.05$ with $n=300$, and (b) $\epsilon = 0$ vs. $\epsilon = 0.1$ with $n=200$.
We observe that under both scenarios, $F_0$ and $F_1$ for $T_\text{frob}$ (the proposed method) are much better separated than $T_\text{ase}$ and $T_\text{eig}$.
% Distributions (both $F_0$ and $F_1$) for $T_\text{frob}$, $T_\text{ase}$, and $T_\text{eig}$ for these two scenarios are reported in Figures \ref{Scena} and \ref{Scenb}.
% Under scenario (a), Figure \ref{Scena}  shows that $F_0$ and $F_1$ are practically conjoined for both $T_\text{ase}$ and $T_\text{eig}$, which implies that neither test statistic can effectively distinguish between $H_0$ and $H_1$.
% In contrast, $F_0$ and $F_1$ for $T_\text{frob}$ are reasonably well separated which implies it can effectively distinguish between  $H_0$ and $H_1$.
% Under scenario (b), $F_0$ and $F_1$ are slightly separated for $T_\text{ase}$ and $T_\text{eig}$, but $F_0$ and $F_1$ for $T_\text{frob}$ are much more strongly separated.
% Under both scenarios, $F_0$ and $F_1$ for $T_\text{frob}$ are much better separated.
This implies $T_\text{frob}$ is much more sensitive to the difference between $H_0$ and $H_1$ than $T_\text{ase}$ and $T_\text{eig}$, and therefore more effective at resolving the hypothesis test.
We note that a more formal and rigorous comparison between the test statistics can be carried out by deriving the asymptotic relative efficiencies, and we consider this as a future research direction \citep{rothe1981some}.

\section{Simulation studies}
\label{sec:simsupp}

\subsection{Sparse RDPG }
\begin{figure}[h!]
\centering
\includegraphics[width=\textwidth]{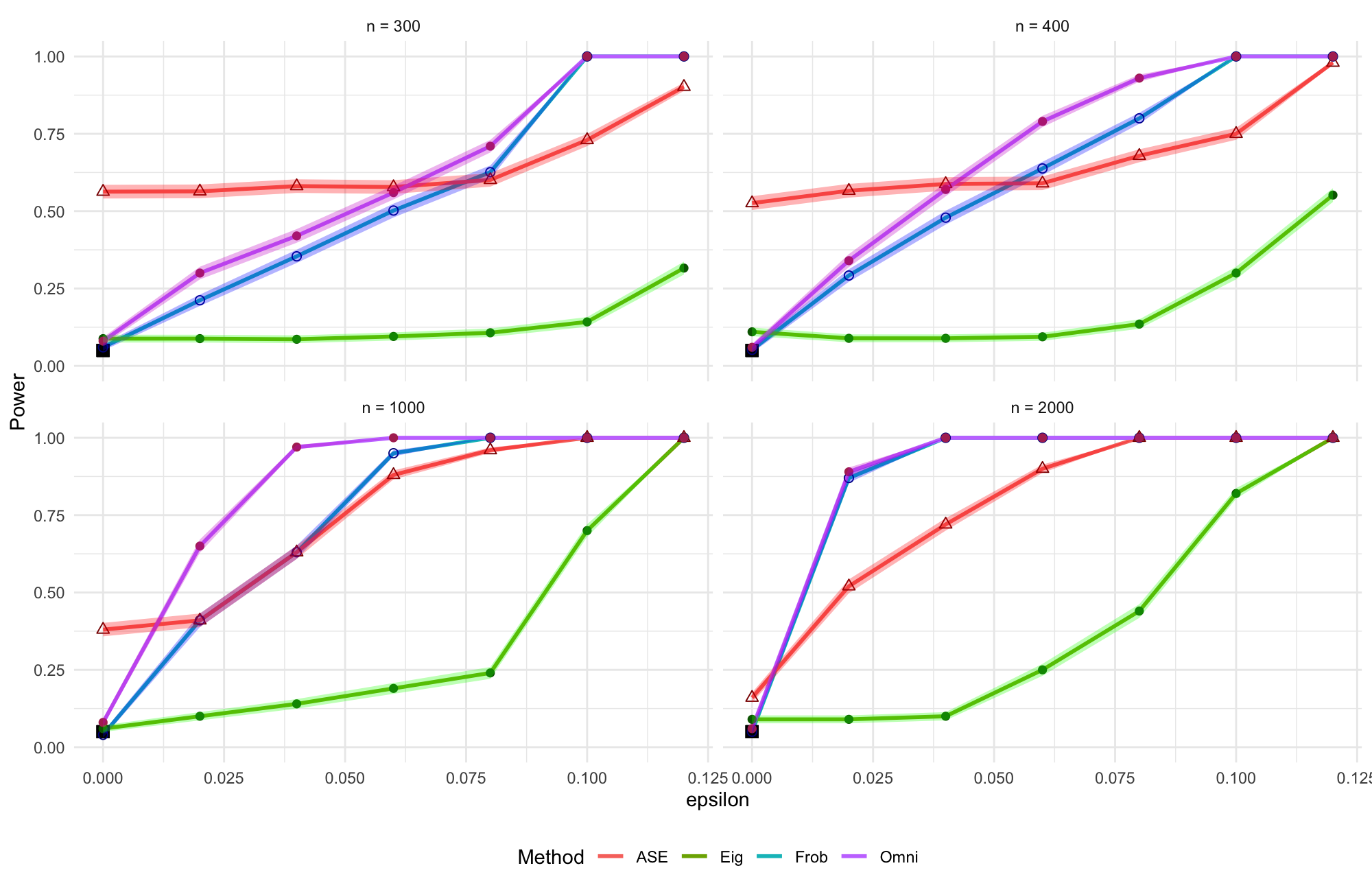}
\caption{\small Rejection rates for sparse RDPG test of equality.
\label{Fin3}}
\end{figure}

% \noindent \textbf{Case 3 (sparse networks):} 
In this simulation, we consider a comparison between sparse networks. The probability matrix $P_1$ and $P_2$ is defined similarly as in Case 1, and then we set $P_1^* = n^{-0.25}P_1$ and $P_2^* = n^{-0.25}P_2$. We compute the rejection rates based on $P_1^*$ and $P_2^*$ for the network sizes $n=300, 400, 1000, 2000$ and $\epsilon = 0$, 0.025, 0.05, 0.075, 0.1, 0.2, and the results are displayed in Figure \ref{Fin3}. We can see that the proposed test $T_\text{frob}$ performs better than $T_\text{ase}$ or $T_\text{eig}$. The method $T_\text{omni}$ works better than $T_\text{frob}$ for smaller networks, but both the $T_\text{frob}$ and $T_\text{omni}$ works similarly for larger networks.

 \subsection{Stochastic Block Model}
\begin{figure}[h]
\centering
\includegraphics[width=0.45\textwidth]{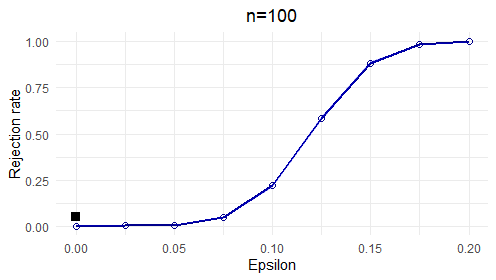}
\includegraphics[width=0.45\textwidth]{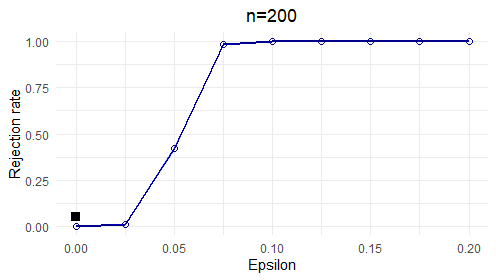}\\
\includegraphics[width=0.45\textwidth]{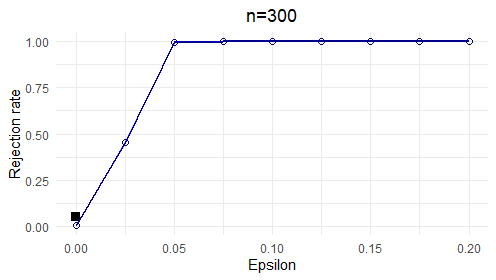}
\includegraphics[width=0.45\textwidth]{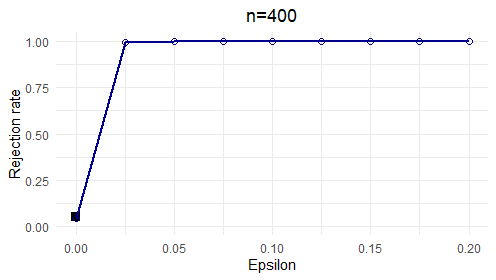}
\caption{\small SBM equality: Rejection rates for deviating alternatives.
\label{plot_sbm}}
\end{figure}

 Under the SBM, we set $P_1 = \omega_{c_i c_j}$ and $P_2 = \omega^{(\epsilon)}_{c_i c_j}$ for the test of equality, where
 $$
 \omega = \begin{pmatrix}
0.4 & 0.2\\
0.2 & 0.4
\end{pmatrix}
\text{ and }
\omega^{(\epsilon)} = \begin{pmatrix}
0.4 & 0.2\\
0.2 & 0.4 + \epsilon
\end{pmatrix},
 $$
 with $n=100, 200, 300, 400$ and $K=2$ communities of equal size.
 We used   $\epsilon = 0, 0.025, 0.05, \ldots, 0.2$, where $\epsilon = 0$ satisfies $H_0: P_1 = P_2$ and the non-zero values of $\epsilon$ satisfies $H_1: P_1 \neq P_2$.
 The results are plotted in Figure \ref{plot_sbm}.
 We observe that the proposed test performs well with low Type-1 error and power increasing to 1 with increasing $\epsilon$ and $n$. 
 Since $T_\text{eig}$ had very high rejection rates under the Chung-Lu model (Figure \ref{plot_chunglu}), we skipped it for the SBM and subsequent models.

 For the scaling case, we used $P_2 = c \times P_1$ with $c=0.5,0.7,0.75,0.8,0.9$ under the null model, and to configure the alternative scenario we used $P_2 = \omega^{(\epsilon)}_{c_i c_j}$ with $\epsilon = 0.2$.
 The results are reported in Table \ref{tab_sbm}, where we again find that the proposed test $T_\text{scale}$ performs well with low Type-1 errors and high power.
 We note one exception for $n=100, c=0.5$ where the type-1 error is unusually high.
 We conjecture that this is due to low sample size, as the type-1 error for higher values of $n$ are zero for $c=0.5$.

\begin{table}[ht]
    \begin{center}
    \begin{tabular}{|c||c|c|c|c|c|c|}
    \hline
    & \multicolumn{5}{c|}{$H_0$ is true} & 
    {$H_1$ is true}\\\hline
     & {$ P_2 = 0.5  P_1$} & {$ P_2 = 0.7  P_1$} & {$ P_2 = 0.75 P_1$} &  {$ P_2 = 0.8  P_1$} & {$ P_2 = 0.9  P_1$} & {$ P_2 \neq c P_1$}\\
    \hline
     $n$ & $T_\text{scale}$ & $T_\text{scale}$ & $T_\text{scale}$ & $T_\text{scale}$ & $T_\text{scale}$ & $T_\text{scale}$\\
     \hline
     100 & $28.30$ & $2.25$ & $0.95$ & $0.30$ & $0.00$ & $100$ \\
    \hline
    200 & $0.00$ & $0.00$ & $0.00$ & $0.00$ & $0.00$ & $100$ \\
    \hline
    300 & $0.00$ & $0.00$ & $0.00$ & $0.00$ & $0.00$ & $100$ \\
    \hline
    400 & $0.00$ & $0.00$ & $0.10$ & $0.05$ & $3.45$ & $1.00$ \\
    \hline
\end{tabular}
\end{center}

    \caption{\small SBM scaling case: Rejection rates (in percentage) from $T_\text{scale}$ using $B=200$ bootstrap iterations and averaged over 2000 Monte Carlo simulations.
        \label{tab_sbm}}
\end{table}

 \subsection{Degree Corrected Block Model}
\begin{figure}[h]
\centering
\includegraphics[width=\textwidth]{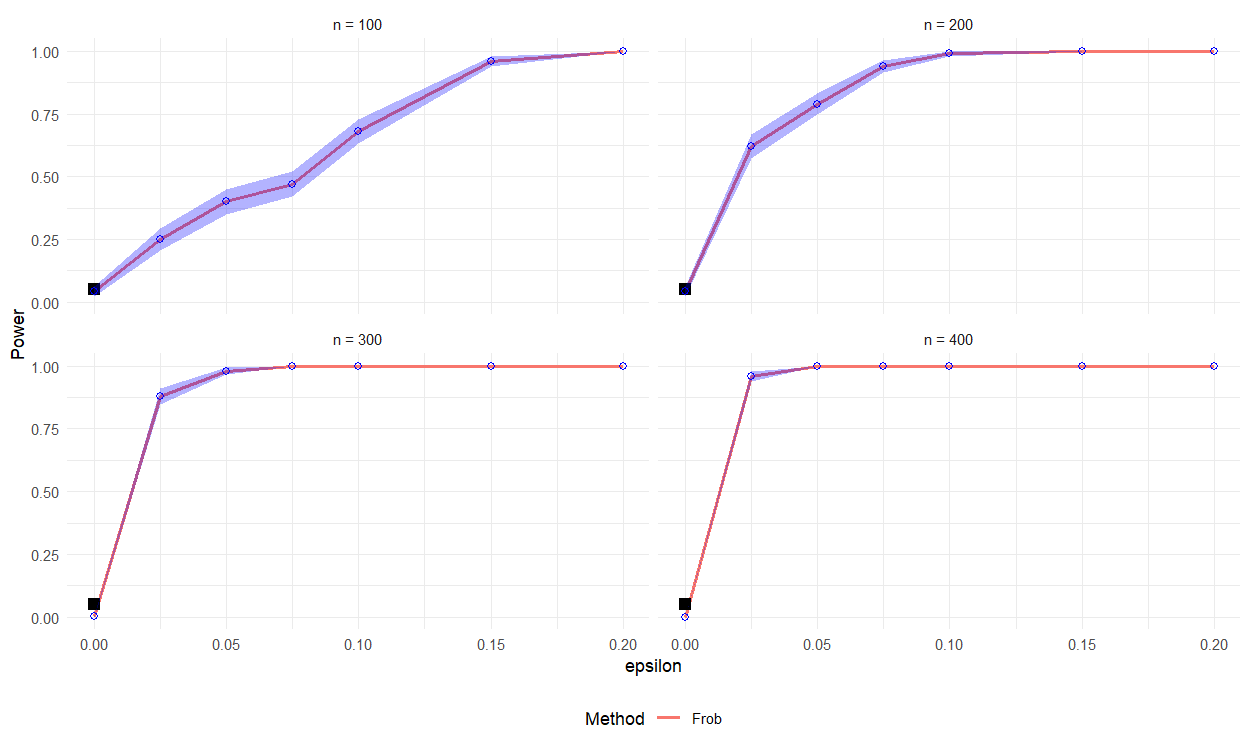}
\caption{\small DCBM equality: Rejection rates for deviating alternatives.
\label{plot_dcbm}}
\end{figure}

Under the DCBM, we used 
\[
\omega \propto \begin{pmatrix}
4 & 2 & 1\\
2 & 4 & 1 \\
1 & 1 & 4 
\end{pmatrix},
\]
$K=3$ unbalanced communities with the community assignment vector sampled from a multinomial distribution with $\pi = (0.25, 0.25, 0.5)$, 
and the resultant matrix of $P(i,j)$'s was then scaled to ensure that the expected network density is $\delta = 0.1$.
We generated $\theta_i$ for $P_1$ from the $Beta(1,5)$ distribution (as under the Chung Lu model). 
Under the DCBM, $P_1$ and $P_2$ can be configured to be different in several ways --- by changing the community structure, by changing the $\omega$ matrix, by changing the degree parameters $\theta$, or a combination of all three.
In this study we kept the community structure $c$ and the block matrix $\omega$ unchanged between $P_1$ and $P_2$, changing only the generation of $P_2$ as $P_2(i,j)=(\theta_i+\epsilon)(\theta_j+\epsilon)$ (as under Chung Lu model) to see how the effect of change in $\epsilon$ in rejection rate calculation. 
For the scaling case, we used $P_2 = c \times P_1$ with $c=0.5,0.7,0.75,0.8,0.9$ under the null model, and to configure the alternative scenario we generated another separate set of parameters $\eta_i\sim\text{Beta}(a=4,b=3)$ and used $P_2(i,j)=\eta_i\eta_j$ (similar to the Chung Lu setup).
The results for the equality case are plotted in Figure \ref{plot_dcbm} and those for scaling are reported in Table \ref{tab_dcbm}. For the scaling case in particular, we observe that the power was 1 but the Type-I error showed a decreasing pattern as $n$ increases.
We plan to analyze this more closely in future research.

% \textbf{Why are these results so different from Somnath?}

\begin{table}[ht]
    \begin{center}
    \begin{tabular}{|c||c|c|c|c|c|c|}
    \hline
    & \multicolumn{5}{c|}{$H_0$ is true} & 
    {$H_1$ is true}\\\hline
     & {$ P_2 = 0.5  P_1$} & {$ P_2 = 0.7  P_1$} & {$ P_2 = 0.75 P_1$} &  {$ P_2 = 0.8  P_1$} & {$ P_2 = 0.9  P_1$} & {$ P_2 \neq c P_1$}\\
    \hline
     $n$ & $T_\text{scale}$ & $T_\text{scale}$ & $T_\text{scale}$ & $T_\text{scale}$ & $T_\text{scale}$ & $T_\text{scale}$\\
     \hline
     100 & $0.5$ & $0.7$ & $0.4$ & $1.2$ & $1.7$ & $100$ \\
    \hline
    200 & $0.8$ & $1.1$& $0.4$ & $1.2$ & $1.9$ & $100$ \\
    \hline
    300 & $0.5$ & $0.2$ & $0.1$ & $0.3$ & $0.2$ & $100$ \\
    \hline
    400 & $0.1$ & $0.0$ & $0.0$ & $0.1$ & $0.0$ & $100$ \\
    \hline
\end{tabular}
\end{center}

    \caption{\small DCBM scaling case: Rejection rates (in percentage) from $T_\text{scale}$ using $B=200$ bootstrap iterations and averaged over 2000 Monte Carlo simulations.
        \label{tab_dcbm}}
\end{table}

\subsection{Model mis-specification}
%$P_1$ is RDPG and $P_2$ is LSM.
In practice, it might not be known which model class the networks were generated from, and there could be model misspecification. In this simulation study, we consider two cases: when we mis-specify both models as RDPG (and use ASE estimator) and when we  mis-specify both models as LSM (and use the LSM estimator).

\subsubsection{Cross-simulation for RDPG}
\begin{figure}[h!]
\centering
\includegraphics[width=\textwidth]{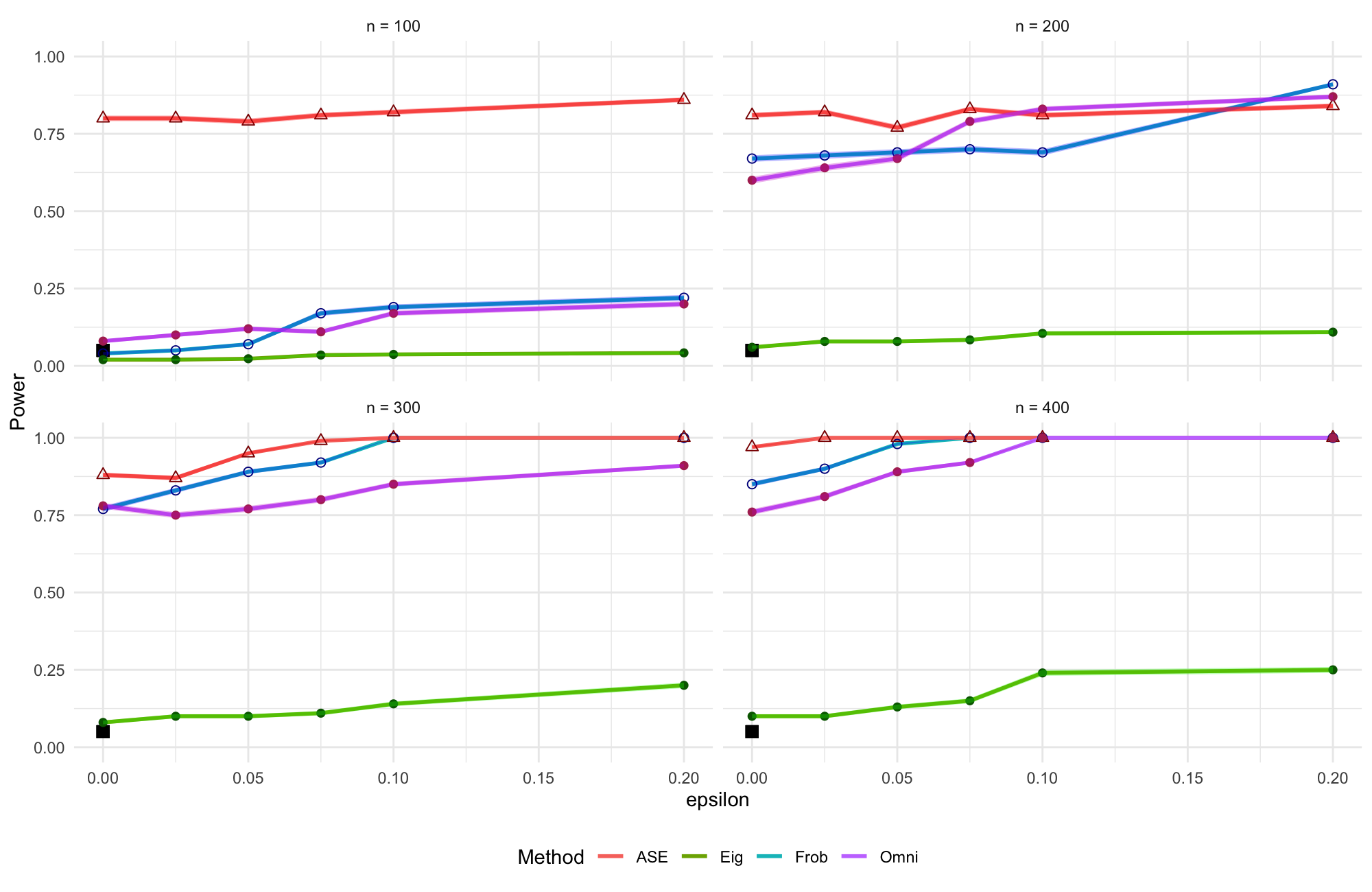}
\caption{\small Model mis-specification: Rejection rates for deviating alternatives.
\label{sim6}}
\end{figure}

In this simulation, we use the latent distance model of \cite{hoff2002latent} to generate the networks, and use RDPG based estimation methods for testing. We used $d=3$, $\alpha = 3$, and sampled the latent positions $z_1, \ldots, z_n \sim N(\mathbf{0}, \mathbf{I})$ independently for $P_1$.
For the equality case, we set $P_2=P_1$ under the null.
To configure the alternative scenario for the equality case, we kept $d$ unchanged, used $\alpha-\epsilon$ instead of $\alpha$, and sampled a second set of latent positions $z_1, \ldots, z_n \sim N(\mathbf{0}, \mathbf{I})$ independently for $P_2$. The results based on 2000 Monte Carlo simulations are displayed in Figure \ref{sim6}.
We observe that none of the candidate methods work well, including the methods proposed in this paper.

\subsubsection{Cross-simulation for LSM}
In this simulation, we consider the networks being generated from two RDPG models. Let $A_1 \sim \text{RDPG}(X_1)$ and $A_2 \sim \text{RDPG}(X_2)$, where $X_1$ and $X_2$ are two latent matrices of dimension $n \times 2$. The rows of $X_1$ and $X_2$ are generated by sampling with replacement from the rows of $M_1$ and $M_2$ respectively with probability vector $\pi$, where $$M_1 = \begin{pmatrix} 0.6 & -0.4\\ 0.6 & 0.4 \end{pmatrix}; \; M_2 = \begin{pmatrix} 0.6 & -0.4 - \epsilon\\ 0.6 & 0.4 + \epsilon \end{pmatrix}; \; \pi  = \begin{pmatrix} 0.4\\ 0.6\end{pmatrix}.$$ The null hypothesis $H_0 : P_1 = P_2$ holds true when $\epsilon = 0$, and the alternative $H_1 : P_1 \neq P_2$ holds true when $\epsilon>0$. We will use LSM approach to fit the model and do the hypothesis test based on that estimate.

\begin{figure}[ht]
\centering
\includegraphics[width=\textwidth]{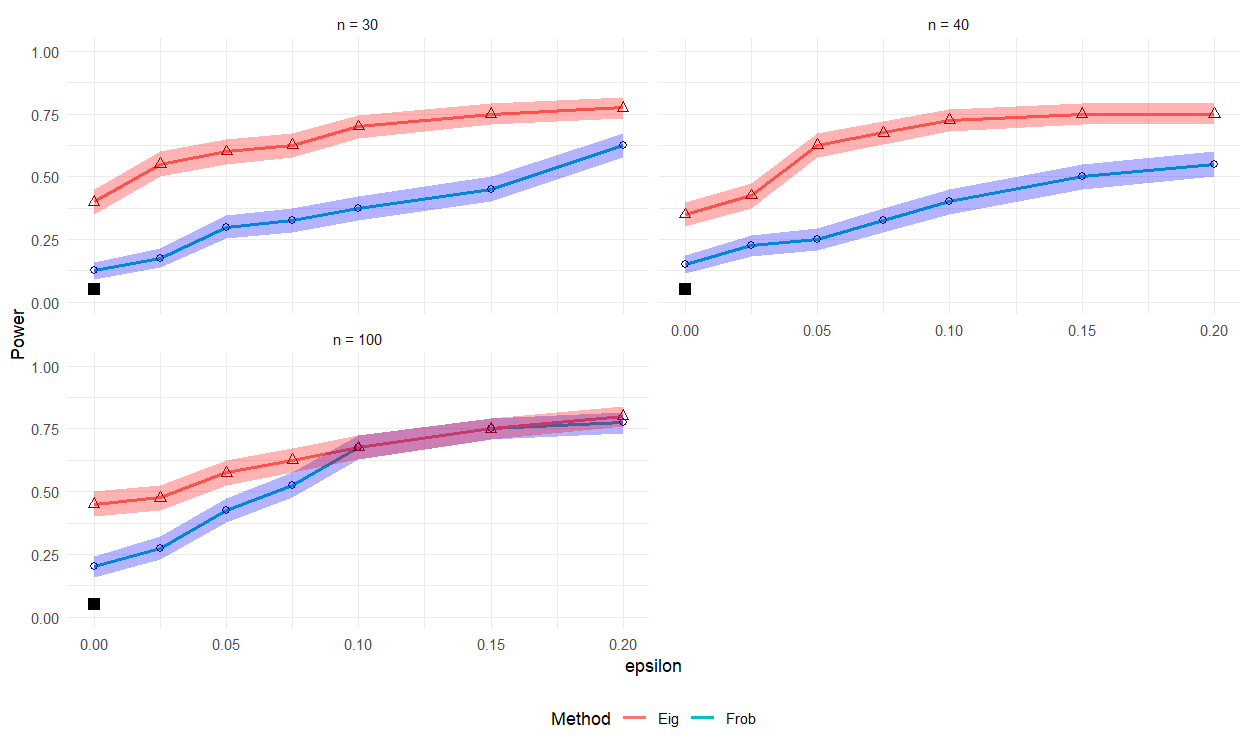}
\caption{\small LSM simulation Rejection rates for deviating alternatives for model mis-specification using both $T_\text{eig}$ and $T_\text{frob}$ using 50 Monte Carlo replicates.
\label{plot:crosssim_LSM}}
\end{figure}

Figure \ref{plot:crosssim_LSM} shows that the power of the test using mis-specified LSM estimate has an increasing trend with $\epsilon$, but has much less power than using RDPG estimation here. Also, to compare between the two different test statistics, $T_\text{frob}$ works better than $T_\text{eig}$ as rejection rates are higher when using $T_\text{eig}$.

% \clearpage

% \clearpage

\end{document}